\documentclass{article}
\usepackage{amssymb}
\usepackage{amsmath}
\usepackage{amsfonts}
\usepackage{graphicx}
\usepackage{color}
\let\UnmodifSec=\section
\renewcommand{\section}{\setcounter{equation}{0}\UnmodifSec}

\newtheorem{definition}{Definition}[section]
\newtheorem{lemma}{Lemma}[section]
\newtheorem{prop}{Proposition}[section]
\newtheorem{proposition}{Property}[section]

\def\dsphere{S^{d-1}}
\def\xx{{\boldsymbol\xi}}

\def\Wl{W_{\lambda}}
\def\wl{w_{\lambda}}
\def\WWl{{\cal W}_{\lambda}}
\def\dimension{\kappa}
\def\p{\dimension}
\def\formdif{{\alpha}}

\def\bC{{\bf C}}
\def\bR{{\bf R}}

\def\bU{{\bf U}}
\def\Im{\mathop{\rm Im}\nolimits}
\def\Re{\mathop{\rm Re}\nolimits}

\def\ch{\mathop{\rm ch}\nolimits}
\def\sh{\mathop{\rm sh}\nolimits}
\def\sign{\mathop{\rm sign}\nolimits}

\def\CC{{\cal C}}
\def\DD{{\cal D}}
\def\EE{{\cal E}}
\def\FF{{\cal F}}

\def\HH{{\cal H}}
\def\II{{\cal I}}
\def\JJ{{\cal J}}
\def\KK{{\cal K}}
\def\LL{{\cal L}}
\def\MM{{\cal M}}
\def\NN{{\cal N}}

\def\SS{{\cal S}}
\def\TT{{\cal T}}

\def\VV{{\cal V}}
\def\WW{{\cal W}}

\def\wh{\widehat}
\def\wt{\widetilde}
\def\ovl{\overline}

\def \vhi{\varphi}
\def \veps{\varepsilon}

\def\HB{\hfill\break}
\def\half{{\scriptstyle{1 \over 2}}}
\def\interior#1{\setbox1=\hbox{$#1$}\rlap{$#1$}\kern0.4\wd1\raise1.1\ht1%
\hbox{$\scriptstyle \circ$}}
\def\bydef{\mathrel{\buildrel \hbox{\scriptsize \rm def} \over =}}
\def\boxit#1#2{\setbox1=\hbox{\kern#1{#2}\kern#1}%
\dimen1=\ht1 \advance \dimen1 by #1 \dimen2=\dp1 \advance \dimen2 by #1
\setbox1=\hbox{\vrule height\dimen1 depth\dimen2\box1\vrule}%
\setbox1=\vbox{\hrule\box1\hrule}%
\advance \dimen1 by .4pt \ht1=\dimen1 \advance \dimen2 by .4pt \dp1=\dimen2
\box1\relax}
\def\endprf{\raise .5ex\hbox{\boxit{2pt}{\ }}}
\def\Ylm{{ Y}_{lM}}

\def\ifundefined#1{\expandafter\ifx\csname#1\endcsname\relax}

\def\beq{\begin{equation}}
\def\endq{\end{equation}}
\def\beqa{\begin{eqnarray}}
\def\endqa{\end{eqnarray}}
\def\x{{\bf x}}
\def\y{{\bf y}}

\def\P{{\bf P}}
\def\Q{{\bf Q}}

\def\dil{{\rm D\,}}
\def\Y{{\bf Y}}

\def\j{{lM}}

\def\twofcn{{F}}
\def\hd{{\underline{\HH^{d+1,n}}}}

\renewcommand{\cosh}{\ch}\renewcommand{\sinh}{\sh}

\def\dateline{\today}

\title{de Sitter tachyons and related topics}
\author{Henri Epstein$^{(a)}$ and Ugo Moschella$^{(a,b)}$\\ \\ $^{(a)}$IHES, Bures-sur-Yvette, France\\
$^{(b)}$DiSat, Universit\`a dell'Insubria, Como, and INFN, Sezione di Milano, Italia}
\date{\dateline}

\begin{document}
\maketitle

\abstract{We present a complete study of a family of tachyonic scalar fields living on the de Sitter universe. 
We show that for an infinite set of discrete values of the negative squared mass the fields exhibit a gauge symmetry and 
there exists for them a fully acceptable local and covariant quantization similar to the Feynman-Gupta-Bleuler quantization of free QED. 
For general negative squares masses we also construct positive quantization where the de Sitter symmetry is spontaneously broken.
{We discuss the sense in which the two quantizations may be considered physically inequivalent even when there is a Lorentz invariant subspace in the second one.}}

%%%%%%%%%%%%%%%%%%%%%%%%%%%%%%%%%%%%%%%

\section{Introduction}
The de Sitter universe occupies  in contemporary cosmology a place of the greatest importance. The astronomical observations  of the last fifteen years have indeed led to the surprising conclusion that the geometry of empty spacetime 
(namely of spacetime deprived of its matter and radiation content) is  more akin to the de Sitter geometry rather than the Minkowski one. 
Furthermore, if the description of the cosmic fluid provided by the standard $\Lambda$CDM cosmological model is correct, 
matter and radiation must in the future progressively
thin out and eventually vanish  letting the cosmological
constant term alone survive. Therefore, the de Sitter geometry is also the one  
which the universe approaches asymptotically.  This change of paradigm has 
triggered a renewed interest in the study of de Sitter quantum field theory, 
both on the foundational side and in view of its applications to cosmology.

In this paper  we present a complete study of a  family of scalar tachyonic quantum fields whose existence was announced in  \cite{yaa}.
They are linear Klein-Gordon quantum fields on the de Sitter manifold whose squared masses are negative and take an infinite set of discrete values as follows:
\begin{equation}
m^2_n = - n(n+d-1), \ \ \ n=0,1,2,\ldots \label{discrete}
\end{equation}
here the letter $d$ denotes the  dimension of de Sitter space-time.

These models are related to the discrete series of representations of the 
de Sitter group and they have largely stayed uncovered until now. Already 
in the early times of de Sitter QFT \cite{borner} the possibility of having 
quantum fields associated to the discrete series had been evoked but this prospect 
was immediately set aside \cite{borner}  because of the alleged problems with 
causality of the corresponding Green functions. 
Later on however, at least two models belonging to the family (\ref{discrete}) 
were drawn to the general attention: the first and more important is the 
massless scalar field \cite{allenfolacci}. This model has been and still is 
the object of a considerable amount of work and plays a crucial role in the 
perturbative calculations of  cosmological power spectra; the general 
understanding is that the massless scalar field is substantially different 
from all other massive fields  because of the necessity of breaking the 
de Sitter symmetry while performing its (canonical) quantization 
\cite{allenfolacci,ford,starobinski,mottola4}.  The second, perhaps less known, model has emerged in 
the study of the conformal sector of quantum gravity \cite{anto,antoniadis,folacci} 
and corresponds to the value $n=1$ (for recent works on the subject see \cite{higuchi,wood1,morrison,wood2} and references therein). Also in this case canonical quantization \cite{folacci} 
points towards the necessity of breaking  the de Sitter symmetry \cite{folacci,wood2, wood3}.

Here we we will show that both the above examples belong to the infinite 
family of QFT's labeled by the masses (\ref{discrete}). 
All those models share an underlying gauge invariance 
related to the existence of polynomial regular solutions of the 
field equations. 
Because of the existence of such gauge invariance, the properties of 
local commutativity, de Sitter invariance and positive semi-definiteness 
\cite{sw,reedsimon} cannot hold simultaneously; in other words 
positive-semi-definiteness and de Sitter invariance of the vacuum expectation values 
of the fields are incompatible requirements, and, as for gauge QFT's in 
Minkowski space \cite{strocchiwightman,strocchibook},  
one is left with the choice of preserving either de Sitter invariance or 
positivity while quantizing the fields. 

In the first case, renouncing the positive-semi-definiteness of the correlation 
functions not only gives rise to a pseudo-Fock space with an indefinite metric 
but also  to a modification of the Klein-Gordon equation by the appearance 
of an anomalous non homogeneous term at the rhs of the field equations:
\beq
(\Box +m^2_n)\phi(x) = \Theta_n(x) .
\label{fe}
\endq
The physical states  
are identified in the pseudo-Fock space  by the 
vanishing of the matrix elements of that anomalous term: 
\beq
\langle \Psi_1,\Theta_n(x)\Psi_2\rangle.
\label{physical}
\endq
This condition identifies a {\em de Sitter invariant and positive physical 
subspace} where the field equation is weakly re-established: it holds in 
the mean value sense between physical states. An analogous phenomenon
happens in the Feynman-Gupta-Bleuler (covariant) quantization of 
Quantum Electrodynamics where one is led to modify Gauss' law by adding a 
term at the rhs: 
\beq
\partial^\nu F_{\mu\nu}=j_\mu + \partial_\mu \partial^\nu A_\nu.
\endq
The physical states are selected by the supplementary condition 
$\partial^\nu A_\nu^-|phys\rangle=0$ and the ordinary Gauss' law holds weakly between physical states.

Nothing of this kind is possible for the tachyonic fields in Minkowski space. 
Also there it is possible 
to find a covariant quantization of the tachyon fields \cite{schroer}
that respects causality but there exists no Poincar\'e invariant positive subspace 
of the pseudo-Fock space and therefore no direct physical interpretation.

The existence of the above-mentioned gauge invariance was discovered in a 
Euclidian framework by A.~Folacci in \cite{FolacciGI}, where a BRST type
quantization was performed. At the moment the relation between
that Euclidian BRST approach and the present Gupta-Bleuler type 
quantization is not completely clear.

In the last part of the paper we explore the alternative possibility and construct all possible  
positive {canonical} quantizations of the tachyon fields with spontaneously broken de Sitter symmetry.  
This is similar to the standard treatment of the massless case  as given in \cite{allenfolacci,mottola5}; we treat all tachyonic fields 
\begin{equation}
m^2_\lambda = - \lambda(\lambda+d-1), \ \ \ \lambda\geq 0, \label{continuum}
\end{equation}
on the same footing. Spontaneous breaking of the de Sitter symmetry 
means that {\em the commutators  are de Sitter invariant distributions} and vanish at space 
like separations while {\em the Wightman functions and the other Green functions  are not invariant}. 
The commutator for a given mass parameter $\lambda$  is  indeed 
unique and is the same in all possible quantizations.  All quantizations representing 
the given commutator are therefore {\em canonical} by construction; but positive semi-definiteness
leads to de Sitter non invariant  correlation functions while the invariant quantizations are necessarily non positive. 
By contrast, in the flat Minkowski space one can find 
positive quantizations of the tachyons which however do not satisfy 
local commutativity \cite{feinberg}. 

Finally we  show that  in the case  $\lambda=n$ it is possible to find under certain circumstances de Sitter invariant subspaces
in the reconstructed pseudo-Fock space of the theory which are however not stable under the application of gauge invariant operators.
This means that also in the case $\lambda=n$ the positive quantization is physically inequivalent to the covariant quantization \cite{morrison,wood2,youssef}.

One general lesson to be drawn from these theories is that, even when the 
cosmological constant is tiny, Minkowskian intuition can be misleading and 
de Sitter QFT is essentially different from the standard relativistic 
(Minkowskian) QFT. Other physical consequences of our models, and possible generalizations to higher spin fields \cite{deser1,deser2} will be 
explored elsewhere. 

\section{Preliminaries}\subsection{de Sitter space-time: notations and basic definitions.}
The easiest way to look at either the real or the complex de Sitter manifold is to visualize them 
as subsets of the complex Minkowski manifold with one 
spacelike dimension more. This allows also a natural 
description of the fundamental tubular domains encoding the spectral 
condition of de Sitter quantum field theory \cite{bgm,Bros,bem}. Let therefore $M_{d+1} = \bR^{d+1}$
be the real
$(d+1)$-dimensional Minkowski space-time and $M_{d+1}^{(c)} = \bC^{d+1}$
be its complexification, where $d \ge 2$.
In a chosen Lorentz frame $\{e_\mu, \ \mu=0,\ldots,d\}$
the scalar product of two events 
$x = 
(x^0, \vec{x})$
and $y = 
(y^0, \vec{y})$ is given by
\beq
x\cdot y =  
x^0y^0- \vec{x}\cdot\vec{y}
= \eta_{\mu\nu} x^\mu y^\nu .
\label{s.1}\endq
The future cone $V_+$, the future light cone $C_+$ and the future and past
tubes $T_\pm$ are defined as follows:
\begin{eqnarray}
V_+ &=& \{x\in M_{d+1}\ :\ x\cdot x >0,\ \ \ x^0> 0\},\\
C_+ &=& \{x\in M_{d+1}\ :\ x\cdot x =0,\ \ \ x^0 \ge 0\},
\label{s.1.1}\\
T_\pm &=& \{x+iy \in M_{d+1}^{(c)}\ :\ y\in \pm V_+\}\ .
\label{s.1.2}\end{eqnarray}
The de Sitter space-times with radius $R=1$ and its complexification 
may be represented as  the following one-sheeted hyperboloids:
\beq
X_d = \{x \in M_{d+1}\ :\ x\cdot x = -R^2=-1\}\ \ \ {\rm and}\ \ \
X_d^{(c)} = \{z \in M_{d+1}^{(c)}\ :\ z\cdot z = -R^2=-1\}\ .
\label{s.2}\endq
The connected real
Lorentz group\footnote{The Lorentz group acts as the relativity group of the 
de Sitter manifold. In this context it may also be called the de Sitter group.} 
$L_+^\uparrow = SO_0(1,\ d,\ \bR)$ leaves
$X_d$ invariant and acts transitively on it. Similarly the complexification  
$L_+(\bC) = SO_0(1,\ d,\ \bC)$  acts transitively on the complex manifold   
$X_d^{(c)}$.
The future and past tuboids $\TT_\pm$ are the intersections of the ambient tubes $T_\pm$ with the complex de Sitter manifold:
\beq
\TT_\pm = \{x+iy \in X_d^{(c)}\ :\ y\in \pm V_+\}\ .
\label{s.2.1}\endq
The de Sitter Lorentzian metric $ds^2_{d}$ is obtained by restriction of
the ambient space-time interval $ds^2$ to $X_d$.
Let us consider in particular the global hyperspherical coordinate system 
\begin{eqnarray}
&&x(t,\x) = \left\{
\begin{array}{l} x^0 =\sinh t \\ \vec x \, \  = \cosh t  \  \x
\end{array}  \right.
\ = \ \left\{
\begin{array}{lcl}
x^0 &=& \sinh t, \\
x^1 &=& \cosh t \sin \theta^1, \\
\vdots \\
x^{d-1} &=&\cosh t \cos \theta^1 \ldots \cos\theta^{d-2}\sin \theta^{d-1},\\
x^{d} &=& \cosh t \cos \theta^1 \ldots \cos\theta^{d-2}\cos \theta^{d-1}.
\end{array}  \right.
\label{s.4}\end{eqnarray}
where $t\in{\bR}$,  $0<\theta^i<\pi$ and $0<\theta^{d-1}<2\pi$.
It follows that
\begin{equation}
ds^2_d = ds^2|_{X_d} = dt^2 - \cosh^2 t \ dl^2.
\end{equation}
where 
\beq
dl^2= {(d\theta^1 )}^2+ \cos ^2\theta ^1 {(d\theta^2)}^2+
\cos ^2\theta ^1\cos ^2\theta ^2 {(d\theta^3)}^2+\ldots
\label{s.13}
\endq
 is the line element on a unit $(d-1)$-dimensional sphere $\dsphere$. The  hyperspherical  Section
$S_0= \{\xi \in C_+\ :\ \xi^0 = 1\}$  of the asymptotic cone $C_+$ is also a copy of $\dsphere$
which we parametrize as in Eq. (\ref{s.4}):
\begin{eqnarray}
&&\xi(\xx) =
\left\{
\begin{array}{lcl}
\xi^0 &=& 1\\
\xi^1 &=& \sin \theta^1 \\
\vdots \\
\xi^{d-1} &=& \cos \theta^1 \ldots \cos\theta^{d-2}\sin \theta^{d-1}\\
\xi^{d} &=& \cos \theta^1 \ldots \cos\theta^{d-2}\cos \theta^{d-1}
\end{array}  \right. .
\label{sphericcone}\end{eqnarray}
In the above coordinates the various scalar products are
written as follows
\begin{eqnarray}
x(t,\x) \cdot x(t',\x') &=&
\sinh t \,\sinh t' - \cosh t\,\cosh t' \ \x \cdot { \x' },
\label{s.10}\\
{x(t,\x) \cdot \xi(\xx)} &=&
\sinh t - \cosh t \ \x \cdot {\xx},\label{opq}
\label{s.11}\\
\xi(\xx) \cdot \xi(\xx') &=& 1 -  { \xx}\cdot { \xx '}.
\label{s.12}\end{eqnarray}
The volume form on the unit sphere is
normalized in the standard way:
\beq
d\x =
\prod_{i=1}^{d-1}  |\cos\theta^i|^{d-1-i}   \,  d\theta^i\ .
\label{s.14}\endq
The invariant measure on $X_d$ is
$dx = 2\delta(x\cdot x+1)\,dx^0\ldots dx^d = (\ch t)^{d-1}\,dt\,d\x$
where $x$ is of the form (\ref{s.4}) and $d\x$ is given by (\ref{s.14}).

\subsection{D'Alembert operators and Klein-Gordon equation}
\label{dal}
The Lie algebra $\LL$ of $L_+^\uparrow$ is the vector space of linear operators
$A$ such that $(Ax)\cdot y+ x\cdot (Ay)=0$ for all $x$ and $y$ in $M_{d+1}$.
Particular elements of $\LL$ are $M_{\mu\nu} = -M_{\nu\mu} = e_\mu\wedge e_\nu$.
The regular faithful representation $\VV$ of $L_+^\uparrow$ in the space of
$\CC^\infty$ functions on $M_{d+1}$ or $X_d$ is given by
\beq
\VV(\Lambda)f(x) = f(\Lambda^{-1}x)\ .
\label{f.4}\endq
The representatives $\MM_{\mu\nu}$ of the generators $M_{\mu\nu}$ are
\beq
\MM_{\mu\nu} f(x) = {d\over dt} f(e^{-tM_{\mu\nu}}x)|_{t=0} =
(x_\mu \partial_\nu - x_\nu \partial_\mu)f(x)\ \ {\rm on\ } M_{d+1} \ .
\label{f.6}\endq

The de Sitter-d'Alembert operator $\Box$  is related to the
d'Alembert operator $\Box_{\rm Mink} = \partial^\mu \partial_\mu$ in the ambient 
Minkowski space $M_{d+1}$
as follows: if $f$ is a $\CC^\infty$ function (or a distribution)
homogeneous of degree 0 on $M_{d+1}$
then $\Box (f|_{X_d}) = (\Box_{\rm Mink} f)|_{X_d}$ (see e.g. \cite{Faraut}).
Let $\dil$ denote the infinitesimal generator of dilations
\beq
(\dil f)(x) = %{d\over dt} f(e^t x)|_{t=0} =
x^\mu \partial_\mu f(x).
\label{f.7}\endq
The following relation holds on $M_{d+1}$:
\beq
\MM^2 = \MM^{\mu\nu}\MM_{\mu\nu} = 2x^2\Box_{\rm Mink} -2(d-1)\dil -2 \dil^2\ .
\label{f.8}\endq
If $f$ and $h$ are $\CC^\infty$ functions on $M_{d+1}$
and $h$ is Lorentz invariant, then
\beq
\VV(\Lambda)(h(x)f(x)) = h(x)f(\Lambda^{-1}x),\ \ \
\MM_{\mu\nu} (h(x)f(x)) = h(x) \MM_{\mu\nu} f(x)\ .
\label{f.9}\endq
This remains true if $h$ is a Lorentz invariant distribution
such as $\delta(x\cdot x +1)$.
In particular the operators $\VV(\Lambda)$, $\MM_{\mu\nu}$, $\MM^2$, commute with
the restriction to invariant submanifolds of $M_{d+1}$. They also
commute with dilations.

Let $f$ be a $\CC^\infty$ function (or  a
distribution) defined on the open set of events spacelike w.r.t. the chosen origin
\begin{equation}
\ovl V^c=\{x \in M_{d+1}\ :\ x\cdot x < 0\};
\end{equation}
if $f$ is homogeneous
of degree 0 (i.e. $Df = 0$) it satisfies
$\MM^2 f(x) = 2x^2\Box_{\rm Mink}f(x)$, hence
$\MM^2 (f|_{X_d}) = -2\Box (f|_{X_d})$. Since any $\CC^\infty$ function
(or any distribution) on $X_d$ has a unique extension homogeneous
of degree 0 in the open set $\ovl V^{\,c}$, we have
\beq
\Box f = -\half \MM^2f\ \ \ \ {\rm on\ } X_d
\label{f.10}\endq
for every $\CC^\infty$ function or distribution $f$ on $X_d$.

Assume now that a  $\CC^\infty$ function $f$ on $\ovl V^c$
is homogeneous of degree $\lambda$ and satisfies the wave equation
$\Box_{\rm Mink}f = 0$. Eq. (\ref{f.8}) implies that
\beq
\MM^2 f(x) = -2\lambda(\lambda+d-1) f(x)
\label{f.11}\endq
and, by (\ref{f.10}),
\beq
(\Box - \lambda(\lambda+d-1)) (f(x)|_{X_d}) = 0
\label{f.12}\endq
i.e. the restriction of $f(x)$ to the de Sitter manifold satisfies the Klein-Gordon equation
with {\em complex} squared mass
\begin{equation}
m^2_\lambda = - \lambda(\lambda+d-1).
\end{equation}
Note also that the same is true for a function $f$
homogeneous of degree $(1-d-\lambda)$ i.e. the Klein-Gordon equation 
(\ref{f.12}) is invariant under the involution
\begin{equation}
\lambda\to 1-d-\lambda.\label{involution}
\end{equation}
This symmetry will play a considerable role in what follows. Conversely,
any distribution solving the de Sitter Klein-Gordon equation (\ref{f.12}) 
has an extension to $\{x \in M_{d+1}:\, x\cdot x < 0\}$ which is homogeneous
of degree $\lambda$ and satisfies the wave equation.
Finally putting together (\ref{f.10}) and (\ref{f.8}) shows that
for every $\CC^\infty$ function on on $\ovl V^c$
\beq
\Box_{dS}(f|X_d) = \left (\Box_{\rm Mink} +(d-1)D +D^2 \right )f \Big | X_d\ .
\label{f.15}\endq

\section{de Sitter Klein-Gordon fields with complex squared mass}
\label{masskg}
This  following two Sections one constitute a 
summary of the  linear Klein-Gordon
quantum field theory on the de Sitter universe.
As regards the massive theories of course this is not new \cite{Th, N, ChT, spindel,GiH, BuD, mottola, allen, bgm,Bros, bem}; our construction 
however also describes fields having complex squared masses and is preliminary for the construction 
of the massless and tachyonic fields of this paper.
We will extensively use the
hyperspherical harmonics described, e.g.
in \cite[Chap. XI]{HTF2} and \cite[Chap. IX]{Vilenkin}, and
the closely connected Gegenbauer polynomials. 
In the following $\{\Ylm = Y_M^{d,l}\}$ are the  {\it real} hyperspherical 
harmonics described in Appendix \ref{gegen}; they are homogeneous 
polynomials of degree $l$ on $\bR^d$. For each $l$ their restrictions
to the sphere $\dsphere$ satisfy
\beq
- \triangle_{\x} \Ylm(\x) = l(l+d-2)\Ylm(\x).
\label{s.18}\endq
In the spherical coordinate system (\ref{s.4}) 
the de Sitter Klein-Gordon equation  takes the form
\begin{equation}
\Box \phi - \lambda(\lambda+d-1) \phi=
\frac{1}{(\cosh t)^{d-1}} \ \partial_t ((\cosh t)^{d-1} \ \partial_t \phi)
- \frac{1}{\cosh^2 t} \ \triangle_{\x} \phi - \lambda(\lambda+d-1)\phi =0 .
\label{s.16}\end{equation}
The parameter $\lambda$ and, consequently, the squared mass $m^2_\lambda = - \lambda(\lambda+d-1)$ are assumed to be complex numbers;  
$m^2_\lambda$ is real and positive in the following special cases:
\begin{eqnarray}
&{\rm either\ \ }&   \lambda = -\frac {d-1}2 + i \rho, 
\ \ \ \Im \rho = 0,\ \ \ m = \sqrt{\left(\frac{d-1}2\right)^2 + \rho^2 }\geq \frac{d-1}{2},
\label{principal}
\\
&{\rm or\ \ }&  \Im \lambda =0, 
\ \ \ 1-d<\Re \lambda <0, \ \ \ 0<m< \frac{d-1}{2}.
\label{reallambda}
\end{eqnarray}

If $\phi(t,\x) = \phi(x(t, \x))$  
 is  a distribution satisfying
equation (\ref{s.16}) and $\varphi$  is any $\CC^\infty$ function on the
 sphere $\dsphere$, a general result asserts that  
the average
\beq
\int_{\dsphere} \phi(t,\x)\,\varphi(\x)\,d\x
\label{s.23}\endq
is a $\CC^\infty$ function of the $t$ coordinate; in particular this is true 
for the function
\beq
(\cosh t)^{\frac{2-d}{2}}\, h(t) = \int_{\dsphere} \phi(t,\x)\,\Ylm(\x)\,d\x.
\label{s.23.1}\endq
Eq. (\ref{s.18}) implies that the projectors ${V}_{lM}$ in $L^2(\dsphere)$
with kernel $\Ylm(\x)\Ylm(\x')$ and, consequently, their sum 
$V_l = \sum_M V_{lM}$ commute with the Laplacian $\Delta_\x$.
As a consequence the operators
\begin{eqnarray}
({\bf V}_{lM}\phi)(t, \x) = \ \int_{\dsphere}  \Ylm(\x)\Ylm(\x') \phi(t, \x') \,d\x',
\label{s.23.2}\\
({\bf V}_l\phi)(t, \x) = \sum_M ({\bf V}_{lM}\phi)(t, \x) = \int_{\dsphere} V_l(\x,\x')\phi(t,x')\,d\x',
\label{s.24.3}
\end{eqnarray}
defined on either $\CC^\infty (X_d)$ or $\DD'(X_d)$ are  
 also projectors and commute with $\Box$.
In the above equation we introduced the kernel 
\begin{align}
%&(V_lf)(\x) = \int_{\dsphere} V_l(\x, \y) f(\y)\,d\y, \cr
&V_l(\x, \x') = \sum_M \Ylm(\x)\Ylm(\x') = 
{(2l+d-2)\Gamma(d/2)\over 2\pi^{d/2}(d-2)}\,C_l^{d-2\over 2}(\x\cdot \x')
\label{s.24.2}\end{align}
which is nothing but a suitably normalized Gegenbauer polynomial (see Appendix \ref{gegen}, (\ref{u.6})). 
%In general, if a distribution $T$ and a test-function $\vhi$ on $X_d$ are given by \begin{align} &T(x) = (\cosh t)^{\frac{2-d}{2}}\, T_0(t)\,\Ylm(\x),\ \ \ T_0(t) = t(-i\sh t),\cr &\vhi(x) = (\cosh t)^{\frac{2-d}{2}}\, \vhi_0(t)\,Y_{l'M'}(\x),\ \ \ \vhi_0(t) = \vhi_1(-i\sh t), \label{s.24}\end{align} then \beq \int_{X_d}T(x)\,\vhi(x)\,dx = \delta_{ll'}\delta_{MM'} \int_\bR T_0(t)\vhi_0(t)\,(\ch t)\,dt = \delta_{ll'}\delta_{MM'} \int_{i\bR} t(z)\vhi_1(z)(idz)\ . \label{s.24.1}\endq
%This also holds when $T$ and $\vhi$ are both  in $L^2(dx)$.
Summarizing, if $\phi$ is a solution of the Klein-Gordon equation and $h(t)$ 
is defined as in Eq.  (\ref{s.23.1})
\beq
({\bf V}_{lM}\phi)(x) = (\ch t)^{2-d\over 2} h(t) \Ylm(\x)\ 
\label{s.25}\endq
is also a solution of the Klein-Gordon equation with a definite value of the 
angular momentum. 

Let us now introduce the complex variable $z=- i \sinh t$,
so that $1-z^2 = \cosh^2 t$, and denote $h(t) = f(-i\sh t)$. 
 Eq. (\ref{s.16}) implies that $f$  has to be a solution of the Legendre differential equation:
\beq
(1-z^2) f''(z) -2zf'(z) + \nu(\nu+1) f(z) -{\mu^2 \over (1-z^2)}f(z) = 0 ,
\label{s.19}\endq
with
\beq
\nu = \lambda+\p\ ,\ \ \ \mu = -(l+\p)\ ,\ \ \ \ \p= {d-2\over 2}\ .
\label{s.20}\endq
Any $\CC^\infty$ solution of Eq. (\ref{s.19})
which is defined for $z = -i\sh t$, $t$ real, extends to a holomorphic function
with singularities only at $z = \pm 1$ or $\infty$. Two linearly 
independent solutions are the so-called
``Legendre functions on the cut'' $\P_\nu^\mu(z)$ and $\Q_\nu^\mu(z)$
(see \cite[p. 143]{HTF1}). These functions are holomorphic in the
cut-plane
\begin{equation}
\Delta_2 =
{\bf C}\setminus (-\infty-1] \cup [1,\infty) .
\label{t.22}\end{equation}
The function of the first kind $\P_\nu^\mu(z)$  respects the symmetry  (\ref{involution}): for all $z\in\Delta_2$ it  satisfies the identity \cite{HTF1}
\beq
 \P_{\lambda +\kappa}^\mu(z) = \P_\nu^\mu(z) =\P_{-\nu-1}^\mu(z)=  \P_{1-d-\lambda +\kappa}^\mu(z).
\label{t.22.1}\endq
The reality conditions
\begin{equation}
\ovl{\P_\nu^\mu(z) }= {\P_{\ovl{\nu}}^{\ovl{\mu}}(\bar z)}, \ \ \
\ovl{\Q_\nu^\mu(z)} = {\Q_{\ovl{\nu}}^{\ovl{\mu}}(\bar z)} \label{reality}
\end{equation}
also hold for all $z\in\Delta_2$. If $\mu+\nu$ and $\mu-\nu-1$ are not non-negative integers,  
$\P_\nu^\mu(z)$ and $\P_\nu^\mu(-z)$
also constitute two linearly independent solutions of Eq. (\ref{s.19}).
Therefore, if $(\lambda-l)$ 
and $-(\lambda+d-1+l)$ are not non-negative integers, the general solution 
with angular momentum $l$ and  ``magnetic'' multi-index $M$ has
the form\footnote{In the sequel we shall mostly ignore the $M$-dependence
of the coefficients $a_{\lambda l M}$ and $b_{\lambda l M}$ in Eq. (\ref{modesab0}).
This will enforce rotational invariance in the chosen frame.}
\begin{equation}
\phi_{\lambda l M}(t,\x) =   (\cosh t)^{-\p}\
[a_{\lambda l M}\, \P^{-l-\p}_{\lambda+\p}(i\sh t)
+b_{\lambda l M}\, \P^{-l-\p}_{\lambda+\p}(-i\sh t)]\ \Ylm({\x}).
\label{modesab0}
\end{equation}
In the following without loss of generality, we restrict our attention to values of $\lambda$
such that 
\beq
\Re \lambda + {d-1\over 2} \ge 0\ .
\label{t.23}\endq

\subsection{Normalization - Covariant commutators}
The first task in quantizing a field on any globally hyperbolic curved space-time is to construct the covariant commutator
\begin{equation}
[\phi(x),\phi(x')] = C(x,x').
\label{comm}
\end{equation}
The commutator is a c-number and has therefore an algebraic character; it does not depend on anything but the Lorentzian  structure of the 
space-time manifold (see e.g. \cite{dimock,wald,moschaeff}). 
In our case, given a complex $\lambda$ we want to construct a  bivariate distribution 
solving the Klein-Gordon equation in each variable, with initial conditions 
set by the the equal-time canonical commutation relations; 
in the  spherical coordinate system 
\begin{eqnarray} 
&& \left[\Box_{x} -\lambda(\lambda+d-1)\right] C_{\lambda}(x,x') = 
\left[\Box_{x'} -\lambda(\lambda+d-1)\right] C_{\lambda}(x,x')=0, \\
&& C_{\lambda}(t, {\x}, t', {\x'})|_{t = t'} = 0,\ \ \ \ 
{\partial \over \partial t'}\, 
C_{\lambda}(t, {\x}, t', {\x'})|_{t = t'} 
= i (\ch t')^{d-1}\,\delta({\x},{\x'})\ .\label{c.1}
\end{eqnarray}
These requirements have, for each $\lambda$, a unique solution which is
an entire function of $\lambda$ and is antisymmetric
and Lorentz invariant in the variables $x,\, x'$.
To find it, we generalize the method of canonical quantization to fields with complex 
squared masses by  defining an  
involution in the space of solutions of Eq. (\ref{s.19}) as follows: 
\begin{eqnarray}
 \phi_{\lambda\, l M}(t,\x)\rightarrow \phi^*_{\lambda\, l M}(t,\x) =  
(\cosh t)^{-\p}\ [\ovl a_{l}\,
\P^{-l-\p}_{\lambda+\p}(-i\sh t)+
\ovl b_{l}\, \P^{-l-\p}_{\lambda+\p}(i\sh t)]\ \Ylm({\x}).
\label{star}
\end{eqnarray}
where $a_{l} = a_{l}(\lambda)$ and $b_{l} = b_{l}(\lambda)$. When $t$ and $m^2_\lambda$ are real, the above involution coincides 
with the complex conjugation  
$
\phi^*_{\lambda\, l M}(t,\x) =  \ovl{\phi_{\lambda\, l M}(t,\x)}; \label{ccc}
$
this follows 
from Eq. (\ref{t.22.1}) for  $\lambda$  satisfying (\ref{principal}) while 
it is obvious for $\lambda$ real. 

Notice  that $\phi^*_{\lambda\, l M}(t,\x)$  is also a solution of  Eq. (\ref{s.19})  
(while the complex conjugate $\ovl{{\phi}_{\lambda\, l M}(t,\x)}$ in general is not; 
only when $m^2_\lambda$ is real). 
We are thus led to define a generalized Klein-Gordon product 
\begin{equation}
(f,g)_{KG} = 
i \int_\Sigma (f ^*\partial_\mu g -g\, \partial_\mu f ^* )  d\Sigma^\mu(x) 
\label{kgprod}\end{equation}
that coincides with the usual Klein-Gordon product 
(see e.g. \cite[p. 49]{HE}) when $m^2_\lambda$ is real.  

If $\lambda-l$ is not a non-negative integer  
and 
\begin{equation}
|a_{l}|^2 - |b_{l}|^2 \not = 0 
\label{condabl}
\end{equation}
the solutions 
$\{ \phi_{\lambda\, l M}(t,\x), \phi^*_{\lambda\, l M}(t,\x) \}$
are well-defined and all together constitute a complete and orthogonal set 
w.r.t. the generalized Klein-Gordon product.  
This can be seen by using the following basic Wronskian relations 
\begin{equation}
{\rm Wr}(\P^\mu_\nu,\Q^\mu_\nu)=
\frac{\Gamma(1+\nu+\mu)}{\Gamma(1+\nu-\mu)}(1-z^2)^{-1}
\end{equation}
which imply that
\begin{eqnarray}
(\phi_{\lambda \, l M},\phi_{\lambda\, l' M'})_{KG}
= \frac {1 }{ \gamma_l }\,(|a_{l}|^2 - |b_{l}|^2) \, \delta_{ll'} \, \delta_{MM'}, \ \ \ (\phi_{\lambda\, l M}, \phi^*_{\lambda \, l' M'})_{KG} =  0,
\end{eqnarray}
where we have defined 
\begin{equation}
\gamma_l(\lambda) = {1\over 2}
\Gamma(l-\lambda) \Gamma(1 +\lambda + l + 2\p). 
\label{gammal}\end{equation}
All in all,   the covariant commutator admits the following series expansion in spherical harmonics
\begin{align}
&  C_\lambda(x,x')  = \sum_{l,M} 
\frac { \gamma_l}{|a_{l}|^2 - |b_{l}|^2 }
\left[\phi_{\lambda\, l M}(t,\x) \phi^*_{\lambda\, l M}(t',\x') 
-\phi_{\lambda\, l M}(t',\x')  \phi^*_{\lambda\, l M}(t,\x) \right] =\cr
& = \sum_{l}   { \gamma_l(\lambda) \left[\P^{-l-\p}_{\lambda+\p}(i\,\sinh t)
\P^{-l-\p}_{\lambda+\p}(-i\sh t')
-\P^{-l-\p}_{\lambda+\p}(i\sh t')\P^{-l-\p}_{\lambda+\p}(-i\sh t) \right]  \over (\ch t)^\p (\ch t')^\p} 
 \sum_M\Ylm({\x})\Ylm({\x'}) \cr &&
\label{tpabcomm}\end{align}
which does not depend on the choice of $a_{l}$ and 
$b_{l}$ as expected.
As mentioned above, the commutator is  an entire function of the 
$\lambda$ variable. Indeed the poles of the coefficients
$\gamma_l(\lambda)$ are exactly compensated by zeros of the time-dependent factors 
and a de Sitter invariant canonical commutator exists also for the integer 
values of the variable $\lambda$. Hence (\ref{tpabcomm}) holds for
all complex $\lambda$.

\subsection{Two-point functions: massive case}
The canonical commutation relations and the
Klein-Gordon equation uniquely determine the commutation rules  
(\ref{tpabcomm}). 
A quantization of the Klein-Gordon field is accomplished by finding a 
Hilbert space representation of the above commutation rules.
This amounts  \cite{moschaeff} to finding a two-point function 
$\twofcn (x,x')$,  interpreted as the "vacuum" expectation value of the field
\begin{equation}
\twofcn (x,x') = \langle \Omega, \phi(x)\phi(x') \Omega\rangle, 
\end{equation}
which  solves the Klein-Gordon equation w.r.t. both $x$ and $x'$.  The 
antisymmetric part of $F(x,x')$ is uniquely determined by the commutator through the 
functional equation 
\begin{equation}
\twofcn (x,x')-\twofcn (x',x) = C(x,x').
\label{functionaleq}
\end{equation}
The two-point function must also satisfy 
the positive semi-definiteness condition \cite{sw,reedsimon}
\begin{equation}
\int \twofcn (x,x')\bar f(x) f(x') dx dx' \geq 0.
\end{equation}
The above conditions still leave a large degree of arbitrariness as regards the symmetric
part of $\twofcn$.

Let us focus for the moment on the standard massive fields $m^2_\lambda>0$. 
In this case the coefficients $\gamma_l(\lambda)$ are all 
strictly positive:
\begin{equation}
\gamma_l (\lambda)=  \left\{\begin{array}{ll}
{1\over 2}\Gamma\left(\frac{d-1}2+l +i \rho \right)
\Gamma\left(\frac{d-1}2 + l-i \rho  \right)>0\, ,  &\lambda = - \frac{d-1}2+i \rho, \ \ \rho \in \bR, 
\\ & \\
{1\over 2} \Gamma\left(\frac{d-1}2 + l+ \rho  \right)
\Gamma\left(\frac{d-1}2 +l-\rho  \right)>0\, , &\lambda = - \frac{d-1}2+ \rho, \ \ \  \rho \in \bR, 
\ \ \ |\rho| <\frac{d-1}2 .
\end{array}\right.
\label{positive1}\end{equation}
%Therefore $\sqrt{\gamma_l(\lambda)}$ is also a positive real number. 
This property allows us to pose
$
a_{l}= \ch \alpha_{l},\, b_{l} = e^{i\beta_{l}}\sh \alpha_{l}
$
with  $\alpha_{l}$ and $\beta_{l}$  real, 
and define the corresponding canonical orthonormal modes as follows:

\begin{equation}
\psi_{\lambda l M}(t,\x) =  \sqrt{\gamma_l(\lambda)}
\ (\cosh t)^{-\p}\, \, 
\left[\cosh \alpha_{l}\,\P^{-l-\p}_{\lambda+\p}(i\sh t)
+e^{i\beta_{l} }{\sinh \alpha_{l} }\,\P^{-l-\p}_{\lambda+\p}(-i\sh t)\right] \, \, \Ylm({\x}).
\label{modesab}
\end{equation}

The two-point function associated with the above choice of modes
(\ref{modesab}) through the standard Fock construction may be explicitly written as a series
\begin{eqnarray}
\twofcn (x,x')  &= & \sum_{lM} \psi_{\lambda\,l M}(t,\x) \label{tpab0}
\ovl\psi_{\lambda\,l M}(t',\x') \\ & = &\sum_{lM}\gamma_l(\lambda)  \phi_{\lambda\,l M}(t,\x) 
\phi^*_{\lambda\,l M}(t',\x') \label{tpab}
\end{eqnarray}
and is manifestly positive-semi-definite \cite{sw,reedsimon} because the coefficients
(\ref{positive1}) are all positive. 
Eq. (\ref{tpab}) admits a further generalization that describes also mixed stated as follows:
\begin{eqnarray}
 \twofcn (x,x') & = &  \sum_{l}  \gamma_l(\lambda) 
(\ch t \ch t')^{-\p}
\left[\ch^2 \alpha_{l} \ \P^{-l-\p}_{\lambda+\p}(i\sinh t)
\P^{-l-\p}_{\lambda+\p}(-i\sh t') +\right.\cr && \cr
&+& \sh^2 \alpha_{l}\ \P^{-l-\p}_{\lambda+\p}(-i\sh t)\P^{-l-\p}_{\lambda+\p}(i\sh t') 
+  \rho_l \ch \alpha_{l}\sh \alpha_{l}  e^{-i\beta_{l}} 
\P^{-l-\p}_{\lambda+\p}(i\,\sinh t)\P^{-l-\p}_{\lambda+\p}(i\sh t') + \cr 
&& \cr &+& \left.\rho_l \ch \alpha_{l}\sh \alpha_{l}  
e^{i\beta_{l}} \P^{-l-\p}_{\lambda+\p}(-i\,\sinh t)
\P^{-l-\p}_{\lambda+\p}(-i\sh t')\right]  
\sum_M\Ylm({\x})\Ylm({\x'})\label{tpabhh}
\end{eqnarray}
where $0\leq \rho_l \leq 1$. The case $\rho_l = 1$ coincides with the pure states (\ref{tpab}).
\vskip 5 pt
\subsection{Into the tachyons}
\label{ttt}
While  (\ref{tpab0})  makes sense as such only for $m^2_\lambda$ positive, Eqs. (\ref{tpab})  and (\ref{tpabhh}) are well defined also for complex values of the  $\lambda$ variable;
furthermore, if $\alpha_l(\lambda)$, $\beta_l(\lambda)$ and $\rho_l(\lambda)$ are well-behaved, the series converge and define meromorphic functions of $\lambda$ with poles at the integers. 

Focusing on the pure states (\ref{tpab}) we see in particular that when $\lambda\to 0$ from the left the coefficient 
$\gamma_0(\lambda)$ explodes; then it becomes negative in the interval $0<\lambda <1$,  
diverging to minus infinity at the endpoints  (see the figure). 
\begin{figure}[h]
\includegraphics[width=0.4\textwidth,height=0.3\textheight]{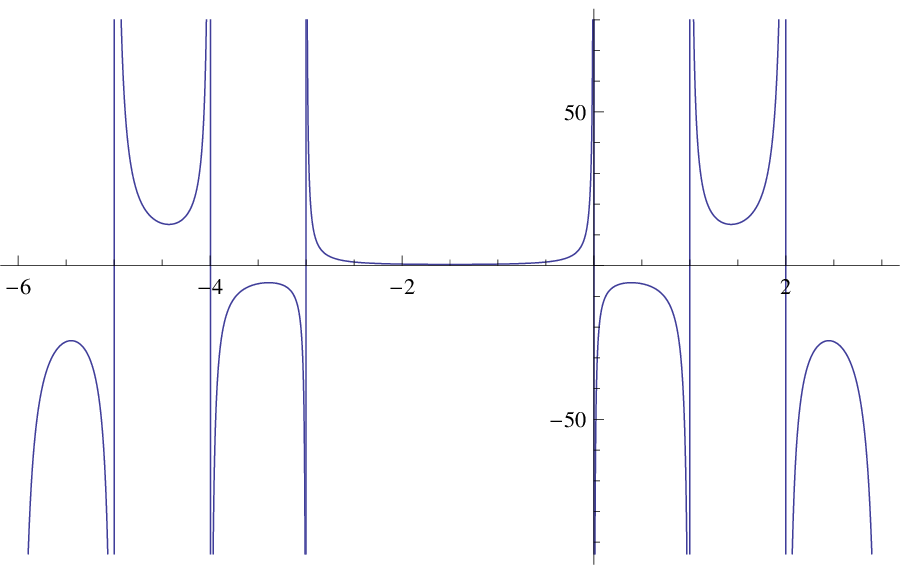}\hfill
\includegraphics[width=0.4\textwidth,height=0.3\textheight]{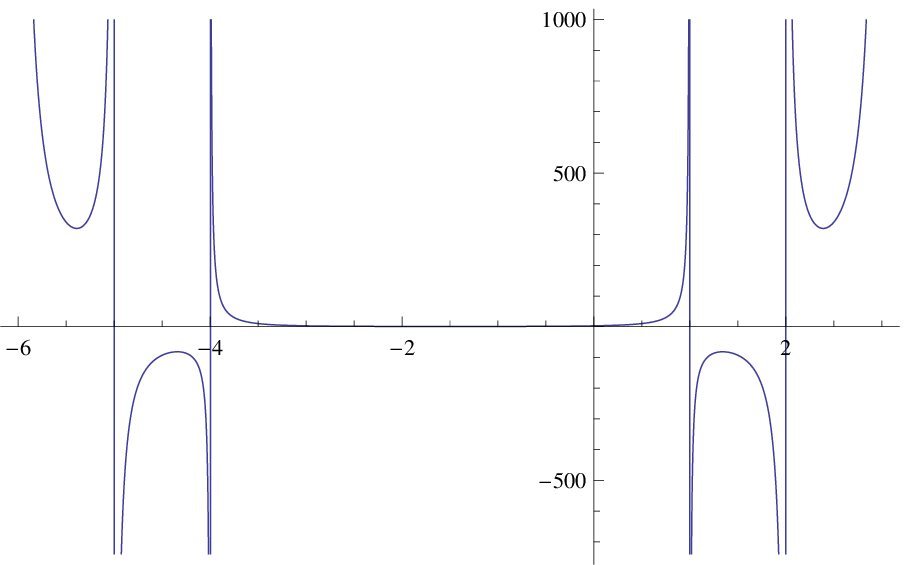}
\label{fig1}
\caption{Plot of $\gamma_0(\lambda)$ (left) and $\gamma_1(\lambda)$ (right) in the four-dimensional case $d=4$.}
\end{figure}
Therefore, in the interval $0<\lambda<1$ the modes $\phi_{\lambda 0 0}$ have negative norm.
Similarly, when $\lambda\to n$ the coefficients $\gamma_0(\lambda), \gamma_1(\lambda),\ldots,\gamma_n(\lambda)$ diverge. 
In the interval $n< \lambda <n+1$ all the $\gamma_l(\lambda)$ such that $n-l$ is even are negative and the corresponding modes $\phi_{\lambda l M}$ have negative norms. The advantage of using the spherical coordinates is that the infrared problem relative to the tachyons manifests itself by altering the behaviour of a finite number of modes, because of the compactness of the spatial sections.

Therefore, if $\alpha_l(\lambda)$, $\beta_l(\lambda)$ and $\rho_l(\lambda)$ are well-behaved, the conclusion is that there is no issue in the convergence of the series Eqs. (\ref{tpab})  and (\ref{tpabhh}) or of the series for the corresponding Green functions \cite{higuchi,wood2}, but the two-point functions defined this way are not positive semi-definite (see \cite{Miao}, footnote 1).  This is true in particular for the choice $\alpha_l=\beta_l=0, \rho_l=1$ corresponding to the maximally analytic invariant (Bunch-Davies) vacuum (see Eq. (\ref{a.6}) below).

We will see  that, while in general it is necessary to fully break the de Sitter symmetry to reestablish a positive quantization, it is precisely the divergence of the coefficients $\gamma_l(n)$ that allows to construct de Sitter invariant quantizations of the tachyon fields for the discrete values of the masses $m^2_n$. The non invariant case will be treated in Section \ref{noninvar} at the end of the paper where 
we will describe all possible non-invariant canonical quantizations with positive metric.

\section{Plane waves and de Sitter invariant two-point functions}
\subsection{Holomorphic plane waves}
Even though Eq. (\ref{tpabhh}) provides through the standard reconstruction
procedure \cite{sw} the most general class of Fock  canonical quantizations
of the de Sitter Klein-Gordon massive free field, we are interested in singling
out among them the de Sitter invariant vacua. To this aim we make appeal to
a remarkable set of solutions of the de Sitter Klein Gordon equation which
may be interpreted as de Sitter plane waves \cite{Bros,Gelfand,Faraut}.
Here is their definition.
For a given  nonzero lightlike vector $\xi\in C_+$ and a complex number $\lambda \in \bC$ let us construct the  homogeneous function
\begin{equation}
z\mapsto (\xi\cdot z)^\lambda.
\label{pwholo}\end{equation}
When considered in the ambient Minkowski spacetime this is a
holomorphic function  in the tubes $T_\pm$ and satisfies the massless Klein-Gordon
equation $\Box_{{\rm Mink}, z}(\xi\cdot z)^\lambda =0$.
When restricted to the de Sitter universe it is holomorphic in the
tuboids  $\TT_\pm$ and  satisfies the massive (complex) Klein-Gordon equation:
\begin{equation}
(\Box_{ z}+ m_\lambda^2)(\xi\cdot z)^\lambda =0. \label{cdskg}
\end{equation}
The parameter $\lambda$ is here unrestricted, i.e. we consider
{\em complex squared masses} $m_\lambda^2$.
The symmetry (\ref{involution}) also implies that, corresponding to the same complex squared mass,
\beq
\left(\Box_z +m_\lambda^2\right)(\xi\cdot z)^{-\lambda-d+1} =0.
\label{p.2abis}
\endq
The boundary values
\beq
(\xi\cdot x)_\pm^\lambda = \lim_{z\in T_\pm, \, z \rightarrow x}
(\xi\cdot z)^\lambda\ ,
\label{p.1}\endq
are homogeneous distributions of degree $\lambda$ on $M_{d+1}$,
and their restrictions to $X_d$, denoted with the same symbols,
are solutions of the de Sitter Klein-Gordon equation
\beq
\left(\Box_x +m_\lambda^2\right)(\xi\cdot x)_\pm^\lambda =0, \ \ \ \ \ \ \ \  \left(\Box_x +m_\lambda^2\right)(\xi\cdot x)_\pm^{1-d-\lambda} =0.
\label{p.2}\endq
All these
objects depend in a $\CC^\infty$ way on $\xi$ and are entire in $\lambda$.

For $f \in \DD(X_d)$, $\xi \in \CC_+\setminus \{0\}$, $\lambda \in \bC$,
the holomrphic plane waves may be used to define the Fourier transforms
as follows \cite{Bros,brosfourier}:
\begin{align}
\tilde{f}_{+}(\xi,\lambda)  = \tilde{f}_{\lambda,+}(\xi)&= \int_{X_d}
(\xi\cdot x )_+^{\lambda}f(x)\,dx,
\label{p.10}\\
\tilde{f}_{-}(\xi,\lambda) = \tilde{f}_{\lambda,-}(\xi)&= \int_{X_d}
(\xi \cdot x )_-^{\lambda}f(x)\,dx.
\label{p.11}\end{align} These definitions obviously
commute with the Lorentz transformations.

The following remarkable property, which will be used later, shows the relationship between the waves  (\ref{pwholo}) and (\ref{p.2abis}):
\begin{lemma}
\label{xiprop}
If $\Re \lambda > -(d-1)/2$
and $z \in \TT_\pm$, then
\beq
(\xi\cdot z)^{\lambda} =
{e^{\pm i \pi\left(\lambda+{d-1\over 2}\right)}\Gamma (\lambda+d-1 )\over
\pi^{d-1\over 2}2^{\lambda +d-1}\Gamma\left(\lambda+{d-1\over 2}\right)}
\int_\gamma (\xi\cdot \xi')^\lambda
(\xi'\cdot z)^{1-d-\lambda} \alpha(\xi'),
\label{p.12}\endq
where $\gamma$ is any smooth $(d-1)$-cycle homotopic to the sphere $S_0 = \{\xi \in C_+\ :\ \xi^0 =1\}$   in $C_+\setminus \{0\}$ and $\formdif$ is the $(d-1)$-differential form
\beq
\formdif(\xi) = (\xi^0)^{-1}
\sum_{j=1}^d (-1)^{j+1} \xi^j\,d\xi^1\wedge\ldots\ \wh{d\xi^j}\wedge
\ldots\ d\xi^d\ .
\label{r.10}\endq
\end{lemma}
{\bf Proof}:  on $C_+\setminus \{0\}$, $\xi^0 = \sqrt{\sum_{j=1}^d (\xi^j)^2}$ and
if $\vhi$ is a smooth function on $C_+\setminus \{0\}$,
\beq
d(\vhi(\xi)\formdif(\xi)) = \left ( (d-1)\vhi(\xi) +
\sum_{j=1}^d \xi^j {\partial\vhi(\xi)\over \partial \xi^j}\right )\,
(\xi^0)^{-1}d\xi^1\wedge\ldots\wedge d\xi^d\ .
\label{r.11}\endq
If $\vhi$ is
homogeneous of degree $(1-d)$ the form
$\vhi\formdif$ is closed. For any  smooth $(d-1)$-cycle $\gamma$
homotopic to $S_0$ in $C_+\setminus \{0\}$ one has that
\beq
(\dil + d-1)\vhi = 0\ \ \Longrightarrow\ \ \
\int_{S_0} \vhi(\xi)\, d\xx =
\int_{\gamma} \vhi(\xi)\,\formdif(\xi)
\label{r.11.1}\endq
where $d\xx$ is the standard measure
on the sphere $S_0$.
In the same case the above integral is invariant under Lorentz
transformations, i.e. for any $\Lambda \in L_+^\uparrow$
\beq
\int_{S_0} \vhi(\xi)\, d\xx = \int_{S_0} \vhi(\Lambda^{-1}\xi)\, d\xx
\ .
\label{r.11.2}\endq
Now, since the integral (\ref{p.12}) is convergent, it follows from the analyticity,
invariance and homogeneity of the rhs that the two sides of
(\ref{p.12}) are equal up to a constant factor.
The constant can be
evaluated by choosing $z=(0,\ldots,0,\pm i)$, $\xi=(1,0,\ldots,\ 0,\ 1)$.
With this choice,
\begin{align}
\int_{S_0} (\xi\cdot \xi')^{\lambda} (\xi'\cdot z)^{1-d-\lambda}
d\xx' &=
{2\pi^{d-1\over 2} e^{\pm {i\pi\over 2}(1-d-\lambda)}\over
\Gamma\left({d-1\over 2}\right)}
\int_0^\pi (1-\cos \theta)^{\lambda} (\sin \theta)^{d-2} d\theta\cr
&= \pi^{d-1\over 2}e^{\pm {i\pi\over 2}(1-d-\lambda)}
{2^{\lambda+d-1} \Gamma\left(\lambda+{d-1\over 2}\right)\over
\Gamma (\lambda + d-1 )}\ ,
\label{p.13}\end{align}
and the assertion of the lemma follows.

\subsection{Maximally analytic vacua}
We will need a small part of the harmonic analysis on the de Sitter space
(see e.g. \cite{Bros}). The main result is the following:
for ${z} \in \TT_-$ and ${z'} \in \TT_+$ one has
\begin{eqnarray}
&& \int_{S_0}(\xi\cdot {z})^{1-d-\lambda}(\xi\cdot {z'})^{\lambda}\,
d\xx = \int_\gamma(\xi\cdot {z})^{1-d-\lambda}(\xi\cdot {z'})^{\lambda}\,
\formdif(\xi)=
\label{r.2}\\
&&={2\pi^{d\over 2}e^{i\pi\left ( \lambda+{d-1\over 2} \right )}\over
\Gamma \left ({d\over 2}\right )}
F\left(-\lambda,\ \lambda+d-1;\ {d\over 2};\ {1-\zeta\over 2}\right ),
\ \ \ \zeta = {z}\cdot {z'}, \ \ \ \lambda\in {\bf C}\, .
\label{r.2.1}\end{eqnarray}
Here as above $S_0 $ is identical to
the unit sphere in $\bR^d$ and $\gamma$ is any smooth $(d-1)$-cycle homotopic to $S_0$
in $C_+\setminus \{0\}$ and $\formdif$ is the $(d-1)$-differential form
(\ref{r.10}).
The equality in (\ref{r.2}) is proven along the lines of the proof of the previous lemma. Lorentz invariance is then used to compute (\ref{r.2.1}).
Both sides of (\ref{r.2}) are entire in $\lambda \in \bC$,
as well as the rhs of (\ref{r.2.1}).

%%%%%%%%%%%%%%%%%%%%%%%%%%%%%%%%%%%%%%%%%%%%%%%%%%%%%%%%%%%%%%%%
If $\lambda$ is not a pole of $\Gamma(-\lambda)\Gamma(\lambda+d-1)$,
${z}\in\TT_-$, ${z'}\in\TT_+$, we denote
\begin{eqnarray}
\Wl({z}, {z'}) &=& \wl({z}\cdot {z'}) = c(\lambda)
\int_{S_0}(\xi\cdot {z})^{1-d-\lambda}(\xi\cdot {z'})^{\lambda}\,
d\xx,
\label{r.3}\\
c(\lambda) &=& {\Gamma(-\lambda)\Gamma(\lambda+d-1)
e^{-i\pi\left ( \lambda+{d-1\over 2} \right )}\over 2^{d+1}\pi^d}\ .
\label{r.4}
\end{eqnarray}
We list three equivalent
expressions for $\wl$:
\begin{align}
\wl({z}\cdot {z'}) &= {\Gamma(-\lambda)\Gamma(\lambda+d-1)\over
(4\pi)^{d/2}\Gamma\left({d\over 2}\right)}
F\left(-\lambda,\ \lambda+d-1\ ;\ {d\over 2};\ {1-\zeta\over 2}\right )
\label{r.0}\\
&=
{\Gamma(-\lambda)\Gamma(\lambda+d-1)\over
(4\pi)^{d/2}\Gamma\left({d\over 2}\right)}
\left ({\zeta+1\over 2} \right )^{-{d-2\over 2}}
F\left(\lambda+{d\over 2},\ -\lambda-{d-2\over 2}\ ;\
{d\over 2};\ {1-\zeta\over 2}\right )
\label{r.0.1}\\
&=
{\Gamma(-\lambda)\Gamma(\lambda+d-1)\over 2 (2\pi)^{d/2}}
(\zeta^2-1)^{-{d-2 \over 4}}\,P_{\lambda +{d-2\over 2}}^{-{d-2 \over 2}}(\zeta),
\ \ \ \zeta = {z}\cdot {z'}\ .
\label{r.1}\end{align}
Eq. (\ref{r.0.1}) follows from (\ref{r.0}) by a standard
identity between hypergeometric functions: \cite[2.1.4 (23) p. 64]{HTF1}.
Here and in the sequel $z^\alpha = \exp(\alpha \log z)$ is defined
as holomorphic in $\bC \setminus \bR_-$ and $(z^2-1)^\alpha$ as
$z^{2\alpha}(1-z^{-2})^\alpha$. The function $z \mapsto (1-z^{-2})^\alpha$
is holomorphic and even in $\bC\setminus [-1,\ 1]$.
The function $z\mapsto (z^2-1)^\alpha$ is holomorphic in
\beq
\Delta_1 = \bC\setminus (-\infty,\ 1]\ ,
\label{r.1.1}\endq
and it is not even unless $\alpha$ is an integer.
$z\mapsto (1-z^2)^\alpha$ is holomorphic and even in the domain
$\Delta_2$ (see (\ref{t.22})).
It satisfies
\beq
(1-z^2)^\alpha = e^{\mp i\pi\alpha}(z^2-1)^\alpha \ \ \
{\rm for}\ \ \ \pm \Im z >0\ .
\label{r.1.3}\endq

It is clear from (\ref{r.0}) that
$\zeta \mapsto \wl(\zeta) = w_{-\lambda-d+1}(\zeta)$
is holomorphic in $\bC\setminus (-\infty,\ -1]$ i.e. everywhere except on the locality cut ({\em maximal analyticity property}).

Recall that the distribution $(\xi\cdot x)_\pm^\lambda$ on $X_d$
depends in a $\CC^\infty$ way on $\xi \in C_+\setminus \{0\}$,
and is entire in $\lambda$. The function
$\Wl({z},{z'})$ has a boundary value $\WWl(x,\ x')$
in the sense of distributions as ${z}$ and ${z'}$ tend to the reals
from $\TT_-$ and $\TT_+$, respectively, and
\beq
\WWl(x, x')  = c(\lambda)
\int_{S_0}(\xi\cdot x)_-^{1-d-\lambda}(\xi\cdot x')_+^{\lambda}\,
d\xx\ .
\label{r.5}\endq
Since $(\xi\cdot x)_\pm^\lambda$ satisfies the
Klein-Gordon equation (\ref{p.2}),
$\Wl$ and $\WWl$ satisfy the same equation
in each of their arguments. Furthermore
\beq
\WWl(x, x') = \WWl(-x', -x)
= \ovl{\WW_{\ovl{\lambda}}(x',x)}
\label{r.6}\endq
(provided $\lambda$ is not a pole of
$\Gamma(-\lambda)\Gamma(\lambda+d-1)$).

Eq. (\ref{r.3}) and its boundary value (\ref{r.5}) are Fourier-type representation
of  the so called {\em Bunch-Davies} aka {\em Euclidean} two-point function,
which occupies such a considerable portion of the de Sitter quantum field theoretical literature.
Its role and relevance for understanding de Sitter physics is still under debate \cite{Polyakov,Polyakov2,mottola2,mottola3,akh}.

\subsection{Invariant vacua in spherical coordinates}
To establish the connection between the above manifestly invariant expressions for the two-point functions, the modes (\ref{modesab})
and the two-point functions (\ref{tpab}) let again $x=x(t,\x)$ be parameterized as in Eq. (\ref{s.4}),
with $\x$ real. For $0<\Im t < \pi$ the event $x$ belongs to forward  tube $\TT_+ $ 
while for $-\pi<\Im t < 0$ it belongs  to the past tube $\TT_- $; for such events 
the following fundamental representation holds (see appendix \ref{plwaves}):
\begin{align}
(\xi\cdot x)^\lambda =2 \pi
\left(\frac{2\pi} {\ch t}  \right)^\p
\;\;  \sum _{l=0}^\infty e^{\pm{i\pi\over 2}(\lambda-l)}
\frac{ \Gamma\left(l-\lambda\right) }
{\Gamma\left(-\lambda  \right)}
\P_{\lambda +\p}^{-l-\p}(\mp i\sh t)
\;\; \sum_{M}  { \Ylm}(\xx)  \Ylm({\x})\ ,\label{a.4} \\ \makebox{for } \ x=x(t,\x) \ \makebox{ with }  0<\pm \Im t < \pi,\ \ \xi \in S_0\ . \nonumber
\end{align}
  Eq. (\ref{a.4}) gives the projection
of the holomorphic plane waves onto the hyperspherical harmonics:
\beq
\Xi_{\lambda\,l\,M}(x) =
{ (-1)^l\Gamma(\lambda+1-l)\over {(2\pi)^{{d\over 2}}\Gamma(\lambda +1)}}
\int_{S_0} (\xi\cdot x)^\lambda\,\Ylm({\xx})\
d\xx \ =  h_{\lambda\,l}(t)\,\Ylm ({\x})
\label{a.12.1}\endq
where $x= x(t,\ \x)$ and $h_{\lambda\,l}(t)$ is given by
\begin{equation}
h_{\lambda\,l}(t) = \left\{\begin{array}{ll}
e^{{i\pi\over 2}(\lambda-l)} (\ch t)^{-\p}
\P_{\lambda +\p}^{-l-\p}(- i\sh t), & 0< \Im t < \pi,\cr &\cr
e^{-{i\pi\over 2}(\lambda-l)} (\ch t)^{-\p}
\P_{\lambda +\p}^{-l-\p}( i\sh t),
& -\pi<\Im t < 0. \end{array}\right.
\label{a.12}\end{equation}
%%%%%%%%%%%%%%%%%
These formulae define $\Xi_{\lambda\,l\,M}$ as a holomorphic function on 
$\TT_+\cup \TT_-$ which has two boundary values on the reals from these two
tuboids. Eqs. (\ref{a.12}) show that these boundary values are actually
continuous. They are not equal unless $\lambda$ is an integer $\ge l$
(see sect.~\ref{invarquant}). Which of the two boundary values is used 
will be clear from the context.
%%%%%%%%%%%%%%%%%%%%%%%%%%%%%%
Note that $\Xi_{\lambda l M}$ is entire in $\lambda$ and that for real $\lambda$,
$\Xi_{\lambda l M} (x) = \ovl{\Xi_{\lambda l M} (\bar x)} $.

The remarkable fact is that the plane waves holomorphic in ${\TT}_\pm$
contain, respectively,  only the pure modes $\P_{\lambda +\p}^{-l-\p}(\mp i\sh t)$
and not a linear combination of them, as it
happens in general for a solution of the de Sitter Klein-Gordon equation.
Inserting the representation (\ref{a.4}) into (\ref{r.3}) and
taking into account the identity $\P_\nu^\mu = \P_{-\nu-1}^\mu$ we get
\begin{align}
\WWl(x,\ x')  &=\frac 1 {
(\ch t)^\p (\ch t')^\p} \sum_{l} \gamma_l(\lambda)
\P^{-l-\p}_{\lambda+\p}(i\sinh t)\P^{-l-\p}_{\lambda+\p}(-i\sh t')  
\sum_M\Ylm({\x})\Ylm({\x'}).
\label{a.6}\\
&= \sum_{l,\ M} \gamma_l(\lambda) \Xi_{\lambda lM}(x)
\Xi_{\lambda lM}(x') \ .
\label{a.6.1}\end{align}
The Bunch-Davies two-point function (\ref{r.5}) therefore coincides with the simplest conceivable choice  in Eq. (\ref{tpab}):
$
\alpha_{l} = \beta_{l}=0.
$
The choice 
$
\alpha_{l} = \alpha, \   \beta_{l} = \pi l
$
corresponds instead to the so-called alpha-vacua.
Indeed, since  $\Ylm({-\x}) = e^{i\pi l}\Ylm({\x})$ we get
\begin{equation}
\twofcn (x,x')
= \ch^2 \alpha \, \WWl(x,\ x') + \sh^2 \alpha \, \WWl(x',\ x) +   
\ch \alpha \sh \alpha \, [\WWl(x,\ - x')  
+ \WWl(-x,\  x')]
\label{tpalpha}\end{equation}
More generally, by setting   $\alpha_l = \alpha$, $\beta_l = \pi l +\beta$ 
we obtain the most general de Sitter invariant vacua first described in \cite{spindel} and studied in \cite{mottola,allen} 
and by several authors afterwards.
Once more, the commutator 
$
C_{\lambda}(x, x')
$
 does not depend on which choice one  makes.

\section{Massless fields and tachyons}
\label{nlimit}

Let $n$ denote a fixed non-negative integer.
We saw in Section (\ref{ttt}) 
the normalization of the modes with angular momentum up to
$l=n$ diverges in the limit $\lambda \to n$ for almost all choices of 
the constants $\alpha_l$ and $\beta_l$.
Correspondingly, all the plane waves $(\xi \cdot x)^n$ are non-normalizable 
zero modes (see Eq. (\ref{a.4})).
The point is that such waves are polynomials of degree $n$, globally 
defined and analytic in the complex de Sitter manifold - actually in 
the whole complex ambient spacetime.
The space of such polynomials is finite dimensional and, as such, not rich 
enough to reconstruct
the infinitely many degrees of freedom of a quantum field.

The aforementioned divergence has a
counterpart in the presence of the factor $\Gamma(-\lambda)$ in the
normalization of the two-point function $\WW_\lambda(x, x')$ 
as given in Eq.  (\ref{r.0}); 
as a function of $\lambda$ \  $\WW_\lambda(x, x')$ has a pole at every nonnegative integer $n$, the  case $n=0$ being the massless
minimally coupled field. If this factor is simply removed, the resulting
two-point function tends to a polynomial in $x\cdot x'$ 
which as such has no quantum feature. On the other hand, we saw that the 
commutator function $C_\lambda(x, x')$ associated to 
$\WW_\lambda(x, x')$ does not have any pole at $n$; it is an entire function of $\lambda$
and removing the factor $\Gamma(-\lambda)$ would spoil its good behavior
at $n$. 
The fact that the unitary irreducible representations of the 
Lorentz group of the discrete series exist precisely for squared masses of the form 
$m^2_n = -n(n+d-1)$ suggests that in spite of, and actually thanks to 
that divergence, an acceptable quantum field theory might exist for such squared masses.

In this and the two following Sections we provide a construction of such theories
with explicit Lorentz invariance analogous
to the Gupta-Bleuler quantization of QED.

\subsection{Regularized two point function at $\lambda = n$.  
Equation of motion}
Suppose $\Re\lambda+d-1 >0$ and let
\begin{align}
& G_{\lambda}(x, x')  \bydef
{1\over \Gamma(-\lambda)}\wl(x\cdot x') =
 \sum_{l=0}^\infty  A_1(\lambda,\ l)\sum_{M}
\Xi_{\lambda l M} (x)\,\Xi_{\lambda l M}(x')\ ,
\label{a.17.1}\\
& A_1(\lambda, l) = {(-1)^l\Gamma(l+\lambda+d-1)\Gamma(\lambda+1)\over
2\Gamma(\lambda+1-l)}\ .
\label{a.17.2}
\end{align}
The identity at the rhs of (\ref{a.17.1}) 
is valid for $0< -\Im t <\pi$, $0<\Im t'<\pi$.
%%%%%%%%%%%%%%%%%%%%%%%%%%%%%%%%%%%%%%%%%%%%%%%%%%%%%
When $\lambda = n$ the sum over $l$ in Eq.
(\ref{a.17.1}) stops at $l=n$.
Eq. (\ref{r.0}) then implies that
\begin{equation}
G_{n}(x,x') = {\Gamma(n+d-1)\over (4\pi)^{d/2}}
F\left ( -n,\ n+d-1\ ;\ {d\over 2}\ ;\ {1-\zeta\over 2}
\right )= {\Gamma(n+1)\Gamma(d-1)\over(4\pi)^{d/2}} C_n^{d-1\over 2}(\zeta),
\  \zeta = x\cdot x'\ .
\label{a.19}\end{equation}
$C_n^{d-1\over 2}$ is a Gegenbauer  polynomial of degree $n$ 
(see Appendix \ref{gegen}) so that $G_{n}(x,\, x')$ is a polynomial of
degree $n$ with real coefficients in $x\cdot x'$.
Eq. (\ref{r.3}) also implies that
\beq
G_{n}({z}, {z'}) = {\Gamma(n+d-1)e^{-i\pi\left(n+{d-1\over 2}\right)}\over
2^{d+1}\pi^d}\int_{S_0} (\xi\cdot {z})^{1-d-n}(\xi\cdot {z'})^n\,
d\xx,\ \ \ \ {z}\in \TT_-\ .
\label{a.19.1}\endq

For $\lambda \sim n$, $\Gamma(-\lambda) \sim (-1)^{n+1}/n!(\lambda-n)$.
We can attempt to remove the pole of $\wl$ at $n$
by defining
\begin{align}
\wh W_{n}({z}, {z'}) = \wh w_{n}({z}\cdot {z'}) &=
\Gamma(-\lambda) \big ( G_{\lambda}({z},{z'})
-G_{n}({z},{z'}) \big ) \big |_{\lambda = n}\cr
&= {(-1)^{n+1}\over n!}{d\over d\lambda} G_{\lambda}({z},{z'})
\big |_{\lambda = n}\ .
\label{a.20}\end{align}
The boundary values $\wh\WW_{n}(x, x')$ are similarly
defined. The regularization that we have just used can be applied to
any function $f(\lambda)$ which has at worst a simple pole at
$\lambda = n$, i.e. we can define
\beq
\wh f(n) = {(-1)^{n+1}\over n!}{d\over d\lambda}
\left . \left [ {f(\lambda)\over \Gamma(-\lambda)}\right ]
\right |_{\lambda = n}\ .
\label{a.20.1}\endq
If $f$ is regular at $n$, the procedure leaves it
unchanged, while if $f(\lambda) = a(\lambda-n)^{-1} + h(\lambda)$, where $h$ is
analytic at $n$, this gives
\beq
\wh f(n) = a\psi(1+n) + h(n)\ .
\label{a.20.2}\endq

Since $G_{n}(x,\ x')$ is a polynomial and is symmetric
in $x$ and $x'$, the commutator function $C_{\lambda}(x,\ x')$
associated to $\WWl(x,\ x')$ (see (\ref{tpabcomm}))
is, as already announced, holomorphic in $\lambda$ near $n$ and
\beq
\wh \WW_{n}(x,\ x') - \wh \WW_{n}(x',\ x) = C_{n}(x,\ x') =
C_{\lambda}(x,\ x') \bigg |_{\lambda =n}\ ,
\label{a.21}\endq
i.e. it needs no subtraction. 
In particular the {\em canonical} equal-time commutation
equations (\ref{c.1}) remain valid for $\lambda=n$. 
Furthermore the commutator solves the right Klein-Gordon equation
\begin{align}
(\Box_{x,x'} &-n(n+d-1))\, C_{n}(x,\ x') = 0.
\label{a.22c}\end{align}

$\wh W_{n}({z},{z'})$ has the
same good properties of analyticity and invariance as $W_{\lambda}({z},{z'})$; 
on the other hand the regularization procedure has introduced an anomaly: 
the two-point function now satisfies an inhomogeneous Klein-Gordon equation
\begin{align}
(\Box_{{z},{z'}} &-n(n+d-1))\,\wh W_{n}({z},{z'}) = {(-1)^{n+1}\over n!}
(2n+d-1)\,G_{n}({z},{z'}). %\cr&= (-1)^{n+1}(4\pi)^{-d/2}  (2n+d-1)\Gamma(d-1)C_n^{d-1\over 2}({z}\cdot {z'})\ .
\label{a.22}\end{align}
The pole removing procedure of (\ref{a.20}) is, of course, not unique, since
any function of $\lambda$ with a simple pole at $n$ could have been
used instead of $\Gamma(-\lambda)$. The result would differ from
$\wh W_{n}({z},{z'})$ by a multiple of $G_{n}({z},{z'})$,
without affecting (\ref{a.21}) or (\ref{a.22}).
%Additional remarks about this procedure are given in Appendix \ref{wexp}.
\section{de Sitter invariant quantization of the tachyons}
\label{invarquant}
Recall that the sesquilinear form defined on the
space $\DD(X_d)$ of test-functions on $X_d$ 
\beq
\langle f, g \rangle_\lambda = \int_{X_d \times X_d}
\ovl{f(x)}\,\WW_{\lambda}(x, x')\,g(x')\,dx\,dx'\ 
\label{a.23b}\endq 
is not positive semi-definite when $\lambda$ belong to the complement of the 
union of the sets (\ref{principal}) and (\ref{reallambda}). 
In particular (\ref{a.23b}) is not positive semi-definite when $\lambda$ is 
real and positive (and of course not integer). 
Furthermore, there exists no Lorentz invariant subspace of 
$\DD(X_d)$ where $\WW_\lambda(x,x')$ is positive semi-definite.

The sesquilinear form 
\beq
\langle f, g \rangle = \int_{X_d \times X_d}
\ovl{f(x)}\,\wh\WW_{n}(x, x')\,g(x')\,dx\,dx'\ 
\label{h.3}\endq
also fails to be positive-semi-definite. It is remarkable that, 
as we will show below, $\wh\WW_{n}(x,\ x')$ is 
positive-semi-definite when restricted to the Lorentz invariant subspace
$\EE_n$ of $\DD(X_d)$ defined as follows
\begin{definition}
\begin{align}
\EE_n &= \left\{f \in \DD(X_d)\ :\ \wt f_{n}(\xi) =
\int_{X_d} f(x)\,(\xi\cdot x)^n\,dx\ \
= 0\ \ \forall \xi \in C_+\right\}
\label{b.10}\\
&= \left\{f \in \DD(X_d)\ :\ \int_\gamma \wt f_{1-d-n,+}(\xi)
(\xi\cdot\xi')^n\, \alpha(\xi) = 0\ \ \ \forall \xi'\in C_+\right\}\ .
\label{b.12}\end{align}
\end{definition}
Here we have used the definitions (\ref{p.10}), (\ref{p.11}) and 
Lemma \ref{xiprop}. Note that $\wt f_{n}= \wt f_{n,+} = \wt f_{n,-}$. 

Eqs. (\ref{a.4}-\ref{a.12}) %and (\ref{a.10})
provide the following representation:
\beq
(\xi\cdot x)^n =
(2\pi)^{d\over 2}
\sum_{l=0}^n {(-1)^l\Gamma(n+1)\over \Gamma(n+1-l)}\sum_M
\Xi_{n l M}(x)\,Y_{lM}(\xx)\ ,
\label{b.30}\endq
The lhs of this equation is a real pseudo-harmonic polynomial in $x$, hence
so is (for $l \le n$)
\beq
\Xi_{n l M}(x) = {(-1)^l\Gamma(n+1-l)\over (2\pi)^{d\over 2}\Gamma(n+1)}
\int_{S_0} (\xi\cdot x)^n\,\Ylm({\xx})\,d{\xx}\ .
\label{b.31}\endq
The vector space space spanned by all polynomials
$x\mapsto (\xi\cdot x)^n$ as $\xi$ runs over $S_0$ (or over 
$C_+$) is a Lorentz invariant subspace of the space
$\hd$ of homogeneous pseudo-harmonic
polynomials of degree $n$ on $M_{d+1}$, which carries an irreducible
representation of the Lorentz group, so that these two spaces
coincide.
The space $\EE_n$ is thus the orthogonal subspace in
$\DD(X_d)$ to $\hd \subset \DD'(X_d)$.
The space $\EE_n$ can be further characterized as follows:

\begin{lemma}
\label{En_cond2}
A function $\vhi \in \DD(X_d)$ belongs to $\EE_n$ if and only if
it satisfies one of the following conditions:\HB
(i) for all integer $l \in [0,\ n]$ and for all $M$
\beq
\int_{X_d} \Xi_{n l M}(x)\,\vhi(x)\,dx = 0\ .
\label{b.40}\endq
(ii) For all $x \in X_d$,
\beq
\int_{X_d} G_n(x, x')\,\vhi(x')\,dx' = 0\ .
\label{b.40.1}\endq
Furthermore
\beq
(\Box+m^2_n) \vhi \in \EE_n\ \ \ \ \forall \vhi\in\DD(X_d)\ ,
\label{b.40.2}\endq
\beq
\vhi \in \DD(X_d),\ \ \ \vhi(-x) = (-1)^{n+1}\vhi(x)\ \ 
\Longrightarrow \vhi \in \EE_n\ .
\label{b.40.3}\endq
\end{lemma}
Part (i) follows from (\ref{b.30}) and (\ref{b.31}).
If $\vhi \in \EE_n$, it satisfies (ii) by (\ref{a.19.1}).
Conversely suppose that $\vhi$ satisfies (ii). Then
by (\ref{a.19.1}), for all $z\in \TT_-$,
\beq
\int_{S_0} (\xi\cdot z)^{1-d-n} \wt \vhi_{n}(\xi)\,d{\xx} =0.
\label{b.40.4}\endq
This implies $\wt \vhi_{n}=0$, i.e. $\vhi\in \EE_n$ by virtue of
Lemma \ref{auxlem2}, proved in Appendix \ref{spaces}.
The last assertions of the lemma follows from the characterization
(\ref{b.10}) of $\EE_n$.

Thus the $\{\Xi_{n l M}\}$ span the space $\hd$
and are a basis of this space, since their number is equal to its
dimension. 

We also need the following definitions:
\begin{definition}
$\wt\EE_n^{(1)}$ is the space of functions $f$ which are $\CC^\infty$
and homogeneous of degree $1-n-d$ on $C_+\setminus\{0\}$ and
satisfy
\beq
\int_\gamma f(\xi) (\xi\cdot\xi')^n\,\alpha(\xi) = 0\ \ \
\forall \xi'\in C_+\setminus \{0\}\ .
\label{kk.10}\endq
\end{definition}
The integral in the lhs is independent of $\gamma$ if $f$
is homogeneous of degree $1-d-n$ so that the condition (\ref{kk.10})
is Lorentz invariant, and so is $\wt\EE_n^{(1)}$.
The integral in (\ref{kk.10}) is a homogeneous
polynomial in $\xi'$. if it vanishes for all
$ \xi'\in C_+\setminus \{0\}$,  by analytic continuation it also vanishes
for all $\xi'\in \bC^{d+1}$ such that $(\xi'\cdot\xi')=0$.
According to (\ref{b.12}),
\beq
\EE_n = \{f\in \DD(X_d)\ :\
\wt f_{1-d-n,+}\in \wt\EE_n^{(1)}\}.
\label{k.11}\endq
\begin{definition} $\wt\EE_n^{(2)}$ is the space of functions $f$ which are $\CC^\infty$
and homogeneous of degree $1-d-n$ on $C_+\setminus\{0\}$ and
satisfy
\beq
\int_\gamma f(\xi) p(\xi)\,\alpha(\xi) = 0
\label{k.12}\endq
for any homogeneous polynomial $p$ of degree $n$ on $\bR^{d+1}$.
\end{definition}It is equivalent to require this for all real or
all complex homogeneous polynomial of degree $n$ on $\bR^{d+1}$.
Since $\xi \mapsto (\xi\cdot\xi')^n$ is such a polynomial,
$\wt\EE_n^{(2)} \subset \wt\EE_n^{(1)}$. Again $\wt\EE_n^{(2)}$ is
Lorentz invariant. Taking in particular $\gamma = S_0$, we see that
$\wt\EE_n^{(2)}$ is the set of $\CC^\infty$ homogeneous $f$ such
that
\beq
\int_{S_0} f(\xi)\,{\xx}^\alpha\,d{\xx} = 0
\ \ \ \forall \alpha \ \hbox{with}\ |\alpha| \le n\ ,
\label{k.13}\endq
where $\alpha$ is a multi-index, 
${\xx}^\alpha =\prod_{j=1}^d \xi_j^{\alpha_j}$ and
$|\alpha| = \sum_{j=1}^d \alpha_j$. Indeed ${\xx}^\alpha$ is the restriction
to $S_0$ of $\xi_0^{n-|\alpha|}\vec{\xi}^\alpha$, and conversely
any monomial in $\xi$ of total degree $n$ is of this form.
Any homogeneous polynomial $q$ of degree $\ell$ in $\vec{\xi} \in \bR^d$
can be expressed as
\beq
q(\vec{\xi}) = \sum_{j=0}^{[\ell/2]} r^{2j}\,h_{\ell-2j}(\vec{\xi}),
\label{k.14}\endq
where $r^2 = \vec{\xi}^2$ and $h_k$ is a harmonic homogeneous polynomial of
degree $k$ for $0\le k\le \ell$ (see e.g. \cite{Vilenkin}).
Since, on the unit sphere $S_0$, the restrictions of the harmonic
homogeneous polynomial of degree $\ell$ are spanned by the hyperspherical
harmonics $Y_M^{d,\ell}$ (see Appendix \ref{gegen}), the space
$\wt\EE_n^{(2)}$ is the set of $\CC^\infty$ homogeneous $f$ of degree
$1-d-n$ on $C_+\setminus\{0\}$ whose restrictions to $S_0$
have expansions into hyperspherical harmonics with vanishing components
of order $\ell \le n$. It is proved in Appendix \ref{spaces} that
in fact $\wt\EE_n^{(2)} = \wt\EE_n^{(1)}$.

%We will now prove
\begin{lemma}
\label{pos1}
For all $f \in\wt\EE_n^{(2)}$,
\beq
\int_{\gamma_1\times \gamma_2}
\ovl{f(\xi)}\,(-1)^{n+1}(\xi\cdot \xi')^n\,\log(\xi\cdot \xi')\,
f(\xi')\,
\alpha(\xi)\,\alpha(\xi') \ge 0.
\label{k.15}\endq
This expression vanishes if and only if $f=0$.
\end{lemma}
First, the lhs of this formula is independent of $\gamma_1$
and $\gamma_2$.
For $j=1,\ 2$,
let $\gamma_j$ and $\gamma_j'$ be smooth compact $(d-1)$-cycles
in $C_+\setminus\{0\}$ such that $\gamma_j-\gamma_j' = \partial B_j$
where $B_j$ is a bounded $d$-chain in $C_+\setminus\{0\}$.
Denote
$\vhi(\xi,\ \xi') =
\ovl{f(\xi)}\,(-1)^{n+1}(\xi\cdot \xi')^n\,\log(\xi\cdot \xi')$,
and $\psi(\xi,\ \xi') =
(-1)^{n+1}(\xi\cdot \xi')^n\,\log(\xi\cdot \xi')\,f(\xi')$.
Then
\begin{align}
&\left (\int_{\gamma_1\times\gamma_2} -\int_{\gamma_1'\times\gamma_2'} \right )
\ovl{f(\xi)}\,(-1)^{n+1}(\xi\cdot \xi')^n\,\log(\xi\cdot \xi')\,
f(\xi')\,\formdif(\xi)\formdif(\xi') =\cr
&\int_{B_1\times\gamma_2} (d_\xi \ \vhi(\xi,\ \xi')\,\formdif(\xi))\,f(\xi')\,
\formdif(\xi') +
\int_{\gamma_1'\times B_2} \ovl{f(\xi)}\,\formdif(\xi)\,
(d_{\xi'} \psi(\xi,\ \xi')\,\formdif(\xi'))\ .
\label{k.16}\end{align}
Applying (\ref{r.11}) we find
\beq
d_\xi \vhi(\xi,\ \xi')\,\formdif(\xi) =
(\xi^0)^{-1}\ovl{f(\xi)}\,(-1)^{n+1}(\xi\cdot \xi')^n\,
d\xi^1\wedge\ldots\wedge d\xi^d\ .
\label{k.17}\endq
Since this is a homogeneous polynomial of degree $n$ in $\xi'$,
and $f \in \wt\EE_n^{(2)}$,
the second integral in (\ref{k.16}) vanishes. The third integral
similarly vanishes.

To prove the inequality (\ref{k.15}) it suffices to do so in
the case $\gamma_1 = \gamma_2=S_0$, in which
$(\xi\cdot \xi')= 1-{\xx}\cdot{\xx'}$,
and
\beq
(-1)^{n+1}(\xi\cdot \xi')^n\,\log(\xi\cdot \xi') =
\lim_{\rho\uparrow 1} (-1)^{n+1}(1-\rho\,\xx\cdot\xx')^n
\log(1-\rho\,\xx\cdot\xx')\ .
\label{k.18}\endq
As the kernel of an operator on functions on $S_0$,
$(\xx\cdot \xx')$ is positive semi-definite since
\beq
\int_{S_0\times S_0} \ovl{h(\xx)}(\xx\cdot \xx')
h(\xx')\,d\xx\,d\xx' =
|V|^2,\ \ \
{V} = \int_{S_0} h({\xx}){\xx}\,d{\xx}\ .
\label{k.19}\endq
Furthermore, $(\xx\cdot \xx')^p$ is also positive semi-definite
for any integer $p \ge 0$ by a standard tensor product argument 
(a sketch of a proof may be found in Appendix \ref{spaces}.)

The inequality (\ref{k.15}) finally follows from the following lemma
\begin{lemma}
\label{pow}
For any integer $n \ge 0$ the power series expansion
\beq
u_n(z) = (-1)^{n+1}(1-z)^n\,\log(1-z) = \sum_{m=0}^\infty u_{n,m} z^m
\label{k.20}\endq
converges for $|z|<1$ and
satisfies $u_{n,m} >0$ for all $m > n$.
\end{lemma}
The convergence is obvious.
Note that $u_n(0)=0$, i.e. $u_{n,0}=0$, for $n \ge 0$.
The statement holds for $n=0$. Suppose it holds for a certain $n \ge 0$.
Then $u'_{n+1}(z) = (n+1)u_n(z) - (z-1)^n$, so that
$u_{(n+1),m}= (n+1)u_{n,(m-1)}/m >0$ for $m>n+1$. This proves the inequality
(\ref{k.15}). 
If this expression vanishes, then $f$ must satisfy, for
all integers $N > n$
\beq
\int_{S_0\times S_0} \ovl{f(\xx')}(\xx'\cdot \xx)^N f(\xx)\,d\xx\,d\xx' = 0\ .
\label{k.21}\endq
Since $(\xx'\cdot \xx)^N$ is a positive kernel this implies
(by Schwartz's inequality)
\beq
\int_{S_0\times S_0}(\xx'\cdot \xx)^N f(\xx)\,d\xx = 0\ \ \ \forall \xx'\in S_0\ ,
\label{k.22}\endq
so that, by Lemma \ref{auxlem1} of Appendix \ref{spaces}, 
$f$ is orthogonal to the homogeneous 
polynomials of degree $N$. Since $f\in \wt\EE_n^{(2)}$ it is also orthogonal
to the polynomials of degree $\le n$, and it must vanish. 
This completes the proof of Lemma \ref{pos1}.

We can now prove the main result of this Section:

\begin{lemma}
\label{physpos}
The restriction of $\wh \WW_n$ to the Lorentz invariant subspace 
$\EE_n$ of  $\DD(X_d)$ is positive semi-definite: 
\beq
\int_{X_d\times X_d} \ovl{f(x)}\,\wh \WW_n(x, x')\,
f(x')\,dx\,dx'\ \ \ge\ \ 0 \ , \ \ \ \ f\in\EE_n.
\label{k.30}\endq
This expression vanishes if and only if $\wt f_{1-d-n,+} = 0$. 
\end{lemma}

\noindent {\bf Proof}.
By (\ref{r.3}, \ref{r.4}) and Lemma \ref{xiprop},
for ${z}\in\TT_-$ and ${z'}\in\TT_+$,
\beq
G_\lambda({z},{z'}) = {\Gamma(\lambda+d-1)^2\over
2^{\lambda+2d}\pi^{3d-1\over 2}\Gamma\left(\lambda+{d-1\over 2}\right)}
\int_{\gamma\times\gamma}({z}\cdot\xi)^{1-d-\lambda}\,(\xi\cdot \xi')^\lambda\,
({z'}\cdot\xi')^{1-d-\lambda}\,\alpha(\xi)\,\alpha(\xi').
\label{k.31}\endq
According to the definition (\ref{a.20}),
\begin{align}
&\wh W_n({z}, {z'}) = K'_n G_n({z}, {z'}) +\cr
&+ K_n
\int_{\gamma\times\gamma} (-1)^{n+1}(-\log({z}\cdot\xi))({z}\cdot\xi)^{1-d-n}\,
(\xi\cdot \xi')^n\,
({z'}\cdot\xi')^{1-d-n}\,\alpha(\xi)\,\alpha(\xi')\cr
&+ K_n
\int_{\gamma\times\gamma} (-1)^{n+1}({z}\cdot\xi)^{1-d-n}\,
(\xi\cdot \xi')^n\,(-\log({z'}\cdot\xi'))
({z'}\cdot\xi')^{1-d-n}\,\alpha(\xi)\,\alpha(\xi')\cr
&+ K_n
\int_{\gamma\times\gamma} (-1)^{n+1}({z}\cdot\xi)^{1-d-n}\,
(\xi\cdot \xi')^n\,\log(\xi\cdot \xi')\,
({z'}\cdot\xi')^{1-d-n}\,\alpha(\xi)\,\alpha(\xi')\ .
\label{k.35}\end{align}
Here
\beq
K'_n = {(-1)^{n+1}\over n!}{d\over d\lambda} \log {\Gamma(\lambda+d-1)^2\over
2^{\lambda+2d}\pi^{3d-1\over 2}\Gamma\left(\lambda+{d-1\over 2}\right)}
\bigg |_{\lambda =n}\ ,\ \ \
K_n = {\Gamma(n+d-1)^2\over
2^{n+2d}\pi^{3d-1\over 2}n!\Gamma\left(n+{d-1\over 2}\right)}\ .
\label{k.36}\endq
Boundary values can be taken on both sides of (\ref{k.35}) as ${z}$
and ${z'}$ tend to the reals. Thus, if $f,\ g \in\SS(X_d)$,
\begin{align}
&\langle \wh \WW_n,\ \ovl{f}\otimes g\rangle =
K'_n \int_{X_d\times X_d} \ovl{f(x)}\,G_n(x,\ x')\,g(x')\,
dx\,dx'\cr
&+K_n
\int_{X_d\times S_0\times S_0} (-1)^{n+1}\ovl{f(x)}\,(-\log(x\cdot\xi)_-)
(x\cdot\xi)_-^{{1-d-n}}\,
(\xi\cdot \xi')^n\,\wt g_+(\xi',\ {1-d-n})\,dx\,d{\xx}\,d{\xx'}\cr
&+K_n
\int_{S_0\times S_0\times X_d} (-1)^{n+1}
\ovl{\wt f_+(\xi,\  {1-d-n})}\,(\xi\cdot \xi')^n\,
(-\log(x'\cdot\xi)_+)\,(\xi\cdot x')_+^{{1-d-n}}\,g(x')\,
d{\xx}\,d{\xx'}\,dx' \cr
&+K_n
\int_{S_0\times S_0}(-1)^{n+1}\ovl{\wt f_+(\xi,\  {1-d-n})}\,(\xi\cdot \xi')^n\,
\log( \xi\cdot \xi')\, \wt g_+(\xi',\ {1-d-n})\,d{\xx}\,d{\xx'}\ .
\label{k.37}\end{align}
Suppose now that $f$ and $g$ belong to
$\EE_n$. Then $\wt f_{{1-d-n},\pm}$ and $\wt g_{{1-d-n},\pm}$ belong to
$\wt \EE_n^{(1)} = \wt \EE_n^{(2)}$.
The first term in (\ref{k.37}) gives a contribution of 0. The second term
gives 0 because $\wt g_+(\xi',\ {1-d-n})$ is integrated against
a homogeneous polynomial of degree $n$ in $\xi'$. The third term in
(\ref{k.37}) similarly gives a contribution of 0. Only the last term
in (\ref{k.37}) survives. By Lemma \ref{pos1}, if $g=f$, this term is
non-negative and vanishes only if $\wt f_{{1-d-n},+} = 0$.
This completes the proof of Lemma \ref{physpos}.
%%%%%%%%%%%%%%%%%%%%%%%%%%%%%%%%%%%%%%%%%%%%%%%%%%%%%%%%%%%%%%%%%%%%%

\subsection{Another proof of the positive-semi-definiteness of
$\wh\WW_n$ on $\EE_n$}
\label{another}
For every $f \in \DD(X_d)$ we denote 
\beq
f_{lM} = {\bf V}_{lM} f,\ \ \ \ \ f_{lM}(x) = \wt f_{lM}(t)\Ylm(\x)\ .
\label{e.20}\endq
The necessary and sufficient condition for $f$ to belong to $\EE_n$
is that
\beq
\int_{X_d} f_{lM}(x)\,\Xi_{lM}(x)\,dx = 0\ \ {\rm i.e.}\ \ 
\int_\bR \wt f_{lM}(t)\,(\ch t)^{d-1-\p} \P_{n+\p}^{-l-\p}(i\sh t)\,dt =0\ \ \ 
\forall l \le n,\ \ \ \forall M\ .
\label{e.52}\endq
Let also $\WW_{\lambda lM} (x,\ x') = {\bf V}_{lM} \WW_\lambda (x.\ x') =
({\bf V}_{lM} \WW_\lambda {\bf V}_{lM})(x.\ x')$ and similarly define
$\wh \WW_{nlM}$, $G_{\lambda lM}$. We restrict $\lambda$ to vary in a small
{\it real} open interval $\omega$ containing $n$.

\noindent (1) Suppose $l>n$, hence also $l> \omega$. Let
$f,\ g \in \DD(X_d)$. For $\lambda \in \omega$, $\gamma_l(\lambda) >0$
and is analytic in $\lambda$, so that the pole removing procedure
does nothing to $\WW_{\lambda lM} (x,\ x')$ and
\begin{align}
&\int_{X_d\times X_d} f(x)\wh\WW_{n lM}(x,\ x') \ovl{g(x')}\,dx\,dx' =
\lim_{\lambda \rightarrow n} \int_{X_d\times X_d} 
f(x)\WW_{\lambda lM}(x,\ x') \ovl{g(x')}\,dx\,dx' \cr
&= \int_{X_d\times X_d} 
\gamma_l(n) f_{lM}(x) \Xi_{n lM}(x)\,
\ovl{\Xi_{n lM}(x')} \ovl{g_{lM}(x')}\,dx\,dx'\ ,
\label{e.102}\end{align}
and this is non-negative if $g=f$.

\noindent (2) If $l\le n$ and $f,\ g\in \EE_n\,$,
\begin{align}
&\int_{X_d\times X_d} f(x)\wh\WW_{n lM}(x,\ x') g(x')\,dx\,dx' =\cr
& {(-1)^{n+1}\over n!}
\int_{X_d\times X_d} f_{lM}(x) \partial_\lambda \Big [A_1(\lambda,\ l)
\Xi_{\lambda lM}(x)\Xi_{\lambda lM}(x') \Big ]\Big |_{\lambda = n}
g_{lM}(x')\,dx\,dx'\ .
\label{e.103}\end{align}
The derivative of the bracket at $\lambda=n$ is a sum of terms each of which
contains at least one undifferentiated factor $\Xi_{nlM}(x)$ or
$\Xi_{nlM}(x')$, and the whole expression is therefore equal to 0.

%%%%%%%%%%%%%%%%%%%%%%%%%%%%%%%%%%%%%%%%%%%%%%%%%%%%%%%%%%%%%%%%%%%%%%

%%%%%%%%%%%%%%%%%%%%%%%%%%%%%%%%%%%%%%%%%%%%%%%%%%%%%%%%%%
%\newpage
\section{Hilbert space structure}
\label{hilbert}
The status of the de Sitter tachyons described up to here exhibits a striking analogy  
with gauge QFT  \cite{strocchiwightman,strocchibook} in that it is impossible to reconcile positive-semi-definiteness  (which would provide a direct probabilistic interpretation of the states),  
Lorentz invariance and local commutativity. 
Furthermore, as in QED, to preserve Lorentz invariance and local commutativity of the fields 
it has been necessary to introduce an anomalous term in the field equations. 

There is now an additional complication that has to be faced: the choice of working in a local and covariant gauge  
supplies only the linear set of local states, i.e. those states that can be obtained 
by applying polynomials of the field operators to the vacuum. 
This set is endowed with a non-positive inner product which encodes the knowledge of the Wightman functions.
In general, to identify the full set of physical states of the theory 
it is necessary to consider limits and take in some sense the closure of the local states;
another piece of information is needed which is not contained in the field equations:
a (Hilbert space) topology  compatible with the Wightman functions has 
to be assigned on the local states.
For example the construction of the charged stated in QED - which should be 
possible at the price of breaking Lorentz invariance - and in QCD - 
which should be impossible, providing this way the yet lacking  proof of quark 
confinement - requires taking limits,  such as the so called 
"particle behind the moon" procedure: 
those fundamental questions cannot be decided without completing the local 
states in a suitable Hilbert topology. 

A general framework to deal with this kind of problems has been introduced 
in \cite{morchiostrocchiihp,MPS1}.
In this Section we will use that scheme to construct Krein spaces 
for the free fields associated to the two-point functions
$\wh \WW_n$; the case $n=0,d=2$  has been already sketched in \cite{BCM} .

\subsection{Preliminaries}

In this subsection $l$ is always $\le n$.
The first preliminary observation is that $(\ch t)^{-l}\,h_{nl}(t)$ is a real 
polynomial in $\sh t$. Indeed, we saw in Eq. (\ref{b.31}) that the 
function $\Xi_{n l M}(x)$
extends to a real pseudo-harmonic polynomial on $M_{d+1}$
and its restriction to $X_d$ satisfies the Klein-Gordon equation. Since
\beq
\Xi_{n l M}(x) = h_{n,l}(t)\,\Ylm ({\x}) =
h_{nl}(t)\,(\ch t)^{-l}\,\Ylm (\vec{x}),
\label{b.41}\endq
and $\Ylm (\vec{x})$ is a polynomial in $\vec{x}$, we infer
that $(\ch t)^{-l}\,h_{nl}(t)$ must be a polynomial in $x^0 = \sh t$.
By taking as before $z = -i\sh t$ and assuming that $\Re t >0$,
$\Im t > 0$ small enough, i.e. $\Im z< 0$, $\Re z >0$ small enough, we have
\begin{align}
&(\ch t)^{-l}\,h_{nl}(t) = %e^{{i\pi\over 2}(n-l)} (\ch t)^{-l -{d-2\over 2}}\P_{n +{d-2\over 2}}^{-l-{d-2\over 2}}( -i\sh t)\cr&= 
e^{{i\pi\over 2}(n-l)}(1-z^2)^{\half(-l -{d-2\over 2})}
\P_{n +{d-2\over 2}}^{-l-{d-2\over 2}}( z)\cr
&={2^{l+{d-2\over 2}}
\Gamma\left(l+{d-1\over 2}\right )\Gamma(n-l+1)\over
\pi^{1/2} \Gamma(n+l+d-1)} i^{n-l}C_{n-l}^{l+{d-1\over 2}}(-i\sh t)\ .
\label{b.42}\end{align}
Since $r^N C_N^\nu(z/r)$ is a polynomial in $z$ and $r^2$,
real if $\nu$ is real, it follows that %$i^{n-l}C_{n-l}^{l+{d-1\over 2}}(-i\sh t)$ is a real polynomial in $\sh t$ so that 
$(\ch t)^{-l}\,h_{nl}(t)$ is a real polynomial
of degree ($n-l$) in $\sh t$. The above formulae show
that $\Xi_{n l M}(x)$ coincides
with the polynomial $\underline{Y_L^{d+1,n}}(x^0,\ \vec{x})$
defined in (\ref{u.15}) up to a constant factor.

The next observation is that there exists a collection of real functions
$g_l \in \SS(\bR)$, with $0 \le l \le n$, such that, denoting
\beq
\chi_{lM}(x) = g_l(t)\,\Ylm ({\x})\ ,
\label{b.50}\endq
the following conditions are satisfied :
\beq
\int_{X_d} \Xi_{n l M}(x)\,\chi_{l'M'}(x)\,dx = \delta_{ll'}\delta_{MM'}
\ \ \  \ \ \ \makebox{ for all } l,\,l'= 0, \ldots,n, \ \ \ \ \makebox{ and all } M, M'\
\label{b.51}\endq
and
\beq
\langle \chi_{lM}, \chi_{l'M'} \rangle=\int_{X_d\times X_d} \ovl{\chi_{lM}(x)}\,\wh \WW_n(x,\ x')\,
\chi_{l',M'}(x')\,dx\,dx' = 0\ \ \ \forall M,\ M'\ .
\label{b.52}\endq
There is a large arbitrariness  in the choice of
the functions $g_l$. They will be chosen to belong to $\DD(\bR)$. 
The problem of their existence can be reformulated as follows: let
\beq
H_{\lambda l} = \int_\bR g_l(t)\,h_{\lambda l}(t)\,(\ch t)^{d-1}\,dt\, ;
\label{b.53}\endq
Eqs.  (\ref{a.12}) and (\ref{a.17.1}), the definition
(\ref{a.20}) of $\wh\WW_n$ and  the orthonormality of the $\{\Ylm\}$, 
imply that the above requirements will be satisfied if
\beq
H_{nl} = 1\ ,
\label{b.54}\endq
and
\beq
\left . 2\Re {\partial \over \partial \lambda} H_{\lambda,l}
\right |_{\lambda=n} =
\left . -{{\partial \over \partial \lambda} A_1(\lambda,\ l)\over
A_1(\lambda,\ l)} \right |_{\lambda=n}\ ,
\label{b.55}\endq
i.e.
\beq
\int_\bR g_l(t)\,\Re \partial_\lambda h_{\lambda,l}(t)\,(\ch t)^{d-1}\,dt
\big |_{\lambda = n} =
\left . -{{\partial \over \partial \lambda} A_1(\lambda,\ l)\over
2A_1(\lambda,\ l)} \right |_{\lambda=n}\ ,
\label{b.56}\endq
We take $\lambda$ real and close to $n$ in these conditions.
If two real distributions $t,\ t' \in \SS'(\bR)$ are
linearly independent, then for arbitrary real numbers $a_1,\ a_2$
there exists a real test function $\vhi \in \SS(\bR)$ such
that $(t, \vhi) = a_1$ and $(t',\vhi) = a_2$.
Thus the existence
of $g_l$ is guaranteed if we show that $t\mapsto h_{n l}(t)$
and $t\mapsto \Re\partial_\lambda h_{\lambda l}(t) |_{\lambda = n}$
are linearly independent. This is done in Appendix \ref{indep}.

%%%%%%%%%%%%%%%%%%%%%%%%%%%%%%%%%%%%%%%%%%%%

By expanding the rhs of Eq. (\ref{a.22}) using (\ref{a.17.1})
we may rewrite it as follows
\begin{align}
(\Box_{x,x'} -n(n+d-1))\,\wh \WW_{n}(x,\ x') &=
\sum_{l=0}^n A(l) \sum_M \Xi_{n l M} (x)\,\Xi_{n l M}(x')\ ,
\label{h.1}\\
A(l) = {(-1)^{n+1}\over n!}(2n+d-1)A_1(n,\ l) &=
{(-1)^{n-l+1}(2n+d-1)\Gamma(l+n+d-1)\Gamma(n+1)\over
2n!\,\Gamma(n+1-l)}\ .
\label{h.2}\end{align}
\subsection{Krein spaces - infrared states }
In the following we abbreviate $\Xi_{n l M}$ as $\Xi_\j$.
 $\JJ$ is the set of all
pairs $l,M$ where $0\le l\le n$ and $M$ is a multi-index associated to $l$; where no confusion arises an element of $\JJ$ will be denoted $j$, $k$, $r$,  etc.  If $j= \{l,M\}$ and $k= \{r,L\}$
are elements of $\JJ$, we denote $\delta_{jk}= \delta_{lr}\delta_{ML}$ and $A(j) = A(l)$. We define an infrared regular part $f_0$ of $f$ by
\beq
f_0 (x) = f(x) - \sum_{j\in \JJ} c_j(f) \chi_j(x)
\label{h.5}\endq
where
\beq
c_j(f) = \int_{X_d} \Xi_j(x)\,f(x)\,dx\ ,\ \ (j\in \JJ).
\label{h.4}\endq
$f_0$ belongs to $\EE_n$, since it is orthogonal
to all the $\Xi_j$. The map $f\rightarrow f_0$ is a projection
onto $\EE_n$ parallel to the subspace spanned by the $\chi_j$,
in particular $\chi_{j0}= 0$.
It follows from the above equations that
\begin{align}
 \langle (\Box +m^2_n)f,\, g\rangle &= \int_{X_d\times X_d}
 \sum_{j\in\JJ} A(j)\,\ovl{f(x)}\,\Xi_j(x)\,\Xi_j(x')\,g(x')\,dx\,dx' =  \sum_{j\in\JJ} A(j)\,\ovl{c_j(f)}\,c_j(g) .
\label{h.6}\end{align}
We are now ready for the construction of the Hilbert-Krein topology. 
In the first step we use the above projection to single out the infrared regular part in 
the inner product (\ref{h.3}) associated to the
Wightman function:
\beq
\langle f,g\rangle = \langle f_0,g_0\rangle +
\sum_{j\in\JJ} \overline{c_j(f)} \langle \chi_j, g\rangle +
\sum_{j\in\JJ} {c_j(g)} \langle f, \chi_j\rangle. %\label{wightman}
\label{h.7}\endq
In the second step we introduce a pre-Hilbert product on the full
space of test functions $\DD(X_d)$ as follows:
\beq
(f,g) = \langle f_0,g_0\rangle + \sum_{j\in\JJ} \overline{c_j(f)} {c_j(g)}
+  \sum_{j\in\JJ} \langle f, \chi_j\rangle\langle \chi_j,g\rangle .%\label{hilbert}
\label{h.8}\endq
The  majorization
\begin{equation}
\left|\langle f,g\rangle\right|^2 \leq (f,f)(g,g)
\label{h.10}\endq
holds on test functions and shows the compatibility of the Hilbert topology with the Wightman functions.
The ideal of the pre-Hilbert product is the subspace
\beq
{\cal{I}} = \{f\in {\EE_n},\  \langle f,f\rangle =0, \
\langle \chi_j,f\rangle =0 \ \  \forall j\in\JJ\}\ .
\label{h.9}\endq
By standard  Hilbert space techniques, taking the  quotient $\DD(X_d)/{\cal{I}}$ and completing it  
we get a Hilbert space 
which we denote ${\KK}^1$. The Wightman inner product can be uniquely
continuously extended to the Hilbert space ${\KK}^1$.
The linear functionals $f\mapsto \langle \chi_j,f\rangle$
are continuous in the topology of ${\KK}^1$ by (\ref{h.10})
hence there is, for each
$j\in\JJ$, a unique vector $v_j \in \KK^1$
such that
\beq
\langle \chi_j,f\rangle = (v_j,f).
\label{h.11}\endq
The vectors $v_j$ are referred to as {\em infrared states}. In the closely similar case of the massless two-dimensional Klein-Gordon field on the Minkowski space-time there exists only one such vector; it corresponds to an infinitely delocalized state in the sense that it cannot be obtained by applying a localized field to the vacuum but only as an infrared limit \cite{MPS1}.  On the contrary, here the vectors $v_j$ can be identified with  specific test functions:
\beq
v_j = {1\over A(j)} (\Box+m_n^2) \chi_j %\label{vj}
\label{h.12}\endq
Let indeed $v'_j$ denote the rhs of (\ref{h.12}).
Since $\Xi_k$ is a solution of the Klein-Gordon equation, $c_k(v'_j) =0$ 
for every $k\in\JJ$, hence $v'_{j0} = v'_j$ and
\beq
(v'_j , f) = \langle A(j)^{-1}( \Box+m^2_n) \chi_j, f_0\rangle
+ \sum_{k\in\JJ} \langle v'_j ,\ \chi_k \rangle
\langle \chi_k,\ f \rangle
\label{h.13}\endq
The first term in the rhs is zero by applying (\ref{h.6}) since
$c_k(f_0) = 0$. For the other terms,
\beq
\langle v'_j , \chi_k \rangle  =
\langle A(j)^{-1}( \Box+m^2_n) \chi_j,\ \chi_k \rangle
= A(j)^{-1}\sum_{r\in\JJ}A(r) \ovl{c_r( \chi_j)} c_r( \chi_k)
=\delta_{jk}
\label{h.14}\endq
since
$
c_r( \chi_j) = \delta_{rj}\, .
%\label{h.15}
$
Therefore $(v'_j ,f) =  \langle \chi_j,\ f\rangle$, i.e. we can take
$v_j = v'_j$.
Note that $v_j \in \EE_n$ and
\beq
(v_j, \chi_k) = 0, \ \ \
\langle v_j,\ \chi_k \rangle = (v_j,v_k) = \delta_{jk}, \ \ \  \langle v_j,v_k\rangle = 0, \ \ \
(\chi_j, \chi_k) = \delta_{jk},
\label{h.16}\endq
and
\beq
\langle v_j, f \rangle = c_j(f),\ \ \ (\chi_j, f)
= \langle v_j,\ f \rangle .
\label{h.17}\endq
The result of this Section may be summarized in the following
\begin{prop}
The Hilbert space ${\mathcal K}^1$ can be written as a direct sum:
\beq
{\mathcal K}^1=
\overline{{\mathcal E}_n^\bot}^{\langle\cdot,\cdot\rangle}\oplus V_n\oplus X_n\ ,
%\label{equ:kreindec}
\label{h.18}\endq
where ${\mathcal E}_n^\bot$ is the subspace of ${\mathcal E}_n$
orthogonal to $V_n$,
$V_n$ is the linear span of the set $\{v_j\}$ and $X_n$ is the linear span
of the set $\{\chi_j\}$.\\
The metric operator $\eta^{(1)}$ representing $\langle\cdot,\cdot\rangle$
in the Hilbert product $(\cdot,\cdot)$ is given by
\begin{equation}
\eta^{(1)} =
\left[\begin{array}{lll}
\mathbb{I} & 0 & 0 \\
0 & 0 & \mathbb{I} \\
0  & \mathbb{I} & 0
\end{array}\right]
\label{h.19}\end{equation}
$\eta^{(1)} \eta^{(1)} = {\mathbb I}$ so that ${\mathcal K}^1$
is a Krein space.
\end{prop}
\subsection{de Sitter Symmetry}
The functions $\Xi_j$, considered as polynomials
on $M_{d+1}$, form a basis of the space of homogeneous pseudo-harmonic
polynomials of degree $n$ on $M_{d+1}$
(see Appendix \ref{gegen}); on this space the quasi-regular representation
of the Lorentz group reduces to an irreducible representation.
Hence there exists a matricial representation $ L_+^\uparrow \ni \Lambda \mapsto S(\Lambda)$
such that 
\beq
\Xi_j(\Lambda^{-1}x) = \sum_{k\in\JJ} S_{jk}(\Lambda)\,\Xi_k(x)
\label{h.50}\endq
for all $j\in\JJ$. The matrix elements $S_{jk}(\Lambda)$ are real polynomials in the
matrix elements of $\Lambda$. Let
\begin{align}
&\delta\chi_j(x,\Lambda) =
\chi_j(\Lambda x)-\sum_{k\in\JJ}S_{jk}^T(\Lambda)\chi_k(x)\ ,
\label{h.51}
\end{align}
The function 
$\delta \chi_j(\cdot\,, \Lambda )$ belongs to $\EE_n$:
\beq
c_k(\delta\chi_j(\cdot , \Lambda ))= \int_{X_d}\delta \chi_j(x,\Lambda)\,\Xi_k(x)\,dx =
\int_{X_d}\chi_j(x) \Xi_k(\Lambda^{-1}x)\,dx
- \sum_{r\in\JJ}S_{rj}(\Lambda)\int_{X_d}\chi_r(x)\Xi_k(x)\,dx =0.
\label{h.53}\endq

\vskip 5pt
Let now $f$ be an element of the ideal $\II$ (see (\ref{h.9}))
and let $f_\Lambda(x) = f(\Lambda^{-1} x)$. Obviously  $f_\Lambda\in\EE_n$ and
$\langle f_\Lambda, f_\Lambda\rangle =0$. Since the product
$\langle \cdot \,,\cdot \rangle$ is positive semi-definite on $\EE_n$,
$\langle f,g\rangle =0$ for all $g\in \EE_n$, hence
\begin{equation}
\langle f, \delta \chi_j(\cdot\, , \Lambda)\rangle =\langle f_\Lambda , \chi_j\rangle = 0
\end{equation}
and therefore also $f_\Lambda \in \II$.
The regular representation of the Lorentz group naturally defined on
$\DD(X_d)$ (see (\ref{f.4})) is thus compatible with the procedure of quotienting
and completing which leads to $\KK^1$, and this provides a natural
continuous representation $U$ of the de Sitter group on $\KK^1$ that 
preserves the sesquilinear form $\langle \cdot \,,\cdot \rangle$.
However $U$ is not unitary.  The following proposition characterizes the 
action of the Lorentz group on the infrared states:
\begin{proposition}
The infrared states span a Lorentz invariant subspace of the Krein 
one particle space $\KK^1$. 
\end{proposition}
Define indeed
\begin{align}
&\delta v_j(x,\Lambda) = \frac 1 {A(j)}(\Box+m^2_n) \delta\chi_j(x,\Lambda) =
v_j(\Lambda x)-\sum_{k\in\JJ}S_{jk}^T(\Lambda)v_k(x)\ .
\label{h.52}
\end{align}
$\delta v_j(\cdot \,,\Lambda)$ manifestly belongs to $\EE_n$. Furthermore, for any $f\in \DD(X_d)$,
\beq
\langle \delta v_j, f\rangle = \frac 1{A(j)} \langle (\Box+m^2_n) \delta \chi_j, f\rangle
=0
\label{h.54}\endq
by applying (\ref{h.6}) and (\ref{h.53}). It follows that
$(\delta v_j, f)=0$ for every $f\in \DD(X_d)$. This means that
\beq
U(\Lambda) v_j = \sum_k S_{kj}^{-1}(\Lambda) v_k = (S^{-1 T}(\Lambda)v)_j\ .
\label{h.55}\endq
\subsection{The role of the infrared operators: field equation and Gupta-Bleuler condition}
The Krein-Fock space $\KK$ is obtained from the ``one-particle''
space $\KK^1$ by the standard construction \cite{reedsimon}. The one-particle scalar
product $\langle f,g\rangle$ extends to a continuous non degenerate sesquilinear
form on $\KK$, also denoted $\langle \cdot ,\cdot \rangle$ and called the Wightman product. The scalar tachyon 
field $\phi(x)$ is  represented in $\KK$ by formulae of the usual kind. In particular $\phi$ may be split into its creation and annihilation parts $\phi(f)= \phi^+(f)+\phi^-(f)$, where
\begin{eqnarray}
&&(\phi(f)^+\Psi_h)^{(n)}(x_1\,,\ldots,\ x_n)
=\frac1 {\sqrt{n}} 
\sum_{j=1}^n f(x_j)h^{(n-1)}(x_1\,,\ldots,\hat x_j,\ldots, x_n),
\label{h.55.1})\\
&&(\phi(f)^-\Psi_h)^{(n)} (x_1\,,\ldots,\ x_n)
= \sqrt{n+1} \int f(x)\wh\WW(x,x')h^{(n+1)}(x',x_1\,,\ldots, x_n). 
\label{h.55.2}\end{eqnarray}
%The two-point function is obviously \beq \langle \phi(x)\Omega,\ \phi(y)\Omega\rangle = \wh\WW_n(x,\ y)\ , \label{h.80}\endq
It follows, as usual, that
\beq
[\phi^+(x), \phi^+(x')] = [\phi^-(x), \phi^-(x')] = 0,\ \ \
[\phi^-(x), \phi^+(x')] = \wh\WW_n(x,x'),
\label{h.82}\endq
so that 
\beq
[\phi(x), \phi(x')] = \wh\WW_n(x,x') - \wh\WW_n(x',x) = C_n(x, x').
\label{h.81}\endq
The infrared operators $\phi(v_j)$ belong to the center of the field algebra, 
i.e. they commute with the operator $\phi(f)$ for every $f\in\DD(X_d)$: 
\beq
[\phi(v_j), \phi(f)] = {1\over A(j)} \int_{X_d\times X_d}
[(\Box+m^2_n)\chi_j(x)]\,C_n(x,x')\,f(x')\,dx' =0
\label{h.88} \endq
since $(\Box+m^2_n)C_n(x,x') =0$.

Using (\ref{h.17}), Eq. (\ref{h.6}) can be rewritten as follows:
%\beq \langle (\Box +m^2_n)f,\ g\rangle = \sum_{j\in\JJ} A(j)\,\ovl{c_j(f)}\, \langle v_j\,,\ g\rangle, \label{h.83}\endq i.e.
\beq
\int_{X_d} \ovl{f(x)} \left\langle (\Box  +m^2_n )\phi(x)\,\Omega,\
\phi(g)\,\Omega\right\rangle dx = \int_{X_d}\ovl{f(x)}\sum_{j\in\JJ} A(j)\,
\Xi_j(x) \langle \phi(v_j)\,\Omega, \phi(g)\,\Omega\rangle\,dx .
\label{h.84}\endq
%We also have \begin{align} &((\Box +m^2_n)f,\ g) = \sum_{j\in\JJ} \langle (\Box +m^2_n)f,\ \chi_j \rangle \langle \chi_j\,,\ g \rangle \cr&= \sum_{j\in\JJ} \sum_{k\in\JJ}A(k) \ovl{c_k(f)}\,c_k(\chi_j)(v_j,\ g) = \sum_{j\in\JJ} A(j) \ovl{c_j(f)}\,(v_j,\ g),\label{h.85}\end{align}i.e.\beq\int_{X_d}\ovl{f(x)} ( (\Box_x +m^2_n)\phi(x)\,\Omega,\\phi(g)\,\Omega) = \int_{X_d}\ovl{f(x)}\sum_{j\in\JJ} A(j)\,\Xi_j(x) ( \phi(v_j)\,\Omega,\ \phi(g)\,\Omega)\ .\label{h.86}\endq
The non-degeneracy of the Wightman product and the factorization property 
of the $n$-point functions then allow us to conclude that the tachyon field 
satisfies the non-homogeneous equation 
\beq
(\Box +m^2_n)\phi(x) = \sum_{j\in\JJ} A(j)\,\phi(v_j)\,\Xi_j(x) \equiv\Theta_n(x) .
\label{h.87}\endq
This allows us  to reinterpret the whole construction as  a 
Gupta-Bleuler quantization of the tachyons: 
the physical states of the theory are annihilated by the destruction part 
of the rhs of Eq. (\ref{h.88}):
\begin{equation}
\Theta(x)^-|{phys}\rangle=0
\label{supp}
\end{equation}
Indeed the physical states in $\KK_1$ are the closure of $\EE_n$ in
the Hilbert space topology and, by (\ref{h.17}), they are exactly the
vectors in $\KK_1$ which are annihilated by all the operators
$\phi(v_j)^-$, $j\in \JJ$. Hence due to the Fock space construction
(see (\ref{h.55.1}), (\ref{h.55.2})), the physical states are those
which are annihilated by all the operators
$\phi(v_j)^-$, $j\in \JJ$. Thus they are also annihilated by
$\Theta(x)^-$ for all $x$, and the converse is also true by
smearing $\Theta(x)^-$ with $\chi_j(x)$. As a consequence, the usual homogeneous field equation holds on physical 
states, similarly to what happens with the Gauss law in QED 
\cite{strocchiwightman,strocchibook};
\begin{equation}
\langle {phys}|\ (\Box + m^2_n)\phi(x) \ |{phys}\rangle=0
\label{equationfield}
\end{equation}

\subsection{Gauge transformations}
There is indeed a gauge symmetry preserving the field equation (\ref{h.87}):
\begin{equation}
\phi(x) \longrightarrow \phi(x) + \sum \lambda_j \Xi_j(x).
\label{gauge}
\end{equation}
The infrared  operators may be used to define the gauge charge operators
\beq
Q_j = {1\over 2i}\Big [ \phi^+(v_j) - \phi^-(v_j)\Big ],
\label{h.90}\endq
that act as generators of the above gauge transformations. Indeed, 
by Eq. (\ref{h.82}),
\beq
[\phi^{\pm}(v_j),\ \phi(f)] = \mp \langle v_j\,\ f\rangle
= \mp c_j(f) =\mp \int_{X_d} \Xi_j(x)\,f(x)\,dx,
\label{h.89}\endq
and therefore
\beq
[Q_j,\ \phi(f)] = ic_j(f),\ \ \ \ [Q_j,\ \phi(x)] = i\Xi_j(x)\ .
\label{h.91}\endq
Is there any conserved current associated to the above charges? 
Suppose that  $f$ is a smooth solution of the Klein-Gordon equation and $\Sigma$
a smooth space-like hypersurface homotopic to the sphere $\{x :\, x^0=0\}$
in $X_d$. Then
\beq
\int_\Sigma f(x) {\buildrel \longleftrightarrow \over
{\partial_\mu}}
C_n(x, y)\,d\Sigma^\mu(x) = if(y)\ .
\label{h.92}\endq
Indeed, since $f$ and $C_n(x,\ y)$ are solutions of the
Klein-Gordon equation, the lhs is independent of $\Sigma$,
and we can choose $\Sigma = \{x\in X_d\ :\ x^0 = y^0\}$. Then
the result follows from the equal-time commutation relations.
(\ref{c.1}). This holds for any mass.
For each $k\in\JJ$ let
\beq
j_{k\mu}(x) = \Xi_k(x){\buildrel \longleftrightarrow \over
{\partial_\mu}} \phi(x) .
\label{h.93}\endq
Then, with $\Sigma$ as above we find
\beq
\left [ \int_\Sigma j_{k\mu}(x)\,d\Sigma^\mu(x)\, , \phi(y) \right ]
= i\Xi_k(y) .
\label{h.94}\endq
The integral in this formula has the same commutator with $\phi$
as $Q_k$. However $j_{k\mu}(x)$ is not a conserved current since
$\phi$ is not a solution of the Klein-Gordon equation. By (\ref{h.87})
\beq
\nabla^\mu j_{k\mu}(x) = -\Xi_k(x)\sum_{j\in\JJ} A(j)\,\phi(v_j)\,\Xi_j(x)\ .
\label{h.95}\endq
The function $x \mapsto \Xi_k(x)\Xi_j(x)$ is the restriction to $X_d$ of
a homogeneous polynomial of degree $2n$ on $M_{d+1}$. We can apply
the following lemma:

\begin{lemma}
Let $q$ be a polynomial of degree $m \ge 0$ on $M_{d+1}$.
There exists a $\CC^\infty$ function $f$ on $X_d$ such that
$\Box_{dS} f = q|X_d$.
\end{lemma}

\noindent {\bf Proof}
If $h$ is a homogeneous pseudo-harmonic polynomial of degree $s > 0$, i.e.
$\Box_{\rm Mink} h = 0$, then, as mentioned in Subsect. \ref{dal},
$(h|X_d) = \Box_{dS}[s(s+d-1)]^{-1}(h|X_d)$. If $h$ is a constant,
we may use (\ref{a.22}) with $n=0$, i.e. if ${z'} \in \TT_+$,
\beq
\Box_x \wh W_0(x,{z'}) = -{\Gamma(d)\over 4\pi^{d\over 2}}\ ,
\label{h.100}\endq
so the statement also holds in this case.
For a general homogeneous polynomial $q$ of degree $m \ge 0$ on
$M_{d+1}$, there exists (see \cite[p. 441]{Vilenkin}) a unique
expansion
\beq
q(x) = \sum_{k=0}^{[m/2]} (x\cdot x)^k h_{m-2k}(x)\ ,
\label{h.101}\endq
where each $h_{m-2k}$ is a pseudo-harmonic homogeneous polynomial of
degree $m-2k$, so that
\beq
q|X_d = \sum_{k=0}^{[m/2]} (-1)^k h_{m-2k}|X_d\ .
\label{h.102}\endq
Therefore the statement of the lemma holds.

By virtue of this lemma, there exist $\CC^\infty$ functions
$\vhi_{kj}$ on $X_d$ such that $\Xi_k\Xi_j = \Box_{dS}\vhi_{kj}$.
We can now define
\beq
j'_{k\mu}(x) = j_{k\mu}(x) + \sum_{j\in\JJ} A(j)\,\phi(v_j)\,
\partial_\mu \vhi_{kj}(x)\ .
\label{h.103}\endq
The current $j'_{k\mu}(x)$ has the same commutation properties as
$j_{k\mu}(x)$, but $\nabla^\mu j'_{k\mu}(x) = 0$.

\subsection{The case $n=1$}
The scalar tachyon whose squared mass is equal to the opposite of the space-time dimension $m_1^2 = -d$ plays a role in the construction of the de Sitter linearized quantum gravity \cite{antoniadis,folacci}. 
In this context the $d+1$ implementable gauge transformations are actually interpretable as describing a residual conformal symmetry. 
Let us summarize here the nice geometrical construction underlying this theory.
The pseudo-harmonic homogeneous polynomials of
degree $n=1$ on $M_{d+1}$ are just the linear homogeneous polynomials
and their space is spanned by $x_0,\ \ldots,\ x_d$. It is easy
to check that $\Xi_{00} = {\rm Const.\ }x_0$, while the other
$\Xi_{lM}$ are proportional to the other coordinate functions
$x_j$, $1\le j\le d$. The set of indices $\JJ$ becomes therefore
identical to the set $\{0,\ 1,\ \ldots,\ d\}$. Its elements
will be denoted with latin letters to avoid confusions with 
(arbitrary) coordinates on $X_d$. 
It is in fact convenient in
this case to renormalize the $\Xi_j(x)$ so that $\Xi_j(x) = x_j$
($0\le j\le d$), and contragrediently renormalize the $\chi^j$ and $v^j$.
We now have the following forms of (\ref{a.19}) and (\ref{h.1})
(see (\ref{a.22}))
\beq
G_1(x,\ x') = {\Gamma(d-1)\over (4\pi)^{d/2}} C_1^{d-1\over 2}(x\cdot x') =
{\Gamma(d)\over (4\pi)^{d/2}}(x\cdot x')\ ,
\label{i.10}\endq
\beq
(\Box_{x,x'} -d)\,\wh \WW_{n}(x,\ x') = 
{(d+1)\Gamma(d)\over (4\pi)^{d/2}}(x\cdot x') = 
\sum_j {(d+1)\Gamma(d)\over (4\pi)^{d/2}} x_j \eta^{jj} x'_j\ ,
\label{i.20}\endq
i.e. $A(l)$ has become $(-1)^l (4\pi)^{-d/2}(d+1)\Gamma(d)$,
and $A(j) = \eta^{jj}(4\pi)^{-d/2}(d+1)\Gamma(d)$.
In particular
\beq
v^j = \eta^{jj}{(4\pi)^{d/2}\over (d+1)\Gamma(d)} (\Box-d)\chi^j\ .
\label{i.30}\endq
In the case $n=1$ the gauge invariance is related to the existence
on $X_d$ of $(d+1)$ linearly independent ``conformal-Killing vector
fields'' (not included in the list of the true Killing vector fields,
which generate the Lorentz transformations).
Their role has been emphasized by \cite{antoniadis} and \cite{folacci}.
A vector field $w$ on $X_d$ is a conformal-Killing vector
field if it satisfies
\beq
\nabla_\mu w_\nu + \nabla_\nu w_\mu = g_{\mu\nu}f,\ \ \ 
f= {2\over d} \nabla^\mu w_\mu\ ,
\label{i.40}\endq
where the last equality follows from the first. If we impose
on $w$ the further requirement that
\beq
w_\mu = \partial_\mu h\ ,
\label{i.50}\endq
where $h$ is some $\CC^\infty$ function on $X_d$, this function
must satisfy 
\beq
\nabla_\mu\nabla_\nu h = {g_{\mu\nu}\over d}\Box h\ .
\label{i.60}\endq
Conversely if a function $h$ satifies the above equation,
then $w_\mu = \partial_\mu h$ satisfies (\ref{i.40}) with
$f = {2\over d}\Box h$. If, furthermore, $\Box h = dh$, then
$f = 2h$. It will now be shown that (\ref{i.60}) is satisfied
by $h = x_j = \Xi_j$, i.e. any of the Minkowskian coordinates.
By symmetry and analyticity it is sufficient to show this for
$x_0$. We already know that $(\Box-d)x_j=0$. We thus need to prove
that
\beq
\nabla_\mu\nabla_\nu x_0 = g_{\mu\nu}x_0\ .
\label{i.70}\endq
By analytic continuation, it suffices to prove this in one
analytic coordinate patch. We choose (as independent coordinates
on $X_d$) $x^0,\ldots,\ x^{d-1}$, with 
$(x^d)^2 = 1+\sum_{\mu=0}^{d-1} \eta_{\mu\mu}(x^\mu)^2$ in the region $x^d>0$.
Greek indices will take the values $0,\ldots,\ d-1$. We continue
to denote $x_\mu = \eta_{\mu\mu}x^\mu$ (no sum).
We then find
\begin{align}
& g_{\mu\nu}(x) = \eta_{\mu\nu} - {x_\mu x_\nu \over x_d^2}\ ,
\label{i.80}\\
& g^{\mu\nu}(x) = \eta^{\mu\nu} + x^\mu x^\nu\ ,
\label{i.81}\\
& \Gamma_{\mu\nu}^\rho(x) = -g_{\mu\nu}(x)\,x^\rho\ .
\label{i.82}\end{align}
This gives:
\beq
\nabla_\mu \partial_\nu x^0 = \nabla_\mu \eta_\nu^0
= -\Gamma_{\mu\nu}^\rho\eta_\rho^0 = -\Gamma_{\mu\nu}^0 = g_{\mu\nu}x^0\ ,
\label{i.83}\endq
i.e. (\ref{i.70}) holds as announced. This shows that, for each 
$j = 0,\ldots.\ d$ the vector field $w_\mu = \partial_\mu x^j$ is
a conformal-Killing vector field.

\section{Positive quantization of tachyons fully breaking the de Sitter symmetry}
\label{noninvar}
In this final section we construct all the positive 
quantizations of the tachyons and show that, for every real $\lambda$ 
including the integers, it is possible to write two-point
functions which satisfy the Klein-Gordon equation 
with squared mass $m_\lambda^2$ and the canonical commutation relations,
and possess the positive semi-definiteness property, but are not de Sitter
invariant. 
This is possible because of the existence of compact spacelike 
sections in the de Sitter space-time 
which reduces the infrared problem to taking care of a finite number 
of degrees of freedom. 
Positivity may be restored while keeping locality at the price of breaking 
the de Sitter invariance. Several authors look at this as a welcome or else an 
unavoidable feature  of the physical world. Others think that it 
is only a gauge artefact.

As before we assume that $\Re \lambda +(d-1)/2 >0$, 
$l$ and $n$ are non-negative integers, $x = (t, \x)$, $z = i\sh t$, $x' = (t', \x')$, $z' = i\sh t'$.
Suppose $\lambda$ is not a non-negative integer. The most general distribution $F$ on $X_d\times X_d$
satisfying the Klein-Gordon equation with squared mass $m^2_\lambda$
has  the form
\begin{align}
F_\lambda(x, x') & = \sum_{lM} F_{\lambda l M}(x, x') = (\ch t)^{-\p} (\ch t')^{-\p} \sum_{lM} V_{lM}(\x, \x') \Big [ 
A \, \P_{\lambda+\p}^{-l-\p}(z)\P_{\lambda+\p}^{-l-\p}(-z') + \cr & +
B\,  \P_{\lambda+\p}^{-l-\p}(-z)\P_{\lambda+\p}^{-l-\p}(z') +
E\, \P_{\lambda+\p}^{-l-\p}(z)\P_{\lambda+\p}^{-l-\p}(z') + 
G\,  \P_{\lambda+\p}^{-l-\p}(-z)\P_{\lambda+\p}^{-l-\p}(-z')
\Big ] .
\label{m.10bis}\end{align}
The coefficients $A$, $B$, $E$, $G$ may depend on $\lambda$, $l$ and $M$.
Imposing the canonical commutation relations (\ref{tpabcomm}) and 
(\ref{functionaleq}) requires that
\beq
A-B = \gamma_{l} (\lambda).
\label{m.20}\endq
We now assume $\lambda$, $t$ and $t'$ are real. Then 
$\P_{\lambda+\p}^{-l-\p}(-z) = \ovl{\P_{\lambda+\p}^{-l-\p}(z)}$ 
(see Eq. (\ref{reality}))  and the condition for $F_{\lambda l M}(x,\ x')$ to be
positive semi-definite is that
\beq
A \ge 0,\ \ \ B \ge 0,\ \ \ G = \ovl{E},\ \ \ |E|^2 \le AB\ .
\label{m.40}\endq
The normalization factor $\gamma_l(\lambda)$ is always positive for standard theories with $m^2_\lambda$ positive, while it may diverge and/or become negative in the tachyonic region of the spectrum (see figure 1). Always supposing that $\lambda$ is not a non-negative integer, let $\veps$ be the sign of $\gamma_l(\lambda)$ i.e. $\veps = \pm 1$, $\veps \gamma_l({\lambda}) >0$. The above
conditions can be satisfied by taking
\beq
\begin{array}{llll}
A = \gamma_l(\lambda)\ch^2 \alpha_l\,,\ \ &
B = \gamma_l(\lambda)\sh^2 \alpha_l\,,\ \ & 
E= \ovl{G} = \gamma_l(\lambda) \rho_l e^{-i\beta_l}\ch \alpha_l\,\sh \alpha_l\ 
& {\rm if}\ \ \veps = 1\cr
A = -\gamma_l(\lambda)\sh^2 \alpha_l\,,\ \ &
B = -\gamma_l(\lambda)\ch^2 \alpha_l\,,\ \ & 
E= \ovl{G} = -\gamma_l(\lambda) \rho_l e^{i\beta_l}\ch \alpha_l\,\sh \alpha_l\ 
& {\rm if}\ \ \veps = -1\ ,
\end{array}
\label{m.50}\endq
with real $\alpha_l$ and $\beta_l$, and $0 \le \rho_l \le 1$.
In general the constants $\alpha_l$, $\beta_l$, and $\rho_l$ may
depend on $\lambda$ and $M$, but we omit this to simplify the notations.
This gives
\begin{align} &F_{\lambda l M}(x, x') = (\ch t)^{-\p} (\ch t')^{-\p}  V_{lM}(\x,\x') \times \cr & \times \Big[  \veps \gamma_l(\lambda)
\Big ( \ch\alpha_l \P_{\lambda+\p}^{-l-\p}(\veps z) 
+e^{i\beta_l} \sh \alpha_l\,\P_{\lambda+\p}^{-l-\p}(-\veps z) \Big ) 
\Big ( \ch\alpha_l \P_{\lambda+\p}^{-l-\p}(-\veps z') 
+e^{-i\beta_l} \sh \alpha_l\,\P_{\lambda+\p}^{-l-\p}(\veps z') \Big ) \cr
&- \veps \gamma_l(\lambda)(1-\rho_l)\ch \alpha_l\,\sh \alpha_l\, 
\Big (e^{-i\beta_l}\P_{\lambda+\p}^{-l-\p}(\veps z)\P_{\lambda+\p}^{-l-\p}(\veps z') 
+ e^{i\beta_l}\P_{\lambda+\p}^{-l-\p}(-\veps z)\P_{\lambda+\p}^{-l-\p}(-\veps z')\Big )
\Big]. 
\label{m.70}\end{align}
This formula encodes every possible canonical quantization of the de Sitter 
scalar tachyons for non-integer real $\lambda$. 
The de Sitter symmetry is fully broken in the sense that there the 
corresponding Hilbert spaces do not contain any invariant subspace.

Let us now consider the case $\lambda \rightarrow n$. We show that the 
parameters $\alpha_l(\lambda)$, $\beta_l(\lambda)$ and $\rho_l(\lambda)$ 
can be chosen as functions of $\lambda$ 
in such a manner that all
$F_{\lambda l M}(x,x')$ have finite limits when $\lambda \rightarrow n$. 
We obtain in this way canonical quantum 
fields corresponding to any real value of $\lambda$ - including the integers. 
A similar limiting procedure has been described  for the massless case $n=0$ 
in \cite{mottola5}.

When $l> n$ there is nothing to do: the normalization factor 
$\gamma_l(\lambda) >0$ is continuous in a small real neighborhood $\omega$ of
$n$.
We can then fix $\alpha_l$, $\beta_l$, and $\rho_l$ in $\omega$ 
independently of $\lambda \in \omega$, and
define $F_{nlM}$ by the formula (\ref{m.70}) with $\lambda =n$
and $\veps=1$.

To discuss the case $l \le n$ we need the general identities 
(\cite[p. 144]{HTF1})
\begin{eqnarray}
&&\P_\nu^\mu(-z) = \P_\nu^\mu(z)\cos\pi(\mu+\nu)
-{2\over \pi}\Q_\nu^\mu(z)\sin\pi(\mu+\nu),
\label{m.80}
\\ && \Q_\nu^\mu(-z) = -\Q_\nu^\mu(z)\, \cos\pi(\mu+\nu)
-{\pi\over 2} \P_\nu^\mu(z) \,\sin\pi(\mu+\nu).
\label{m.81}\end{eqnarray}
which imply, for $l\le n$ (but not for $l>n$):
\beq
\P_{n+\p}^{-l-\p}(z) = (-1)^{n-l}\P_{n+\p}^{-l-\p}(-z),\ \ \ \ 
\Q_{n+\p}^{-l-\p}(z) = (-1)^{n-l+1}\Q_{n+\p}^{-l-\p}(-z).
\label{m.83}\endq
Taking $\lambda$ real with
$0< s=|\lambda -n|<1$,
we will let $\lambda$ tend to $n$ keeping the sign of
$\lambda - n$ fixed. For every $l\le n$  we denote
\beq
\veps = \sign((-1)^{n-l+1}(\lambda-n)), \ \ \ \ \lambda = n+\veps\tau s,
\ \ \ \tau = (-1)^{n-l+1}\ .
\label{n.90}\endq
(this notation omits, for simplicity, the $l$ dependence of $\tau$ and 
$\veps$) . Then
\beq
0< \veps\gamma_l(\lambda) = {\pi \Gamma(1+l+n+2\p +\veps\tau s) \over
2\sin(\pi s)\Gamma(1+n-l+\veps\tau s)}  
\sim {\Gamma(1+l+n+2\p)\over 2(n-l)!\,s}\ \ \ 
{\rm as}\ \ s\rightarrow 0.
\label{n.91}\endq
We choose:
\beq
e^{\alpha_l} = {1\over A \sqrt{s}},\ \ \ \ 
\beta_l = (n-l+1)\pi-2ABs,\ \ {\rm hence}\ \ \ 
e^{i\beta_l} = \tau e^{-2iABs},\ \ \ \ \rho_l = e^{-2A^2 C s^2}.
\label{n.100}\endq
Here $A>0$,  $B\in \bR$, and $C \ge 0$ are (new) constants which again may 
depend on $l$ and also on $M$. 
By using (\ref{m.80}) we get
\begin{align}
&
\ch(\alpha_l)\, \P_{\lambda+\p}^{-l-\p}(\veps z) +
e^{i\beta_l} \sh(\alpha_l)\,\P_{\lambda+\p}^{-l-\p}(-\veps z)=
%\label{n.101} 
\cr
 &= \left [\ch(\alpha_l) - e^{-2iABs}\sh(\alpha_l)\,
\cos\pi(s) \right ]\P_{n+\veps\tau s+\p}^{-l-\p}(\veps z) 
+ {2\over\pi}e^{-2iABs}\sh(\alpha_l)\,\sin\pi(\veps\tau s)
\Q_{n+\veps\tau s+\p}^{-l-\p}(\veps z)\cr 
& \sim \sqrt s \left[(A+iB)\P_{n+\p}^{-l-\p}(\veps z) - 
{\veps (-1)^{n-l}\over A} \Q_{n+\p}^{-l-\p}(\veps z)
+ {\rm O}(s)\right]\
\label{n.102}\end{align}
when $s \rightarrow 0$. 
The behavior of the last term in (\ref{m.70}) as $s\rightarrow 0$
is straightforward. Finally
\begin{align}
& F_{nlM}((t, \x), (t', \x')) =
\lim_{s\rightarrow 0}F_{\lambda lM}((t,\x), (t',\x')) =\cr
&= {\Gamma(1+l+n+2\p)\over 2(n-l)!}
(\ch t)^{-\p}\,(\ch t')^{-\p} \Bigg \{\cr
&
\left [(A+iB)\P_{n+\p}^{-l-\p}(z) - 
{(-1)^{n-l}\over A}\Q_{n+\p}^{-l-\p}(z)\right ]\,
\left [(A-iB)\P_{n+\p}^{-l-\p}(-z') - 
{(-1)^{n-l}\over A}\Q_{n+\p}^{-l-\p}(-z')\right ]\cr
&+ C\P_{n+\p}^{-l-\p}(z)\P_{n+\p}^{-l-\p}(-z') \Bigg \}\,
\Ylm({\x})\,\Ylm({\x'}),\ \ \ \ 
z = i\sh t,\ \ \ z' = i\sh t'.
\label{n.110}\end{align}
We have repeatedly used (\ref{m.83}). 
If the constants $A$, $B$, and $C$ have been chosen independently of 
$\veps$, the limit $F_{nlM}$ is
independent of $\veps$. 
For real $t$ and $t'$,   
$\P_{n+\p}^{-l-\p}(-z') = \ovl{\P_{n+\p}^{-l-\p}(z')}$
and $\Q_{n+\p}^{-l-\p}(-z') = \ovl{\Q_{n+\p}^{-l-\p}(z')}$,
so that 
$F_{nlM}(x, x')$ is a positive semi-definite kernel, as it must be as the limit
of the positive semi-definite kernel $F_{\lambda lM}(x,x')$.
Again by  using (\ref{m.83}) we obtain
\begin{align}
&F_{nlM}(x, x') - F_{nlM}(x', x) = 
C_{nlM}(x,x') = \cr
&{\Gamma(1+l+n+2\p)\over (n-l)!}(\ch t)^{-\p}(\ch t')^{-\p}
\Big [ \P_{n+\p}^{-l-\p}(z) \Q_{n+\p}^{-l-\p}(z')
- \P_{n+\p}^{-l-\p}(z') \Q_{n+\p}^{-l-\p}(z) \Big ]\Ylm(\x)\Ylm(\x').\cr &
\label{n.112}\end{align}
This last expression, as expected, is independent of $A$, $B$, $C$, 
$\veps$ and $M$. It gives an explicit form for 
$C_{nlM}(x, x') = C_{\lambda lM}(x,x')|_{\lambda=n}$, $l \le n$.

Therefore the general solution depends on three constants $A$, $B$, and $C$ 
(for each $lM$ with $l\le n$) that are only required to be real
with $A>0$ and $C\ge 0$. Making an arbitrary choice for each $lM$ produces an
$F_{nlM}(x,\ x')$ given by (\ref{n.110}) for $l\le n$, and by
(\ref{m.70}) (with $\lambda =n$ and $\veps=1$) for $l> n$.
Under mild
conditions the series $F_n(x,x') = \sum_{lM}F_{nlM}(x,x')$ 
will converge to a distribution with the following properties:
\begin{enumerate}
\item $(\Box_x + n(n+d-1))F_n(x,\ x') = (\Box_{x'} + n(n+d-1))F_n(x,\ x')=0$, 
\item  $F_n(x,\ x')$ is a positive semi-definite kernel,
\item
 $F_n(x,\ x')-F_n(x',\ x) = C_n(x,\ x')$.
\end{enumerate}
Therefore $F_n(x,x')$ can be taken as the two-point function
of a free field with squared mass $m_n^2 = -n(n+d-1)$, which
however is not Lorentz invariant and is not maximally analytic.
Moreover $F_n(x,\ x')$ is the limit from $\lambda > n$ and
from $\lambda < n$ of two-point functions with similar properties.

A particularly simple choice is to let the two-point function $F_{nlM}$ 
coincide, for $l>n$, with the covariant 
maximal analytic $\widehat W_{nlM}$ i.e. to choose the infrared regular part of 
the two-point function according to the  Bunch-Davies prescription 
(\ref{a.6}) :
\begin{align}
&F_{nlM}(x,\ x') = \WW_{nlM}(x,\ x') = \wh \WW_{nlM}(x,\ x') =\cr
&= \gamma_n(l)(\ch t)^{-\p}(\ch t')^{-\p}
\P_{n+\p}^{-l-\p}(i\sh t)\P_{n+\p}^{-l-\p}(-i\sh t') \Ylm(\x)\Ylm(\x')\ ,
\ \ l>n\ .
\label{n.120}\end{align}
%%%%%%%%%%%%%%%%%%%%%%%%%%%%%%%%%%%%%%%%%%%%%%%%%%%%%%%%%%%%%%%%%%%
In particular Allen and Folacci's de Sitter breaking quantization of the 
massless scalar field and Folacci's quantization of the $n=1$ conformal 
field belong to this family.
Whatever choice is made of the remaining part, the resulting $F_n$ cannot 
determine the same sesquilinear form on the space
$\EE_n$ as $\wh \WW_n$.
Indeed if $f \in \EE_n$, Eq. (\ref{e.52}) implies that 
for $l \le n$,
\beq
\int_{X_d \times X_d} f(x)F_{nlM}(x,\ x')\ovl{f(x')}\,dx\,dx' =
 {\Gamma(1+l+n+2\p)\over 2(n-l)!} \Big |
\int_\bR {1\over A}(\ch t)^{d-1-\p}\wt f_{lM}(t)\,\Q_{n+\p}^{-l-\p}(i \sh t)\,
dt \Big |^2 .
\label{e.141}\endq
Since $\P_{n+\p}^{-l-\p}$ and $\Q_{n+\p}^{-l-\p}$ are linearly independent,
it is possible to find (many) $\wt f_{lM} \in \DD(\bR)$ such that
(\ref{e.52}) holds while the rhs of (\ref{e.141}) is not 0.
This points towards the {\em physical inequivalence}
of the two constructions. 
The polynomials in the operators $\phi(f)$, $f\in \EE_n$, which can
be regarded as the observable (gauge invariant) quantities, 
do not have the same expectation values in the two theories.

On the other hand let $f\in \DD(X_d)$ satisfy $f(x) = (-1)^nf(-x)$.
Since  $\Ylm(-\x) = (-1)^l\Ylm(\x)$, %$V_{lM}(-\x, -\y) = V_{lM}(\x, \y)$, 
it follows that $f_{lM}(-x) = (-1)^nf_{lM}(x)$ and hence 
$\wt f_{lM}(t) = (-1)^{n-l}\wt f_{lM}(-t)$. Since, for $l\le n$,
$\Q_{n+\p}^{-l-\p}(i \sh t) = (-1)^{n-l+1}\Q_{n+\p}^{-l-\p}(-i \sh t)$,
\beq
\int_\bR (\ch t)^{d-1-\p}\wt f_{lM}(t)\,\Q_{n+\p}^{-l-\p}(i \sh t)\,dt = 0\ \ \ 
\forall l\le n\ .
\label{e.150}\endq
If we assume that $f$ also belongs to $\EE_n$ we get from (\ref{e.141})
\beq
\int_{X_d \times X_d} f(x)F_{nlM}(x,\ x')\ovl{f(x')}\,dx\,dx' = 0\ \ \ 
\forall l\le n\ .
\label{e.151}\endq
Let us suppose that the choice (\ref{n.120}) has been made,
i.e. $F_{nlM} = \wh \WW_{nlM}$ for all $l > n$. Then $F_n$ and 
$\wh \WW_n$ define the same (Lorentz invariant) sesquilinear form on the
subspace
\beq
\FF_n = \{f \in \EE_n\ :\ f(x) = (-1)^nf(-x)\ \ \ \forall x \in X_d\}\ .
\label{e.152}\endq
In this particular case, the Hilbert space determined by 
the non-invariant sesquilinear form $F_n$
nevertheless contains a subspace equipped with a unitary representation
of the Lorentz group.
Let $\NN$ denote the null-space of $\wh \WW_n$ as a positive-semi-definite 
sesquilinear form on $\EE_n$. According to Lemma \ref{physpos}, the map
$f \mapsto \wt f_{1-d-n, +}$ defines an injective map of $\EE_n/\NN$ into
$\wt \EE_n^{(2)}$. It is easy to find elements $f$ of $\FF_n$ such
that $\wt f_{1-d-n, +} \not= 0$.
On the completion of $\wt \EE_n^{(2)}$ (with respect to the scalar product 
(\ref{k.15})), the natural
representation of the Lorentz group is irreducible and belongs to the
discrete series with squared mass $m_n^2$ (see \cite[chap. X]{Vilenkin},
\cite{Molcanov}).
Therefore the completions of the images of $\EE_n$ and $\FF_n$ under the same 
map must be the same, i.e. the whole completion of $\wt \EE_n^{(2)}$.
The sesquilinear form $\wh \WW_n$ determines (by the usual
quotienting and completion process) the same Hilbert space, whether
one starts from $\EE_n$ or $\FF_n$, and the sesquilinear form $F_n$
determines the same Hilbert space starting from $\FF_n$.
The unitary representation of
the Lorentz group acting there is irreducible with squared mass $m_n^2$.

\section*{Acknowledgements}
We thank Jacques Bros, Antoine Folacci and Richard Woodard for discussions and correspondence. U.~M. thanks the IHES for their kind hospitality.

\newpage
\appendix

\section{Appendix. Gegenbauer polynomials and hyperspherical
harmonics}
\label{gegen}

This appendix follows \cite[Chap. XI]{HTF2},
\cite[Chap. IX]{Vilenkin}, and \cite[Chap. IV]{Szego}, to which
we refer the reader for more details.
For a positive integer $n$ and a complex $\nu$, the
Gegenbauer polynomial $C_n^\nu$ %\equiv P_n^{(\nu)}$
is defined as
\begin{align}
C_n^\nu(z) &= {\Gamma(n+2\nu)\over
\Gamma(n+1)\Gamma(2\nu)} F \left ( -n,\ n+2\nu\ ;\
\nu+\half\ ;\ {1-z\over 2}\right )
\label{u.2}\\
%&= {2^{\half-\nu}\Gamma(\half)\Gamma(n+2\nu)(z^2-1)^{{1\over 4}-{\nu\over 2}}\over\Gamma(n+1)\Gamma(\nu)}P_{n-\half+\nu}^{\half-\nu}(z)\label{u.2.1}\\
&= {2^{\half-\nu}\Gamma(\half)\Gamma(n+2\nu)
(1-z^2)^{{1\over 4}-{\nu\over 2}}\over
\Gamma(n+1)\Gamma(\nu)}
\P_{n-\half+\nu}^{\half-\nu}(z)\ .
\label{u.3}\end{align}
Here $\P_\nu^\mu$ is the ``Legendre function on the cut'' mentioned in
Sect.~\ref{masskg}.
Also:
\beq
\sum_{n=0}^\infty w^n C_n^\nu(z) = (1-2zw+w^2)^{-\nu} =
\sum_{n=0}^\infty (-w)^n C_n^\nu(-z),\ \ \
C_n^\nu(-z) = (-1)^n C_n^\nu(z)\ .
\label{u.3.1}\endq
As a consequence of this, if $\tau,\ z \in \bC$, $\tau^n C_n^\nu(z/\tau)$
is a homogeneous polynomial of degree $n$ in $\tau$ and $z$
and is even in $\tau$, so that it is a polynomial in $z$ and $\tau^2$.
In particular $C_0^\nu(z) = 1$ and $C_1^\nu(z) = 2\nu z$.

Let $E_N$ denote the $N$-dimensional Euclidean space %, i.e.$\bR^N$
equipped with the scalar product $x\cdot y = \sum_{j=1}^N x_jy_j$ and $S^{N-1}$ the unit sphere
in $E_N$ equipped with the measure
$d\sigma(x) = 2\delta(1-x\cdot x)\,dx\ldots dx_N$, with total
measure $\Omega_N = 2\pi^{N/2}/\Gamma(N/2)$. We suppose $N\ge 2$.
We denote $e_j$ the vector with components $e_{jk} = \delta_{jk}$.
A function $f$ on $E_N$ is called harmonic if $\Delta f =0$.
%$\Delta = \partial\cdot\partial$.
For a fixed integer $l \ge 0$, the vector space $\HH^{N,l}$ of
complex harmonic polynomials homogeneous of degree $l$ on $E_N$ has
dimension
\beq
h(N,\ l) = {(2l+N-2)(N+l-3)!\over (N-2)!l!}\ .
\label{u.4}\endq
$\HH^{N,l}$ is a Hilbert space with the scalar product
\beq
(f,\ g) = \int_{S^{N-1}} \ovl{f(x)}g(x)\,d\sigma(x)\ ,
\label{u.5}\endq
and the natural (quasi-regular) representation of the rotation
group of $E_N$ reduces to a unitary irreducible (for $N\ge 3$)
representation on $\HH^{N,l}$. An orthonormal basis of $\HH^{N,l}$ is
called a system of hyperspherical harmonics. For any such basis
$\{S_j\}_{1\le j\le h(N,\ l)}$ the following identity holds for $\x, \y \in S^{N-1}$
(\cite[11.4 (2). p. 243]{HTF2}):
\begin{align}
\sum_{j=1}^{h(N,\ l)} S_j^*(\x) S_j(\y) &=
\sum_{j=1}^{h(N,\ l)} S_j(\x) S_j^*(\y) %={h(N,\ l)\over \Omega_N}{C_l^{N-2\over 2}(x\cdot y) \over C_l^{N-2\over 2}(1)}
= {(2l+N-2)\Gamma(N/2)\over 2\pi^{N/2}(N-2)}\,C_l^{N-2\over 2}(\x\cdot \y)
%\cr & x\cdot x = y\cdot y =1,\ \ \\Omega_N = {2\pi^{N/2}\over \Gamma(N/2)}\ .
\label{u.6}\end{align}
The corresponding identity for $x\cdot x \not= 1$ and
$y\cdot y \not= 1$ is obtained by homogeneity by
applying the above formula to $x/\sqrt{x\cdot x}$ and
$y/\sqrt{y\cdot y}$ and using
$S_j(x) = (x\cdot x)^{l/2}S_j(x/\sqrt{x\cdot x})$. In particular, denoting $r =\sqrt{x\cdot x}$ and
$s = \sqrt{y\cdot y}$
\beq
r^l s^l C_l^{N-2\over 2}(x\cdot y/rs)
\label{u.7}\endq
is a homogeneous  polynomial in $x$, $y$, $r$, $s$ (since
it is even in $r$ and in $s$ as remarked above), harmonic in $x$ and in $y$.

A particular orthonormal basis $\{\Y_K^{N,l}\}$ of $\HH^{N,l}$ is labelled
by the multi-indices $K$ of the form
\beq
K = (k_1,\ldots,\ \pm k_{N-2}),\ \ \
l=k_0 \ge k_1 \ge \ldots \ge k_{N-2} \ge 0,
\label{u.8}\endq
where the $k_j$ are integers:
\beq
\Y_K^{N,l}(x) = A_K^{N,l} \prod_{j=0}^{N-3} r_{N-j}^{k_j-k_{j+1}}
C_{k_j-k_{j+1}}^{{N-j-2\over 2}+k_{j+1}}\left ({x_{N-j}\over r_{N-j}}\right )
\times (x'\pm ix)^{k_{N-2}}\ .
\label{u.9}\endq
Here $r_k^2 = x^2 +\ldots+x_k^2$. We will rather use a real
basis $\{Y_K^{N,l}\}$ given by
\begin{align}
Y_K^{N,l} &= \Y_K^{N,l}\ \ \ {\rm if}\ \ k_{N-2} = 0\ ,\cr
Y_{(k_1,\ldots,\ +k_{N-2})}^{N,l} &= {1\over \sqrt{2}} \Y_{(k_1,\ldots,\ +k_{N-2})}^{N,l}
+ {1\over \sqrt{2}} \Y_{(k_1,\ldots,\ -k_{N-2})}^{N,l}\ \ \ {\rm if}\ \ k_{N-2}>0\ ,\cr
Y_{(k_1,\ldots,\ -k_{N-2})}^{N,l} &= {1\over i\sqrt{2}} \Y_{(k_1,\ldots,\ +k_{N-2})}^{N,l}
- {1\over i\sqrt{2}} \Y_{(k_1,\ldots,\ -k_{N-2})}^{N,l}\ \ \ {\rm if}\ \ k_{N-2}>0\ .
\label{u.10}\end{align}
The positive constants $A_K^{N,l}$ ensure the orthonormality of these bases.
For $l=0$, $Y_0^{N,0}$ is the constant $\Omega_N^{-1/2}$.
For $l>0$, in the case $K = 0$ (i.e. $k_j=0$ for all
$j\ge 1$) $Y_0^{N,l} = \Y_0^{N,l}$ depends only on $x_N$ and $r_N$,
and $Y_K^{N,l}(e_N) =0$ if $K \not= 0$. In the case $l=1$, $K$ can
take the values $K_0\,,\ldots,\ K_{N-1}$ where
$K_0= (0,\ldots,\ 0)$., $K_1 = (1,\ 0,\ldots 0)$, etc.,
$K_{N-2}= (1,\ldots,\ +1)$, $K_{N-1}= (1,\ldots,\ -1)$.
It is clear that $Y_{K_j}^1(x)$ is proportional to $x_{N-j}$ and the
constant factor is determined by the orthonormality property, so
that, finally, $Y_{K_j}^1(x) = (N/\Omega_N)^{1/2}x_{N-j}$, $1\le j \le N$.
In this case we may abbreviate the notation $Y_{K_j}^1$ to $Y_{N-j}^1$.

Let $K$ be a multi-index as in (\ref{u.8}), $m$ and $l$ be integers
such that $m \ge l\ge k_1$ and $L=(l,K)$ be
the multi-index in $N+1$ variables given by
\beq
L = (l_0 = m,\ l_1 = l,\ldots,\ l_j = k_{j-1},\ldots,\
\pm l_{N-1} = \pm k_{N-2})\ .
\label{u.11}\endq
Then
\beq
\Y_{(l,K)}^{N+1,m}((x,\ldots,x_{N+1})) = {A_{(l,K)}^{N+1,m}\over A_K^{N,l}}
r_{N+1}^{m-l} C_{m-l}^{{N-1\over 2}+l}\left({x_{N+1}\over r_{N+1}}\right )
\Y_K^l((x,\ldots,x_N))\ .
\label{u.12}\endq
The same holds for the corresponding $Y$.

Let us choose $N=d$. Let $E_{d+1}^{(c)}$ be the complexified of $E_{d+1}$,
i.e. $\bC^{d+1}$ equipped with the same scalar product as $E_{d+1}$.
Let $J$ be the bijection of $M_{d+1}^{(c)}$ onto $E_{d+1}^{(c)}$ defined by
\beq
(Jx)_{d+1} = ix^0,\ \ \ (Jx)_j = x^j\ \ \ \hbox{for}\ 1\le j \le d,
\label{u.13}\endq
\beq
(Jx\cdot Jy)|_{E_{d+1}^{(c)}} = -(x\cdot y)|_{M_{d+1}^{(c)}}\ .
\label{u.13.1}\endq
If $p$ is a homogeneous pseudo-harmonic polynomial on $M_{d+1}$ (i.e.
$\Box_{\rm Mink} p =0$), it extends to a homogeneous pseudo-harmonic
polynomial on $M_{d+1}^{(c)}$ and $x \mapsto p(J^{-1}x)$ restricted to
$E_{d+1}$ is a homogeneous harmonic polynomial on $E_{d+1}$. Thus,
by analytic continuation, we find that the space
$\underline{\HH^{d+1,m}}$ of homogeneous pseudo-harmonic polynomials
of degree $m$ on $M_{d+1}$ is of dimension $h(N+1,m)$ and carries
an irreducible representation of the Lorentz group on $M_{d+1}$.
A basis of this space is provided by
\beq
 Y_L^{d+1,m}(x^1,\ldots,x^d,\ ix^0).
\label{u.14}\endq
Rewriting $L$ as $(l,K) = (l,\ k_1,\ldots,\ \pm k_{d-2})$
with $m\ge l\ge k_1 \ge \ldots \ge k_{d-2} \ge 0$, it follows
that a basis of $\underline{\HH^{N+1,m}}$ is provided by the
polynomials
\beq
\underline{Y_L^{d+1,m}}(x^0,\ \vec{x}) = {A_{(l,K)}^{N+1,m}\over A_K^{N,l}}
i^{l-m}(-(x\cdot x))^{m-l\over 2}C_{m-l}^{{d-1\over 2}+l}
\left({ix^0\over (-(x\cdot x))^{1/2}}\right )
Y_K^l(\vec{x})\ .
\label{u.15}\endq
As noted before, this is a real polynomial in $x$, whatever the
determination chosen for $(-(x\cdot x))^{1/2}$.

\section{Appendix. Expansion of plane waves into hyperspherical
harmonics}
\label{plwaves}
In this appendix we derive the expansion of the de Sitter plane wave
$z \mapsto (z\cdot \xi)^\lambda$ into hyperspherical harmonics,
where $z\in \TT_\pm$ and $\xi \in C_+\setminus \{0\}$ (see (\ref{a.4})).
We also derive a similar expansion for
$(\xi,\ \xi') \mapsto (\xi\cdot \xi')^\lambda$, where both $\xi$ and
$\xi'$ belong to $C_+\setminus \{0\}$.

For $x \in X_d$ and $\xi \in S_0$ we use the coordinates $x= x(t,\ \x)$
and $\xi = \xi(\xx)$ given in (\ref{s.4}) and (\ref{sphericcone}).

\subsection{Waves}

The representation
\begin{equation}
{a}^{\lambda}\, =
\frac{1}{{\Gamma}\left(-\lambda\right)}\int
_{0}^{\infty}\frac{du}{u}\   u^{-\lambda}\, e^{-u\,a} .
\label{gamma}
\end{equation}
is valid for ${\Re}\, \lambda <0$ and
${\Re}\, a>0$. Let $x+iy \in \TT_+$ and $\xi \in S_0$. Then
\begin{equation}
\Re (-i (x +iy)\cdot \xi) = y\cdot \xi > 0,
 \label{pwave}
\end{equation}
so that
\begin{equation}
{((x +iy)\cdot \xi) }^{\lambda}\, =
\frac{e^{\frac {i\pi\lambda}2}}{{\Gamma}\left(-\lambda\right)}\int
_{0}^{\infty}\frac{du}{u}\   u^{-\lambda}\, e^{i u (x +iy)\cdot \xi } .
\label{gamma1}
\end{equation}
If we consider a purely imaginary event such as
\begin{eqnarray}
&&x \left(t+\frac {i\pi}2,\,\x\right) = \left\{
\begin{array}{l} x^0 =\sinh \left(t+\frac {i\pi}2\right) = i \cosh t \\
\vec x = \cosh  \left(t+\frac {i\pi}2\right)\x = i \sinh t \  \x
\end{array}  \right.
\label{imagtube}
\end{eqnarray}
we get
\beq
\left[x\left(t+\frac {i\pi}2,\, \x\right) \cdot \xi(\xx)\right]^\lambda =
\frac{e^{\frac {i\pi\lambda}2}}{{\Gamma}\left(-\lambda\right)}
\int_{0}^{\infty}\frac{du}{u}\
u^{-\lambda}\, e^{- u \cosh t + u  \sinh t \ \x \cdot {\xx}}
\label{opq1}
\endq
Recall now the  well-known Neumann expansion of the exponential plane wave
in terms of Gegenbauer polynomials  (see e.g. \cite[7.10.1 (5) p. 64]{HTF2}) :
\begin{equation}
 e^{i\gamma z} = \left(- \frac 2 z  \right)^\nu \Gamma(\nu)
\sum_{l=0}^{\infty}(-i)^l (\nu+l)C_l^\nu(\gamma)J_{\nu+l}(-z).
\label{opq1.1}\end{equation}
By setting $\nu = \frac{d-2}{2}$ and taking
advantage of the well-known expansion of $C_l^\nu({ \x}\cdot \xx)$ in
terms  of the hyperspherical harmonics (see Appendix \ref{gegen}):
\begin{equation}
e^{u \sinh t \ { \x }\cdot { \xx}} = (2\pi)^\frac{d}{2}
\sum _{l=0}^\infty i^{l+\frac{d-2}{2}}\ (u\sinh t)^{-\frac{d-2}{2}}
J_{l+\frac{d-2}{2}} (i u \sinh t)
\ \sum_{M}  \Ylm(\x)  \Ylm(\xx).
\end{equation}
By inserting this expression in the above integral  representation
 we get that
\begin{align}
& \left[x\left(t+\frac {i\pi}2,\, \x\right) \cdot \xi(\xx)\right]^\lambda = \cr
& =
\frac{i^{\lambda}}{ (2\pi)^\frac{d}{2} {\Gamma}\left(-\lambda\right)
(\sinh t)^{\frac{d-2}{2}}} \sum _{l=0}^\infty
\int_{0}^{\infty}\frac{du}{u}\   u^{-\lambda-\frac{d-2}{2}}\, e^{- u \cosh t }
   I_{l+\frac{d-2}{2}} ( u \sinh t)
\  \sum_{M}  \Ylm(\x)  \Ylm(\xx).
\label{opq2}\end{align}
The integral at the RHS is the Mellin transform of a product
that can be evaluated by the Mellin-Barnes integral. This is
a way to directly check \cite[7.8 (9) p. 57]{HTF2} :
\begin{align}
& \int_{0}^{\infty}\frac{du}{u}\   u^{-\lambda-\frac{d-2}{2}}\, e^{- u \cosh t }
   I_{l+\frac{d-2}{2}} ( u \sinh t) = \cr
& =
  (\sinh t)^{\lambda + \frac{d-2}{2}} \int_{0}^{\infty}\frac{dv}{v}\
v^{-\lambda-\frac{d-2}{2}}\, e^{- \frac {v \cosh t}{\sinh t}}
I_{l+\frac{d-2}{2}} ( v) = \Gamma(l-\lambda)
P^{-l-\frac{d-2}2}_{\lambda+\frac{d-2}{2}}(\cosh t).
\end{align}
Therefore
\begin{align}
& \left[x\left(t+\frac {i\pi}2,\, \x\right) \cdot \xi(\xx)\right]^\lambda = \cr
& =
\frac{e^{\frac {i\pi\lambda}2} (2\pi)^\frac{d}{2} }{{\Gamma} \left(-\lambda\right)
\left(-i \cosh\left( t+\frac{i\pi}2\right)\right)^{\frac{d-2}{2}}}
\sum _{l=0}^\infty \Gamma(l-\lambda)P^{-l-\frac{d-2}2}_{\lambda+\frac{d-2}{2}}
\left(-i \sinh\left( t+\frac{i\pi}2\right)\right)
\  \sum_{M}  \Ylm(\x)  \Ylm(\xx)\cr
& =
\frac{ (2\pi)^\frac{d}{2} }
{\Gamma(-\lambda)
\left(\cosh\left( t+\frac{i\pi}2\right)\right)^{\frac{d-2}{2}}}
\sum _{l=0}^\infty \Gamma(l-\lambda) e^{\frac {i\pi(\lambda-l)}2}
\P^{-l-\frac{d-2}2}_{\lambda+\frac{d-2}{2}}
\left(-i \sinh\left( t+\frac{i\pi}2\right)\right)
\  \sum_{M}  \Ylm(\x)  \Ylm(\xx).
\label{opq3}\cr &\end{align}
By analytic continuation this remains valid if $t +i\pi/2$ is replaced
by any complex $t$ with $0< \Im t <\pi$, i.e. such that
$x(t,\ \x) \in \TT_+$. As a consequence of (\ref{opq3}),
\beq
\int_{S_0} (x(t,\ \x)\cdot \xi)^\lambda \Ylm(\xx)\,d\xx =
{e^{\frac {i\pi(\lambda-l)}2} (2\pi)^\frac{d}{2}\Gamma(l-\lambda)\over
\Gamma(-\lambda) (\ch t)^{d-2\over 2}}
\P_{\lambda+{d-2\over 2}}^{-l-{d-2\over 2}}(-i\sh t)\,\Ylm(\x)
\label{opq4}\endq
holds for all $l$, $M$, $0<\Im t<\pi$, and $\lambda$ such that $\Re \lambda <0$.
But since both sides  of this equation are entire in $\lambda$, it remains
valid for all $\lambda$. Since, moreover, any $C^\infty$ function on
the sphere $S_0 \simeq \dsphere$ has a convergent expansion into
hyperspherical harmonics, the validity of (\ref{opq3}) also extends
to all $\lambda$.

\subsection{The kernel $(\xi\cdot \xi')^\lambda$}

Once again, if $-(d-1)/2 < \Re \lambda < 0$,
\begin{equation}
{(\xi\cdot \xi') }^{\lambda}\, =
\frac{1}{{\Gamma}\left(-\lambda\right)}
\int_{0}^{\infty}\frac{du}{u}\   u^{-\lambda}\,
e^{- u  + u  \ \xx \cdot {\xx'}}\ .
\label{opq10}\end{equation}
Again using (\ref{opq1.1}), setting $\nu = (d-2)/2$, and using
the expansion of $C_l^\nu({ \xx}\cdot \xx')$ in
terms  of the hyperspherical harmonics, we obtain
\begin{equation}
e^{u  \, { \xx }\cdot { \xx'}} = (2\pi)^\frac{d}{2}
\sum _{l=0}^\infty \ u^{-\frac{d-2}{2}} I_{l+\frac{d-2}{2}} ( u)
\ \sum_{M}  \Ylm(\xx)  \Ylm(\xx').
\end{equation}
Inserting this into the integral representation (\ref{opq10})
we get
\beq
\left(\xi \cdot \xi'\right )^\lambda =
\frac{ (2\pi)^\frac{d}{2} }{{\Gamma}\left(-\lambda\right)}
\sum _{l=0}^\infty \int_{0}^{\infty}\frac{du}{u}\
u^{-\lambda-\frac{d-2}{2}}\, e^{- u  } I_{l+\frac{d-2}{2}} ( u)
\  \sum_{M}  \Ylm(\x)  \Ylm(\xx).
\label{opq12}\endq
The integral
 can be evaluated by the Mellin-Barnes integral:
\beq
\int_{0}^{\infty}\frac{du}{u}\   u^{-\lambda-\frac{d-2}{2}}\, e^{- u  }
 I_{l+\frac{d-2}{2}} ( u ) = 2^{\frac d 2 +\lambda+l}
\Gamma(l-\lambda){\pi}^{-\frac 12}
\frac{\Gamma(\frac{2\lambda +d-1 }{2})}{\Gamma (\lambda +d+l-1)}
\endq
 Therefore
\beq
\left(\xi \cdot \xi'\right )^\lambda =
\frac{ (4\pi)^\frac{d-1}{2} }{{\Gamma}\left(-\lambda\right)}
\sum _{l=0}^\infty 2^{\lambda+l} \Gamma(l-\lambda)
\frac{\Gamma(\frac{2\lambda +d-1 }{2})}{\Gamma (\lambda +d+l-1)}
\  \sum_{M}  \Ylm(\xx)  \Ylm(\xx')\ .
\label{opq15}\endq

\newpage
\section{Appendix. Spaces of functions}
\label{spaces}
\begin{lemma}
\label{auxlem1}
Let $N\ge 0$ be an integer and $\vhi$ be a $\CC^\infty$ function
on the sphere $S_0 = C_+ \cap \{\xi\ :\ \xi^0 = 1\}$,
such that, for every $\vec{u} \in \bR^d$,
\beq
\int_{S_0} (\vec{u}\cdot \vec{\xi})^N\,\vhi(\vec{\xi})\,d\vec{\xi} =0.
\label{sp.1}\endq
Then
\beq
\int_{S_0} p(\vec{\xi})\,\vhi(\vec{\xi})\,d\vec{\xi} =0
\label{sp.2}\endq
for every homogeneous polynomial $p$ of degree $N$ on $\bR^d$.
\end{lemma}

\noindent {\bf Proof}. Let $\alpha = (\alpha_1,\ldots,\ \alpha_d)$
be a multi-index with $|\alpha|=N$. Applying
$(\partial/\partial\vec{u})^\alpha$ to the lhs of (\ref{sp.1})
shows that $\vhi$ is orthogonal to the monomial $\vec{\xi}^\alpha$.

\begin{lemma}
\label{auxlem2}
Let $\mu\in \bC$ not be a non-negative integer,
and $\vhi$ be a $\CC^\infty$ function
on the sphere $S_0$.
Assume that,
for all $z \in \TT_+$ (or for all $z \in \TT_-$),
\beq
\int_{S_0} (z\cdot \xi)^{\mu}\,\vhi(\xi)\,d\vec{\xi} = 0.
\label{sp.10}\endq
Then $\vhi =0$.
\end{lemma}

\noindent{\bf Proof.}
If $z\in \TT_+$, i.e. $z \in M_{d+1}$, $z\cdot z = -1$ and
$\Im z \in V_+$, then (\ref{sp.10}) is equivalent to
\beq
\int_{S_0} \left (1 - \sum_{j=1}^d {z_j\over z_0}\xi_j \right )^{\mu}\,
\vhi(\xi)\,d\vec{\xi} = 0\ .
\label{sp.12}\endq
Since $(z_j/z_0)$ are independent variables in $\TT_+$, the condition
(\ref{sp.12}) is equivalent (by analytic continuation) to
the following:
\beq
\int_{S_0} \left (1-\sum_{j=1}^d u_j\xi_j \right )^{\mu}\,
\vhi(\xi)\,d\vec{\xi} = 0
\label{sp.13}\endq
for all complex $(u_1,\ldots,\ u_d)$ in the domain
\beq
\{(u_1,\ldots,\ u_d) \in \bC^d\ :\ \sum_{j=1}^d|u_j|^2<1\}
\label{sp.14}\endq
Indeed this domain is convex, hence connected, and contains
points in the image of $\TT_+$ under the map
$z\mapsto \vec{z}/z_0$ (e.g. $z = ie_0$).
Differentiating with respect to the $u_j$ and setting $u_j=0$
we find
\beq
\int_{S_0} \xi^\alpha\,\vhi(\xi)\,d\vec{\xi} = 0
\label{sp.15}\endq
for any multi-power $\alpha$, hence $\vhi = 0$. The same proof
holds if $\TT_+$ is replaced by $\TT_-$.

\begin{lemma}
\label{auxlem3}
Let $N \ge 0$ be an integer, and $\vhi$ be a $\CC^\infty$ function
on the sphere $S_0$ such that
\beq
\int_{S_0} (z\cdot \xi)^N\,\vhi(\xi)\,d\vec{\xi} = 0
\label{v.1}\endq
for all $z \in C_+$. Then
\beq
\int_{S_0} p(\vec{\xi})\,\vhi(\xi)\,d\vec{\xi} = 0
\label{v.2}\endq
for all polynomial $p$ of degree $\le N$ on $\bR^d$.
\end{lemma}

As a consequence of this lemma, $\EE_n^{(1)} \subset \EE_n^{(2)}$.
Indeed if $f \in \EE_n^{(1)}$, it satisfies (\ref{v.1}) with
$N=n$, hence (\ref{v.2}), so that $f\in \EE_n^{(2)}$. Therefore
$\EE_n^{(1)} = \EE_n^{(2)}$.

\noindent{\bf Proof of Lemma \ref{auxlem3}.}
By analytic continuation and homogeneity, (\ref {v.1}) is equivalent to
\beq
\int_{S_0} (1-\vec{u}\cdot \vec{\xi})^N\,
\vhi(\vec{\xi})\,d\vec{\xi} = 0
\label{v.3}\endq
for all $\vec{u} \in \bC^d$ such that $\sum_{j=1}^d u_j^2 = 1$.
The assertion of the lemma is obvious if $N=0$. Assume it has been
proved for $N<n$, where $n \ge 1$ is an integer. Let $\vhi$ be such
that (\ref{v.3}) holds for $N =n$. If $\vec{u}$ remains
in an open set of the sphere  where $u_d \not=0$,
we can take $u_1,\ldots,\ u_{d-1}$
as independent coordinates, with $u_d^2 = 1-\sum_{j=1}^{d-1} u_j^2$.
Denoting $I_N$ the lhs of (\ref {v.3}) and
differentiating it with respect to $u_j$, ($1 \le j\le d-1$),
we find
\beq
{\partial \over \partial u_j} I_n =
\int_{S_0} n \left ( -\xi_j + {u_j \xi_d\over u_d} \right)
(1-\vec{u}\cdot \vec{\xi})^{n-1}\,
\vhi(\vec{\xi})\,d\vec{\xi} = 0
\label{v.4}\endq
and hence
\beq
\left ( 1 -
{1\over n}\sum_{j=1}^{d-1} u_j{\partial \over \partial u_j} \right ) I_n =
\int_{S_0} \left ( 1- {\xi_d \over u_d} \right )
(1-\vec{u}\cdot \vec{\xi})^{n-1}\,
\vhi(\vec{\xi})\,d\vec{\xi} = 0
\label{v.5}\endq
There is nothing special about the index $d$, hence, for all $j$,
\beq
\int_{S_0} (u_j - \xi_j)
(1-\vec{u}\cdot \vec{\xi})^{n-1}\,
\vhi(\vec{\xi})\,d\vec{\xi} = 0\ .
\label{v.6}\endq
If $n=1$ this reduces to
\beq
\int_{S_0} (u_j - \xi_j)\,\vhi(\vec{\xi})\,d\vec{\xi} = 0\ .
\label{v.6.1}\endq
for $1\le j \le d$. Taking $u_j=0$ shows that $\vhi$ is orthogonal
to all homogeneous polynomials of degree 1, then $I_1= 0$
gives that $\int \vhi =0$.
Suppose $n \ge 2$.
Again taking $u_1,\ldots,\ u_{d-1}$
as independent coordinates, and differentiating (\ref{v.6}) with
respect to $u_j$ for $j<d$, we get
\beq
\int_{S_0} \left ( 1- \vec{u}\cdot \vec{\xi} +
(n-1)\left [ -(u_j\xi_j)\left (1+{\xi_d\over u_d} \right)
+ u_j^2 {\xi_d\over u_d} +\xi_j^2 \right ] \right )
(1-\vec{u}\cdot \vec{\xi})^{n-2}\,
\vhi(\vec{\xi})\,d\vec{\xi} = 0\ .
\label{v.7}\endq
Summing this over $j<d$ gives
\beq
\int_{S_0} \left ( d-1 + (n-1)\left ( 1+{\xi_d\over u_d} \right) \right )
(1-\vec{u}\cdot \vec{\xi})^{n-1}\,
\vhi(\vec{\xi})\,d\vec{\xi} = 0\ .
\label{v.8}\endq
Combining this with (\ref{v.5}) gives
\beq
\int_{S_0} (1-\vec{u}\cdot \vec{\xi})^{n-1}\,
\vhi(\vec{\xi})\,d\vec{\xi} = 0\ .
\label{v.9}\endq
By the induction hypothesis, $\vhi$ is orthogonal to any
polynomial in $\vec{\xi}$ of degree $<n$, hence, by expanding
(\ref{v.3}),
\beq
\int_{S_0} (\vec{u}\cdot \vec{\xi})^n\,
\vhi(\vec{\xi})\,d\vec{\xi} = 0\ .
\label{v.10}\endq
By homogeneity this holds for all $\vec{u} \in \bC^d$. 
By Lemma \ref{auxlem1}, this implies that $\vhi$ is orthogonal to any
homogeneous polynomial of degree $n$ in $\vec{\xi}$, and finally
to any polynomial of degree $\le n$. This completes the
proof of Lemma \ref{auxlem3}.

We end this appendix with an explanation of the positivity
of the kernel $(\vec{\xi}\cdot \vec{\xi'})^p$ on the unit
sphere $S_0$. This is obvious for $p=1$, and, for $p>1$,
it follows from the following classical result: the pointwise
product of two positive-semi-definite kernels, if it is defined,
is also positive-semi-definite. In the case of square matrices
this product is the Schur product (see e.g.
\cite[6.3.1, p.82]{Mehta}) and the same proof can be given
in the following simple case:
\begin{lemma}
Let $A(x,\ x')$ and $B(x,\ x')$ be two $\CC^\infty$ functions
with compact support
on $V\times V$, where $V$ is a smooth manifold, $\mu$ a positive
measure on $V$ with smooth density. Assume that $A$ and $B$ are
positive-semi-definite, i.e. 
$\int_V \ovl{f(x)}A(x,\ x')f(x')\,d\mu(x)\,d\mu(x') \ge 0$
for all smooth $f$ with compact support, and similarly for $B$. 
Then the same is true for  
$C(x,\ x') =  A(x,\ x')B(x,\ x')$.
\end{lemma}

\noindent {\bf Proof} (sketch).
$A$, $B$ and $C$ are the kernels of Hilbert-Schmidt operators
on $L^2(V,\ \mu)$. Hence
\beq
A(x,\ y) = \sum_n a_n \ovl{\vhi_n(x)}\,\ \vhi_n(y),\ \ \
B(x,\ y) = \sum_n b_n \ovl{\psi_n(x)}\,\ \psi_n(y),
\label{v.20}\endq
where $a_n \ge 0$, $b_n \ge 0$, and the $\vhi_n$ (resp. $\psi_n$)
are an orthonormal sequence of functions on $V$; they are smooth
with compact support
since $a_n\vhi_n(y) = \int_V \vhi_n(x)\,A(x,\ y)\,d\mu(x)$.
We can approximate $A$ by $A_N$ (resp.$B$ by $B_N$) obtained
by stopping the expansions (\ref{v.20}) at $n=N$ and set $C_N=A_NB_N$.
Let $f$ be a $\CC^\infty$ function with compact support on $V$.
\beq
\int_{V\times V} \ovl{f(x)}\,C_N(x,\ y)\,f(y)\,d\mu(x)\,d\mu(y) =
\sum_{n=1}^N\sum_{m=1}^N a_nb_m
\big |\int_V f(x)\vhi_n(x)\psi_m(x)\,d\mu(x)\big |^2 \ge 0\ .
\label{v.21}\endq
Taking the limit of the lhs as $N \rightarrow \infty$ shows
that $C$ is positive-semi-definite.

This lemma can easily be generalized (e.g. to the case of two distributional
kernels that happen to be pointwise multipliable, such as Wightman
functions) by a regularization-deregularization procedure.

\section{Appendix. Existence of the functions $g_l$}
\label{indep}
We need to prove, for real $\lambda$ (close to $n$),
the linear independence of the functions
\beq
y\mapsto \vhi(y,\ n)\ \ \ {\rm and}\ \ \ y\mapsto
\Re {\partial\over\partial \lambda}\,\vhi(y,\ \lambda) ,
\big |_{\lambda = n}\ ,\ \ \ y\in\bR,
\label{ba.70}\endq
where
\beq
\vhi(y,\ \lambda) = e^{{i\pi\over 2}(\lambda-l)}(1+y^2)^{-{l\over 2} -{d-2\over 4}}
\P_{\lambda+{d-2\over 2}}^{-l-{d-2\over 2}}(-iy) ,
\label{ba.71}\endq
It has already been seen that $\vhi(y,\ n)$ is real, hence
we must check that $\Re \vhi(y,\ \lambda)$ and
$\Re {\partial\over\partial \lambda}\,\vhi(y,\ \lambda)$ are linearly
independent at $\lambda = n$.
The formula \cite[3.2 (22), p. 126]{HTF1} gives
\beq
\vhi(y,\ \lambda) = e^{{i\pi\over 2}(\lambda-n)} \Big [
\vhi_1(y,\ \lambda) + \vhi_2(y,\ \lambda) \Big ],
\label{ba.73}\endq
\begin{align}
&\vhi_1(y,\ \lambda) = {e^{{i\pi\over 2}(n-l)} 2^{-l-\p}\pi^{1/2}\over
\Gamma\left({1+l-\lambda\over 2} \right )
\Gamma\left({2+2\p+l+\lambda\over 2} \right )}
F\left ( {l-\lambda\over 2}\ ,\ {1+2\p+l+\lambda\over 2}\ ;\
{1\over 2}\ ;\ -y^2 \right)
\label{ba.74}\\
& \vhi_2(y,\ \lambda) = {2iy\,e^{{i\pi\over 2}(n-l)} 2^{-l-\p}\pi^{1/2}\over
\Gamma\left({l-\lambda\over 2} \right )
\Gamma\left({1+2\p+l+\lambda\over 2} \right )}
F\left ( {1+l-\lambda\over 2}\ ,\ {2+2\p+l+\lambda\over 2}\ ;\
{3\over 2}\ ;\ -y^2 \right)
\label{ba.75}\end{align}
\beq
\partial_\lambda\vhi(y,\ \lambda)|_{\lambda = n} =
{i\pi\over 2} \vhi(y,\ n) + \partial_\lambda\vhi_1(y,\ \lambda)|_{\lambda = n}+
\partial_\lambda\vhi_2(y,\ \lambda)|_{\lambda = n}\ .
\label{w.25}\endq

\subsection{Case $n-l\ge 0$ even : $n-l = 2N\ge 0$, $N$ integer}
In this case, $\vhi_1(y,\ \lambda)$ is real, $\vhi_2(y,\ \lambda)$
and $\partial_\lambda\vhi_2(y,\ \lambda)$ are
pure imaginary, and $\vhi_2(y,\ n,) = 0$. Let
$\lambda = n+2s$ ($s$ real), so that
$\partial_\lambda = \half\partial_s$.
\beq
\vhi_1(y,\ \lambda) = U_1(\lambda)\,V_1(y,\ \lambda),
\label{w.31}\endq
\beq
U_1(\lambda) = {(-1)^N 2^{-l-\p}\pi^{1/2}\over
\Gamma\left({1\over 2}-N-s \right )
\Gamma\left(1+\p+l+N+s \right )}
\label{w.32}\endq
\beq
V_1(y,\ \lambda) = F \left (-N-s\ ,\ {1\over 2}+\p+l+N+s\ ;\
{1\over 2}\ ;\ -y^2 \right)
\label{w.33}\endq
Denoting $V_{1,1}(y,\ \lambda)$ the sum of terms of order $\le N$
in the expansion of the rhs of (\ref{w.33}) in powers of $(-y^2)$,
and $V_{1,2}(y,\ \lambda)$ the sum of terms of order $\ge N+1$, we have
\begin{align}
& \partial_\lambda V_{1,1}(y,\ \lambda)|_{\lambda=n} =
\sum_{m=0}^N{\Gamma(N+1)\Gamma \left({1\over 2}+\p+l+N+m\right )
\Gamma\left({1\over 2}\right )\,y^{2m}\over
2\Gamma(N+1-m)\Gamma \left({1\over 2}+\p+l+N\right )
\Gamma\left({1\over 2}+m\right )\Gamma(m+1)}\times\cr
&\times \left [ \psi(N+1)+ \psi\left({1\over 2}+\p+l+N+m\right )
-\psi(N+1-m) -\psi \left({1\over 2}+\p+l+N\right )\right ]\ .
\label{w.35}\end{align}
\begin{align}
&\partial_\lambda V_{1,2}(y,\ \lambda)|_{\lambda=n} = \cr
&= \sum_{m=N+1}^\infty
{(-1)^{N+1} N! (m-(N+1))!\over m!}{\Gamma \left({1\over 2}+\p+l+N+m\right )
\Gamma\left({1\over 2}\right )\,(-y^2)^{m}\over
2\Gamma \left({1\over 2}+\p+l+N\right )\Gamma\left({1\over 2}+m\right )}\ .
\label{w.36}\end{align}
Since
\begin{align}
&\vhi(y,\ n) = \vhi_1(y,\ n) = U_1(n)\,V_1(y,\ n)\ ,\cr
&\Re \partial_\lambda\vhi(y,\ \lambda)|_{\lambda=n} =
\partial_\lambda U_1(\lambda)|_{\lambda=n}\,V_1(y,\ n) +
U_1(n)\,\partial_\lambda V_1(y,\ \lambda)|_{\lambda=n}\ ,
\label{w.37}\end{align}
proving that $\vhi(y,\ n)$ and $\Re \partial_\lambda\vhi(y,\ \lambda)|_{\lambda=n}$
are linearly independent is equivalent to proving that
$V_1(y,\ n)$ and $\partial_\lambda V_1(y,\ \lambda)|_{\lambda=n}$
are linearly independent. This is obvious since $V_1(y,\ n)$ is
a polynomial of degree $N$ in $y^2$ while the expansion of
$\partial_\lambda V_1(y,\ \lambda)|_{\lambda=n}$ in powers of $y^2$
has non-zero terms of higher degree.

\subsection{Case $n-l > 0$ odd : $n-l-1 = 2N\ge 0$, $N$ integer}
In this case, $\vhi_2(y,\ \lambda)$ is real, $\vhi_1(y,\ \lambda)$
and $\partial_\lambda \vhi_1(y,\ \lambda)$ are
pure imaginary, and $\vhi_1(y,\ n,) = 0$. Setting $\lambda = n+2s$ ($s$ real),
\beq
\vhi_2(y,\ \lambda) = y\,U_2(\lambda)\,V_2(y,\ \lambda)\ ,
\label{w.61}\endq
\beq
U_2(\lambda) =
{(-1)^{N+1}2^{-l-\p+1}\pi^{1/2}\over
\Gamma\left ({1\over 2}-N-s\right )\Gamma(1+\p+l+N+s)}\ ,
\label{w.62}\endq
\begin{align}
V_2(y,\ \lambda) &=
F\left ( -N-s\ ,\ {3\over 2}+\p+N+l+s\ ;\
{3\over 2}\ ;\ -y^2 \right)
\label{w.64}\\
&= V_{2,1}(y,\ \lambda) + V_{2,2}(y,\ \lambda)\ .
\label{w.65}\end{align}
where $V_{2,1}(y,\ \lambda)$ is the sum of the terms of order $\le N$
in the expansion of (\ref{w.64}) in powers of $(-y^2)$, and
$V_{2,2}(y,\ \lambda)$ is the sum of the terms of order $\ge N+1$.
\begin{align}
&\partial_\lambda V_{2,1}(y,\ \lambda)|_{\lambda=n} =
\sum_{m=0}^N {\Gamma(N+1)\Gamma\left({3\over 2}+p+l+N+m\right )
\Gamma\left({3\over 2}\right )\, y^{2m}\over
2\Gamma(N+1-m)\Gamma\left({3\over 2}+p+l+N\right )
\Gamma\left({3\over 2}+m\right )\Gamma(m+1)} \times\cr
&\times \left [ \psi(N+1) + \psi\left({3\over 2}+p+l+N+m\right )
-\psi(N+1-m)-\psi\left({3\over 2}+p+l+N\right ) \right ]\ .
\label{w.66}\end{align}
\begin{align}
&\partial_\lambda V_{2,2}(y,\ \lambda)|_{\lambda=n} = \cr
&= \sum_{m=N+1}^\infty
{(-1)^{N+1} N! (m-(N+1))!\over m!}{\Gamma \left({3\over 2}+\p+l+N+m\right )
\Gamma\left({3\over 2}\right )\,(-y^2)^{m}\over
2\Gamma \left({3\over 2}+\p+l+N\right )\Gamma\left({3\over 2}+m\right )}\ .
\label{w.67}\end{align}
In the same way as in the preceding subsection, it follows that
$\vhi(y,\ n)$ and $\Re \partial_\lambda\vhi(y,\ \lambda)|_{\lambda=n}$
are linearly independent.

\newpage
\section{Appendix. Expansions of $W_\lambda({z_1},{z_2})$ and
$\wh W_n({z_1},{z_2})$}
\label{wexp}

\subsection{Formulae for $w_\lambda$}
In this appendix we regard $w_\lambda({z_1}\cdot{z_2})$ as a function
of $\lambda$, $z = (1+\zeta)/2 = -({z_1}-{z_2})^2/4$, and a complex
variable $\alpha$ eventually to be set equal to $(d-2)/2$.
\begin{align}
w_\lambda({z_1}\cdot {z_2}) &=
{\Gamma(-\lambda)\Gamma(\lambda+2\alpha +1)\over
(4\pi)^{\alpha+1}\Gamma(\alpha+1)}
F\left(-\lambda,\ \lambda+2\alpha+1;\alpha+1\ ;\ 1-z\right )\ ,
\label{fo.1}\\
&=
{\Gamma(-\lambda)\Gamma(\lambda+2\alpha+1)
\Gamma ( -\alpha)\over
(4\pi)^{\alpha+1}\Gamma ( \lambda+\alpha +1)
\Gamma ( -\lambda-\alpha )}
F\left(-\lambda,\ \lambda+2\alpha+1;\ \alpha+1;\ z\right ) +\cr
&+
{\Gamma ( \alpha )
z^{-\alpha}\over (4\pi)^{\alpha+1}}
F\left(\lambda+\alpha+1,\ -\lambda-\alpha\,;\
1-\alpha\,;\ z\right )\ .
\label{fo.2}\end{align}
We are interested in the behavior of $w_\lambda$ in the neighborhood of
$z=0$, which is made explicit by (\ref{fo.2}) unless $\alpha$ is
an integer. This happens if
$d$ is a positive even integer and one attempts to set $\alpha = (d-2)/2$.
This is the case on which we will focus our attention. In this appendix
$n$ will denote a fixed non-negative integer.

In addition to the well-known identity
\beq
\Gamma(u)\Gamma(1-u) = \pi/\sin(\pi u)
\label{fo.3}\endq
we will need the following formulae, where
$\psi(u) = \Gamma'(u)/\Gamma(u)$ and $p \ge 0$ is an integer:
\begin{align}
{\psi(u) \over \Gamma(u)} &= {\Gamma'(u)\over \Gamma(u)^2}
= -{\partial \over \partial u} \Gamma(u)^{-1} =
-{\partial \over \partial u} \pi^{-1}\sin(\pi u) \Gamma(1-u)\cr
&= -\cos(\pi u) \Gamma(1-u) + \pi^{-1}\sin(\pi u) \Gamma'(1-u),\cr
\left . {\psi(u) \over \Gamma(u)} \right |_{u= -p} &= -p!\,\cos(p\pi)
= (-1)^{p+1} p!\ ;
\label{fo.4}\end{align}
\begin{align}
&
{\partial \over \partial u}\left [ {\psi(u) \over \Gamma(u)} \right ]
= \pi\sin(\pi u)\Gamma(1-u) +2\cos(\pi u) \Gamma'(1-u)
- \pi^{-1}\sin(\pi u) \Gamma''(1-u)\ ,\cr
&
{\partial \over \partial u}\left .
\left [ {\psi(u) \over \Gamma(u)} \right ]
\right |_{u= -p} = 2\cos (p\pi) \Gamma'(1+p) =
2(-1)^p p!\,\psi(1+p)\ ,
\label{fo.5}\end{align}
and, for any $x\in \bC$,
\beq
{\Gamma(-x+p)\over \Gamma(-x)}  =
{(-1)^p \Gamma(x+1)\over  \Gamma(x-p+1)}\ .
\label{fo.6}\endq

\subsection{Pole removal procedure at $\lambda = n$}
In sect. \ref{nlimit} $\wh w_n({z_1}\cdot {z_2})$ was defined
by applying a ``pole removal procedure'' (PRP) to
$w_\lambda({z_1}\cdot {z_2})$. This procedure can be applied to
any meromorphic function $f$ that has (at worst) a simple pole
at $n$:
\beq
\wh f(n) = \left . {(-1)^{n+1}\over n!}{\partial \over \partial \lambda}
\left ( {f(\lambda) \over \Gamma(-\lambda)} \right )
\right |_{\lambda= n}\ .
\label{pr.1}\endq
Taking
\beq
f(\lambda) = {a\over \lambda -n} +h(\lambda),
\label{pr.2}\endq
where $h$ is holomorphic at $n$, one gets
\beq
\wh f(n) = a\psi(1+n) + h(n)\ .
\label{pr.3}\endq
In particular if $a=0$, i.e. if $f$ is actually holomorphic at $n$,
then $\wh f(n) = f(n)$, i.e. the procedure does nothing.
Operating with another function than $\Gamma(-\lambda)$ would simply
change the coefficient of $a$.

\subsection{The case of even dimension}
\label{even}

We fix an integer $N \ge 0$.
If $\alpha$ tends to $N$, the first form (\ref{fo.1})
is regular, but each of the two terms in (\ref{fo.2}) has a pole in
$\alpha$. There must therefore
be a cancellation of the poles in the two terms of (\ref{fo.2}).
This may also be understood by representing the value of (\ref{fo.1})
by a Cauchy integral (in the variable $\alpha$) along a small circle
centered at $N$. On this contour both terms in (\ref{fo.2}) are regular
and the two hypergeometric functions may be expanded into their power
series at 0. We will use a more explicit way of exhibiting the
cancellation, but the preceding remark shows that the forthcoming
expansions are uniformly convergent in $\bU \bydef \{z\ :\ |z| < 1\}$.

Expanding the two hypergeometric functions in powers of $z$ shows
that, near $\alpha = N$,
\beq
w_\lambda({z_1}\cdot {z_2}) = \sum_{m=0}^{N-1} c_m(\lambda,\ \alpha) z^{m-\alpha}
+ g(\lambda,\ \alpha){f_1(\lambda,\ \alpha,\ z) +f_2(\lambda,\ \alpha,\ z)
\over \sin(\pi\alpha)}\ ,
\label{e.1}\endq
where (using (\ref{fo.6}))
\beq
c_m(\lambda,\ \alpha) =
{(-1)^m\Gamma(\alpha-m) \Gamma(\lambda+\alpha+1+m)
\Gamma(-\lambda-\alpha +m) \over (4\pi)^{\alpha+1}
\Gamma(\lambda+\alpha+1) \Gamma(-\lambda-\alpha)\,m!}\ ,
\label{e.2}\endq
\begin{align}
&g(\lambda,\ \alpha) = {\pi\over (4\pi)^{\alpha+1}
\Gamma(\lambda+\alpha+1) \Gamma(-\lambda-\alpha)}\ ,\cr
&f_1(\lambda,\ \alpha,\ z) = -\sum_{m=0}^\infty
{z^m \Gamma(-\lambda+m)
\Gamma(\lambda+2\alpha +1+m) \over \Gamma(\alpha+1+m)\Gamma(m+1)}\ ,\cr
&f_2(\lambda,\ \alpha,\ z) = \sum_{m=0}^\infty
{z^{m+N-\alpha}
\Gamma(\lambda+\alpha+1 +m+N) \Gamma(-\lambda-\alpha +m+N) \over
\Gamma(1-\alpha+m+N) \Gamma(m+N+1)} \ .
\label{e.3}\end{align}
All of these functions are regular in $\alpha$ near $N$, and
$f_1(\lambda,\ N,\ z) +f_2(\lambda,\ N,\ z) =0$. Hence
\beq
\left. w_\lambda({z_1}\cdot {z_2}) \right |_{\alpha = N} =
\sum_{m=0}^{N-1} c_m(\lambda,\ N) z^{m-N}+
g(\lambda,\ N) \left [
{\partial_\alpha f_1(\lambda,\ N,\ z)
+\partial_\alpha f_2(\lambda,\ N,\ z) \over
\pi\cos(\pi N)} \right ]\ .
\label{e.4}\endq
With the help of (\ref{fo.6}) this gives
\begin{align}
&w_\lambda({z_1}\cdot {z_2}) =  \sum_{m=0}^{N-1}
{z^{m-N}\Gamma(N-m) \Gamma(\lambda +N+1+m) \over
(4\pi)^{N+1}\Gamma(\lambda +N-m+1)m!} +\cr
& + \sum_{m=0}^\infty
{(-1)^mz^m \Gamma(\lambda +2N +1+m)\over
(4\pi)^{N+1} \Gamma(1+m) \Gamma(N+1+m) \Gamma(\lambda-m+1)}\times\cr
&\Big [ -\log z + \psi(1+m) -\psi(\lambda + 2N+1+m)
+\psi(N+1+m) -\psi (-\lambda +m) \Big ]\ .
\label{e.10}\end{align}

\subsubsection{Applying the pole removal procedure at $\lambda=n$}
Only the terms in (\ref{e.10}) containing $\psi(-\lambda +m)$,
with $m \le n$ fail to be analytic in $\lambda$ at the value $n$.
All terms in the infinite series with $m > n$ vanish at $\lambda = n$.
Using (\ref{fo.4}), (\ref{fo.5}), and (\ref{fo.6}),
it is straightforward to apply the procedure to the relevant terms,
and we obtain:
\begin{align}
&
\wh w_n({z_1}\cdot {z_2}) = \sum_{m=0}^{N-1}
{z^{m-N}\Gamma(N-m) \Gamma(n +N+1+m) \over
(4\pi)^{N+1}\Gamma(n +N+1-m)\Gamma(m+1)} +\cr
& + \sum_{m=0}^n
{(-1)^mz^m \Gamma(n +2N +1+m)\over
(4\pi)^{N+1} \Gamma(1+m) \Gamma(N+1+m) \Gamma(n-m+1)}\times\cr
& \times \Big [ -\log z + \psi(1+m) +\psi(N+1+m)
-2\psi(n+2N+1+m) -\psi (1+n) \Big ]\ .
\label{k.6}\end{align}
Of course this identity extends to the whole cut-plane $\bC\setminus \bR_-$.
By using (\ref{fo.6}), the coefficient of $-\log z$ can be rewritten as
\beq
{\Gamma(n+2N+1)\over (4\pi)^{N+1}\Gamma(N+1)\Gamma(n+1)}
F \left ( -n,\ n+2N+1\ ;\ N+1\ ;\ z \right )\ .
\label{k.7}\endq
Thus, for even dimension $d \ge 2$,
\beq
\wh w_n({z_1}\cdot {z_2}) = z^{1-{d\over 2}} A(z,\ n,\ d) -\log(z) B(z,\ n,\ d)
+ C(z,\ n,\ d),
\label{k.8}\endq
where $A$, $B$, $C$ are polynomials in $z$:
\begin{align}
A(z,\ n,\ d) &= \sum_{m=0}^{{d\over 2}-2}
{z^{m}\Gamma \left ({d\over 2}-1-m \right)
\Gamma \left (n +{d\over 2}+m \right) \over
(4\pi)^{d\over 2}\Gamma(n +{d\over 2}-m)m!}\ ,\cr
B(z,\ n,\ d) &=
{\Gamma \left ({d-1\over 2}\right )\over
4 \pi^{d+1\over 2}}
C_n^{d-1\over 2}(1-2z)\ ,\cr
C(z,\ n,\ d) &=
\sum_{m=0}^n
{(-1)^mz^m \Gamma(n +d-1+m)\over
(4\pi)^{d\over 2} \Gamma(1+m) \Gamma\left ({d\over 2}+m \right)
\Gamma(n-m+1)}\times\cr
& \times \left [\psi(1+m) +\psi \left ({d\over 2}+m \right )
-2\psi(n+d-1+m) -\psi (1+n) \right ]\ .
\label{k.9}\end{align}

%\end{document}

\end{document}